\documentclass[12pt] {article}		% Must use LaTeX 2e
\usepackage{epsfig}
\newlength{\figurewidth}
\normalsize
\newcommand{\be}{\begin{equation}}
\newcommand{\ee}{\end{equation}}
\newcommand{\bea}{\begin{eqnarray}}
\newcommand{\eea}{\end{eqnarray}}
\newcommand{\gsim}{\hbox{ \raise3pt\hbox to 0pt{$>$}\raise-3pt\hbox{$\sim$} }}
\newcommand{\lsim}{\hbox{ \raise3pt\hbox to 0pt{$<$}\raise-3pt\hbox{$\sim$} }}
\newcommand{\mathbold}[1]{\mbox{\boldmath $\bf#1$}}
\begin{document}
%--
%  Title
%--

\begin{titlepage}
\hfill
%\begin{minipage}[t]{2.0in}
%\begin{flushleft}
%\bf
%KEK Preprint 98-187 \\
%DPNU 98-42 \\
%KUDP 98-01 \\
%TUAT-HEP 98-02 \\
%KINDAI-HEP 98-01 \\
%November 1998 \\
%H \\
%\end{flushleft} 
%\end{minipage}

%\vskip 2.0cm
\vskip 4.0cm
\begin{center}
\large \bf
A Handy Tool for History Keeping of Geant4 Tracks\\ 
and its Application to Studies of
Fundamental Limits\\
on PFA Performance
\end{center}

\vskip 1.5cm

\begin{center}
\large 
$Sumie$ $Yamamoto^a$, 
$Keisuke$ $Fujii^{b}$\footnote{
Corresponding authhor.\\
E-Mail address: keisuke.fujii@kek.jp\\
TEL: +8-29-864-5373\\
FAX: +8-29-864-2580
}, and 
$Akiya$ $Miyamoto^b$,
\end{center}

\vskip 0.7cm

\begin{center}
$^a$ School of High Energy Accelerator Science, 
 The Graduate University for Advanced Studies (Sokendai),
 Tsukuba, 305-0801, Japan\\ 
$^b$ High Energy Accelerator Research Organization (KEK),
 Tsukuba, 305-0801, Japan\\ 
\end{center}

\vskip 1cm

\begin{abstract}
It is widely recognized that good jet energy resolution is 
one of the most important requirements to the detectors
for the future linear $e^+e^-$ collider experiments.
The Particle Flow Analysis (PFA) is currently under intense studies
as the most promising way to 
achieving the best attainable resolution.
In order to clarify the fundamental limits on the jet energy resolution
with the PFA, we have developed a set of C++ classes that facilitates
history keeping of particle tracks within the framework of Geant4.
In this paper this software tool is described and applied to
a generic detector model so as to identify fundamental limiting factors
to the PFA performance. 
\end{abstract}
\vfil
Keywords: Geant4, History Keeping, PFA\\
PACS code: 07.05.Tp, 02.70.Lq

\end{titlepage}

%--
%  Text
%--
%\baselineskip 24pt 

%  ------------
%  Introduction
%  ------------
\section{Introduction}

The experiments at the International Linear Collider\cite{Ref:ILC}
will open up a novel possibility to reconstruct all the final states
in terms of fundamental particles (leptons, quarks, and gauge bosons)
as viewing their underlying Feynman diagrams.
This involves identification of heavy unstable particles
such as $W$, $Z$, $t$, and even yet undiscovered new
particles such as $H$ through jet invariant-mass measurements.
The goal is thus to achieve an jet invariant-mass resolution
comparable to the natural width of $W$ or $Z$\cite{Ref:JLC-I}. 
High resolution jet energy measurements will thus be crucial,
necessitating high resolution tracking and calorimetry
as well as an algorithm to make full use of available information
from them.
With a currently envisaged tracking system that aims at a momentum
resolution of $\sigma_{p_T} / p_T = 5 \times 10^{-5} p_T [{\rm GeV/c}]$ or better,
tracker information will be much more accurate than that from calorimetry
for charged particles.
This implies that the best attainable jet energy resolution should be achieved 
when we use the tracker information for charged particles and
the calorimeter information only for neutral particles.
This requires separation of calorimeter clusters due to
individual particles and in the case of charged particle clusters
their one-to-one matching to the corresponding tracks detected in the tracking system.
This is the so-called Particle Flow Analysis (PFA) currently under intense studies\cite{Ref:PFA}.

For the PFA, it is hence desirable to have a highly granular calorimeter that
allows separation of clusters due to a densely packed jet of particles. 
In practice the performance of PFA depends not only on the hardware design
of the detector system consisting of the tracker and the calorimeter
but also on a particular algorithm one
employs to separate calorimeter signals due to neutral particles from
those due to charged particles.
Various realistic algorithms are currently being tested by various groups\cite{Ref:PFA}
and it is probably fair to say that they are still immature. 
Discussions on the realistic PFA is hence beyond the scope of this paper.
Instead we concentrate, in this paper, on clarifying the fundamental limitations
to the PFA performance that would be achieved with an ideal algorithm,
so as to set the ultimate goal for the PFA and 
to help identifying key factors for algorithm improvements.
Our studies have been carried out using a full
Monte Carlo simulator called JUPITER\cite{Ref:acfa,Ref:Jupiter}
based on Geant4\cite{Ref:geant4toolkit}
with a newly developed tool for history keeping of Geant4 tracks
together with a smearing and reconstruction package called SATELLITES\cite{Ref:Jupiter}.
Both JUPITER and SATELLITES are run under a modular analysis 
framework called JSF\cite{Ref:JSF} which is based on
ROOT\cite{Ref:root}.

The paper is organized as follows.
%We begin with listing up possible limiting factors that
%determine the PFA performance.
We begin with defining our problems and concept of so-called
{\it Cheated PFA}.
We then describe our software tool to keep history of particle tracks
(Geant4 tracks) traced through a detector in Geant4
with emphasis put on design philosophy.
The subsequent  section is devoted to their applications
to studies of fundamental limits on the PFA performance
for three processes: $e^+e^- \to q \bar{q}$, $e^+e^- \to ZZ$, 
and $e^+e^- \to ZH$
generated with the Pythia6 event generator\cite{Ref:pythia}.
%which is followed by presentation of test results and discussions.
Finally section \ref{Sec:conclusions} summarizes our
achievement and concludes this paper.

%  ------------
%  Statement of the Problems
%  ------------
%\section{Statement of the Problems and Concept of Cheated PFA}
\section{Concept of Cheated PFA}

% - Limiting factors
% - Dependence on calorimeter structure
% 	- cell structure (hit merge)
%	- layer structure (sampling frequency, thickness, material)
% 	- B, R, tower or clx ?
% - Dependence on E, processes
%	- e+e- --> qqbar, ZH

The purpose of this paper is to clarify what limits the jet energy resolution
so as to help optimizing PFA
as well as to know the ultimate jet energy resolution attainable with
an ideal particle flow algorithm,
thereby setting the performance goal for the realistic PFA.
In principle Monte Carlo simulations allow us to use so-called
Monte-Carlo truth and enable us to unambiguously separate
calorimeter hits due to different incident particles,
thereby performing perfect clustering.
By linking so-formed calorimeter clusters to corresponding 
charged particle tracks in the tracking system again using
Monte-Carlo truth, we can achieve the situation with the perfect PFA.
Hereafter we call this Cheated PFA (CPFA) since it involves cheating
by using Monte-Carlo truth, which is impossible in practice.

\subsection{Perfect Clustering and Perfect Track-Cluster Matching}

\begin{figure}[ht]
\begin{center}\begin{minipage}{\figurewidth}
\centerline{
\epsfig{file=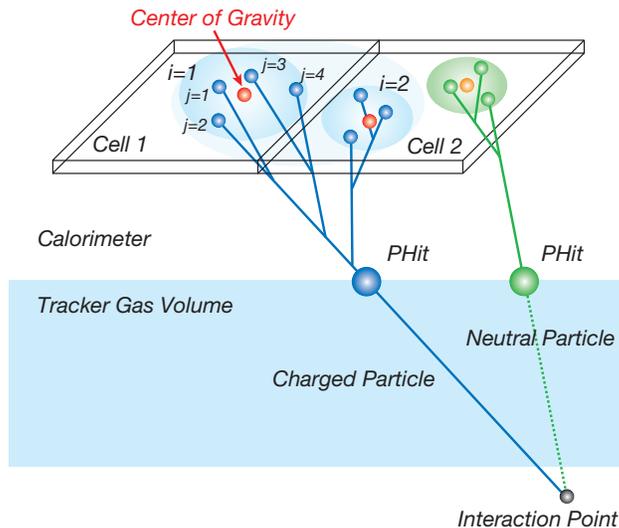,height=7cm}}
\caption[Fig:cpfa]{\label{Fig:cpfa} \small \it
Schematics showing the cheated PFA concept. 
Only two cells in a single sampling layer of the calorimeter are shown 
to simplify the picture though in practice much more cells are expected to be hit
over different sampling layers.
}
\end{minipage}\end{center}
\end{figure}

For the CPFA, 
the history of Geant4 tracks should be kept on a track-by-track basis
starting from a primary track at the interaction point.
The history of all the secondary tracks together with the original one
should be recorded until they hit  any one of pre-registered boundaries
beyond which particles may start showering.
At such a boundary we create a virtual hit called {\tt PHit}.
Calorimeter hits by Geant4 tracks in a particle shower 
will then be tagged with this {\tt PHit}. 
By collecting all the calorimeter hits with the same {\tt PHit}
we can hence form a calorimeter cluster without any confusion
(see Fig.\ref{Fig:cpfa}).
%
% ------------------
%  Fig.1
% ------------------
%

Since the {\tt PHit} carries the information of its parent track,
one-to-one matching between the calorimeter cluster and
its corresponding charged particle track in the tracking system
is possible.
Once matched, we just lock the calorimeter cluster as linked to a
charged track and just use the tracker information.
Calorimeter clusters with no matching charged tracks are hereafter
called neutral PFOs, while all the charged tracks are called
charged PFOs regardless of
whether there are corresponding calorimeter clusters or not.

It is also important to record the mother-daughter correspondence 
for particles decayed in a tracking volume
so as to estimate their effects on the PFA performance.
The mother-daughter correspondence is book-kept together with
the other information on the daughter track such as its particle ID, 
position, and momentum,
in a so-called {\tt BreakPoint} object which is created at the beginning
of each track.
The information stored in the {\tt BreakPoint} objects will be used
to follow particle decays observed as kinks or V$^0$s 
in the tracking volume and to
assign correct particle masses to charged PFOs.
This book-keeping comprises the major role of the history keeper.

\subsection{Infinite Calorimeter Segmentation}

For a realistic calorimeter design, 
the granularity of the calorimeter, or equivalently the cell size, is
finite and hence the signals created by shower particles
stemming from different parent particles sometimes merge into a single hit
degrading the cluster separation capability.
In order to investigate to what extent this limits the PFA performance,
we need to know the performance expected for perfect cluster separation.
It is, however, impracticable to implement infinitely fine segmentation
even in a Monte Carlo detector simulator.
In order to overcome this drawback,
we exploit the following trick.

In each hit cell, say cell $i$, 
we separately store the energy sum of hits originating from the same {\tt PHit}:
\begin{eqnarray}
E^{c}_i & = & \sum_{j} \, E_{i, j}  
\nonumber
\end{eqnarray}
and their center of gravity: 
\begin{eqnarray}
\mathbold{x}^{c}_i & = & \sum_{j} \, E_{i, j} \, \mathbold{x}_{i, j}  \, / \, E^{c}_i  ,
\nonumber
\end{eqnarray}
instead of using the cell center as the hit position.
In the above expression $E_{i, j}$ and $\mathbold{x}_{i, j}$ are the energy deposit 
and the position of $j$-th hit in cell $i$ with the same {\tt PHit}.
Denoting the total energy of the cluster originating from the same {\tt PHit} by
\begin{eqnarray}
E^{c} & = & \sum_{i} \, E^{c}_{i}   ,
\nonumber
\end{eqnarray}
we can then calculate its cluster center as
\begin{eqnarray}
\mathbold{x}^{c} & = & \sum_{i} \, E^{c}_{i} \, \mathbold{x}^{c}_{i} \, / \, E^{c}  
~~ = ~~ \sum_{i}  \sum_{j} \, E_{i, j} \, \mathbold{x}_{i, j} \, / \, \sum_{i} \, \sum_{j} \, E_{i, j}   ,
\nonumber
\end{eqnarray}
showing that the center of gravity calculated this way 
precisely coincides the one would-be obtained when the segmentation is infinitely fine.
It should be also emphasized here that 
hits from different {\tt PHit}s make multiple centers of gravity in the same cell,
which can be later separated even though they are in the same cell,
thereby realizing the infinite segmentation in effect.

%  ------------
%  Tool Design
%  ------------
\section{Tool Design}
%\subsection{Guideline}

Before coding our tool for history keeping, we set the following guideline
to fulfill the required functionality discussed in the last section:
a) there must be a versatile mechanism to register user-defined physical volumes 
	whose specified boundaries can be used to define a {\tt PHit} that marks the 
	source point of a particle shower,
b) whether a track is allowed to create a {\tt PHit} or not depends on 
	whether the track originates from any pre-created {\tt PHit} or not,  
	which should be checked on a track-by-track basis at the beginning of its tracking, 
c) the history keeping is to be done on the track-by-track basis by creating
	a {\tt BreakPoint} object at the beginning of each track if there is no {\tt PHit}
	from which the track stems, and
d) the history keeping should be realized making maximum use of 
	existing Geant4 facilities within the framework of JUPITER,
e) JUPITER should produce Monte-Carlo truths (i.e. exact hits) and their smearing
	should be done later in SATELLITES as needed.

\begin{flushleft}
The following is a sketch of the tool design we adopted according to the guideline:
\end{flushleft}
\begin{enumerate}
\item The history keeping is to be done on the track-by-track basis using {\tt J4TrackingAction}
	that inherits from {\tt G4UserTrackingAction}. 
	Its {\tt PreUserTrackingAction} method is hence called
	at the beginning of a new track.
	The {\tt PreUserTracingAction} method serially invokes {\tt PreTrackDoIt} method of
	each offspring of {\tt J4VSubTrackingAction} pre-registered to the {\tt J4TrackingAction}
	object. 
	Likewise, its {\tt Clear} method serially invokes {\tt Clear} method of
	individual offsprings of {\tt J4VSubTrackingAction}.  
\item {\tt J4VSubTrackingAction} is an abstract class that serves as a base class for 
	user-defined sub-actions 
	taken by {\tt J4TrackingAction} thereby extending the 
	{\tt G4UserTrackingAction} functionality.
	It has a method called {\tt Clear} to reset the object state. 
\item {\tt J4HistoryKeeper} is implemented as a derived class from
	{\tt J4VSubTrackingAction}
	and, in its {\tt PreTrackDoIt} method,
	scans through a collection of pre-registered {\tt J4PHitKeeper} objects corresponding
	to a collection of bounding surfaces.
	It then  creates a {\tt J4BreakPoint} object 
	if none of them has been
	hit by any ancestors of the new track, 
\item {\tt J4PHitKeeper} also inherits from {\tt J4VSubTrackingAction}.
	Its {\tt PreTrackDoIT} method  checks if this new track is 
	stemming from any pre-created {\tt PHit}, and, if not, resets its sate
	to allow creation of a new {\tt PHit}.
	When its corresponding boundary is hit by the current track,
	a {\tt PHit} object is created, if it is allowed,
	to tag subsequent daughter tracks possibly created in a shower.
\end{enumerate}

%\subsection{Jupiter, Uranus, and Satellites}

\subsection{Extension of  G4UserTrackingAction}

The {\tt G4UserTrackingAction} class provides one with a handy tool 
to perform a user-defined action on a track-by-track basis.
In its original form, however, it allows only a single action.
In order to extend its functionality to accept multiple user-defined actions,
we have introduced the concept of {\tt SubTrackingAction} as sketched above.
\begin{flushleft}
\underline{{\tt J4VSubTrackingAction}}
\end{flushleft}
The abstract base class {\tt J4VSubTrackingAction} has the following methods: 
\begin{description}
	\item {\tt void PreTrackDoIt(const G4Track *aTrack = 0) = 0~}\\
		which is pure virtual and to be implemented by
		its derived class to take a sub-tracking action for the given track.
	\item {\tt void Clear()~}\\
		which does nothing, and to be overridden in the derived class as needed.
\end{description}
This class just specifies the interface and requires its users to implement
the methods listed above.

\begin{flushleft}
\underline{{\tt J4TrackingAction}}
\end{flushleft}
The {\tt J4TrackingAction} class is a singleton inheriting from {\tt G4UserTrackingAction}.
It has, among others, an STL vector as a data member to store pointers to objects 
derived from the {\tt J4VSubTrackingAction} class.
Its major methods include the following:
\begin{description}
	\item {\tt static J4TrackingAction *GetInstance()~}\\
		which returns the pointer to the single instance of {\tt J4TrackingAction}.
	\item {\tt void Add(J4VSubTrackingAction *stap)~}\\
		which registers a user-defined object derived from {\tt J4VSubTrackingAction}.
		When {\tt *stap} has already been registered, the pre-registered one is erased
		and the new entry is appended.
	\item {\tt void PreUserTrackingAction(const G4Track *aTrack)~}\\
		which loops over the registered offsprings of {\tt J4VSubTrackingAction}
		and invokes their {\tt PreTrackDoIt} methods.
	\item {\tt void Clear()~}\\
		which loops over the registered offsprings of {\tt J4VSubTrackingAction}
		and invokes their {\tt Clear} methods.
\end{description}

\subsection{P-Hits and P-Hit Keeper}

A {\tt PHit} is a generic name for a {\tt Pre-Hit} or a {\tt Post-Hit}, which
stands for a virtual hit created
on a boundary of a {\tt G4PhysicalVolume}
beyond which particle showering is expected.
The {\tt PHit} creation is done in the user-overridden {\tt ProcessHits} method
of a user-defined virtual detector derived from
{\tt G4SensitiveDetector} corresponding to the physical volume.
Notice that {\tt PHit}s are created for all kinds of particles, even neutrinos, that pass through
the boundary.
One {\tt PHit} class is defined inheriting from the {\tt J4VTrackHit} class
for each such boundary.
The {\tt J4VTrackHit} class carries basic track hit information such as
track ID, particle ID, position, momentum, TOF, energy deposit, etc. and 
setters and getters to access them.
An individual {\tt PHit} class has a data member to store {\tt PHit} ID and
a static data member to store the current {\tt PHit} ID, 
which can be retrieved by a static method to mark calorimeter hits as needed.

\begin{flushleft}
\underline{{\tt J4PHitKeeper}}
\end{flushleft}
The {\tt J4PHitKeeper} class inherits from {\tt J4VSubTrackingAction}.
It serves as a base class for a {\tt PHitKeeper} class defined for
an individual {\tt PHit} class corresponding to a boundary beyond which
particle showering is expected.
The {\tt J4PHitKeeper} class has data members to store i) the current incident track ID 
({\tt fInTrackID}) that is expected to create or has already created a {\tt PHit}, 
ii) the track ID ({\tt fTopTrackID}) of the next track to be processed, if any,  
after the offsprings from the {\tt PHit} are exhausted,
and iii) a flag ({\tt fIsPHitCreated}) to tell whether a {\tt PHit} has been created or not.
The major methods of {\tt J4PHitKeeper} are listed below:
\begin{description}
	\item {\tt void PreTrackDoIt(const G4Track *)~}\\
		implements the corresponding base class pure virtual method
		so as to reset {\tt fInTrackID} and {\tt fTopTrackID} to {\tt INT\_MAX}
		and {\tt fIsPHitCreated} to {\tt FALSE} upon encountering a new track
		which has a track ID smaller than {\tt fTopTrackID}.
	\item {\tt G4bool IsNext()~}\\
		returns {\tt FALSE} if a {\tt PHit} has already been created.
		If not, it updates {\tt fInTrackID} and {\tt fTopTrackID} and
		returns {\tt TRUE} to tell the caller (the {\tt ProcessHit} method of the 
		sensitive detector defining the virtual boundary) 
		that a new {\tt PHit} is to be created.
	\item {\tt void Reset(G4int k = INT\_MAX)~}\\
		resets {\tt fInTrackID} and {\tt fTopTrackID} to {\tt k}.
	\item {\tt G4bool IsPHitCreated()~}\\
		returns {\tt fIsPHitCreated}, which is {\tt TRUE} if a {\tt PHit} has been created,
		and {\tt FALSE} otherwise.
\end{description}
The algorithm of {\tt J4PHitKeepr} heavily depends on Geant4's default track stacking scheme,
which is worth explaining here for readers unfamiliar to it.
By default Geant4 uses two types of track stacks, a Primary Stack ({\it PS}) 
and a Secondary Stack ({\it SS}).

At the beginning of each event, primary particles $1, \cdots, n$ are pushed into {\it PS}.
According to the "last in first out" rule, the top entry, track $n$, is popped out for tracking.
Notice that there remains $n-1$ tracks in {\it PS} at this point.
All the secondary particles produced while track $n$ is being processed are
pushed into {\it SS}.
Let us assume that there will be $m$ secondary particles stacked into {\it SS} 
by the time track $n$ is disposed of.
All of these $m$ secondary particles in {\it SS} are moved to {\it PS} upon the death
of track $n$ and numbered serially as track $n+1, \cdots, n+m$.
Notice that there are $n+m-1$ tracks in {\it PS} at this point since
track $n$ has been popped out and disposed of.

The key point is to bookmark the secondary track which is to be created just
after the creation of a {\tt PHit} by the track which has been being processed,
track $n$ in the present case.
The track ID with the bookmark will be ${\tt fTopTrackID} = n+k'+1$ 
where $k' (\le m)$ is the number of secondary particles in {\it SS} at the time of
the {\tt PHit} creation.
Further {\tt PHit} creation is to be forbidden until it becomes necessary.

The top of the stack, track $n+m$, is popped out and to be processed as before.
Track $n+m$ will produce further $m'$ secondary particles to be pushed into
{\it PS} upon its death and to be numbered as track $n+m+1, \cdots, n+m+m'$.

This procedure is repeated and after some time 
all the secondary particles originating from the track created the last
{\tt PHit} will be disposed of and
the next track to be popped out from {\it PS} will have a track ID that is
smaller than that of the last bookmarked one, {\tt fTopTrackID}.
This signals a new incident track which is allowed to create a new {\tt PHit}.
By repeating this procedure until all the tracks in {\it PS} are exhausted,
we can mark all the calorimeter hits with corresponding {\tt PHit}s.

\subsection{Break Points and History Keeper}

The purpose of the history keeper is to allow us to trace back to
kink and V$^0$ particles that decay before entering calorimeters
so as to correctly link clusters to tracks.
As sketched above, the history keeper is implemented as a
{\tt J4VSubTrackingAction} so as to create
a {\tt J4BreakPoint} object for each new track until
a {\tt PHit} is created on any of the pre-registered boundaries
beyond which particle-shwering is expected.

\begin{flushleft}
\underline{{\tt J4BreakPoint}}
\end{flushleft}
The {\tt J4BreakPoint} class has data members to store
the information about a track at its starting position such as
track ID ({\tt fTrackID}), parent track ID ({\tt fParentID}), 
charge, particle ID,  time, position, 4-momentum, etc..
In addition it has a static data member called {\tt fgBreakPointMap},
which is an STL map that links track ID to a {\tt J4BreakPoint} object.
Besides the getters to these data members, 
{\tt J4BreakPoint} has the methods listed below:
\begin{description}
	\item {\tt static J4BreakPoint *GetBreakPoint(G4int trackID)~}\\
		returns the pointer to the {\tt J4BreakPoint} object
		corresponding {\tt trackID}.
	\item {\tt static void Clear()~}\\
		clears the track-to-break-point map.
\end{description}

\begin{flushleft}
\underline{{\tt J4HistoryKeeper}}
\end{flushleft}
The {\tt J4HistoryKeeper} class is a singleton that inherits from {\tt J4VSubTrackingAction}.
It has an STL vector ({\tt fPHitKeepers})
as a data member to store registered {\tt J4PHitKeeper}s that
correspond to boundaries beyond which particle-showering is expected.
As sketched above, 	it scans through these pre-registered {\tt J4PHitKeeper} objects
to make sure that none of them has a {\tt PHit}, 
and then creates a {\tt J4BreakPoint} object.
The major methods of {\tt J4HistoryKeeper} are listed below:
\begin{description}
	\item {\tt static J4HistoryKeeper *GetInstance()~}\\
		returns the pointer to the single instance of {\tt J4HistoryKeeper}.
	\item {\tt void PreTrackDoIt(const G4Track *)~}\\
		implements the corresponding base class pure virtual method.
		It scans through the pre-registered {\tt J4PHitKeeper} objects
		in {\tt fPHitKeepers} to make sure that none of them has a {\tt PHit}
		by calling their {\tt IsPHitCreated()} method.
		It then creates a {\tt J4BreakPoint} object.
	\item {\tt void Cleart()~}\\
		calles {\tt J4BreakPoint::Clear()}.
	\item {\tt void SetPHitKeeperPtr(J4PHitKeeper *phkp)~}\\
		pushes back the input {\tt J4PHitKeeper} pointer into {\tt fPHitKeepers}.
\end{description}

\begin{flushleft}
\underline{{\tt S4BreakPoint}}
\end{flushleft}
Upon the completion of Monte Carlo truth generation by JUPITER,
each {\tt J4BreakPoint} object is copied to its SATELLITE dual,
an {\tt S4BreakPoint} object.
The {\tt S4BreakPoint} object inherits from ROOT's {\tt TObjArray} and
stores pointers to its daughter {\tt S4BreakPoint}s, if any.
It has additional methods such as
\begin{description}
	\item {\tt void LockAllDescendants()~}\\
		which flags all of its descendants as locked. This functionality proves handy to
		avoid double counting of energies.
	\item {\tt TObject *GetPFOPtr()~}\\
		which returns the pointer to its corresponding Particle Flow Object (PFO), if any.\\
	\item {\tt void SetPFOPtr(TObject *pfop)~}\\
		which is the setter corresponding to {\tt GetPFOPtr} to be
		invoked from a PFO maker.
\end{description}

\section{Tool Usage}

What a tool user has to do for the history keeping is as follows:
\begin{itemize}
\item Inheriting {\tt G4SensitiveDetector},
	create a sensitive detector class, say {\tt J4XXSD}, 
	that corresponds to 
	a boundary on which a {\tt PHit} object ({\tt J4XXPHit}) is to be created 
	for each particle that is expected to produce a shower
	beyond that boundary.
\item Inheriting {\tt J4PHitKeeper},
	create a {\tt J4XXPHitKeeper} as a singleton to bookkeep {\tt J4XXPHit}s.
\item In {\tt J4XXSD}'s constructor, do
\begin{verbatim}
	  J4XXPHitKeeper *phkp = J4XXPHitKeeper::GetInstance();
	  J4TrackingAction::GetInstance()->Add(phkp);
	  J4TrackingAction::GetInstance()->Add(J4HistoryKeeper::GetInstance());
	  J4HistoryKeeper::GetInstance()->SetPHitKeeperPtr(phkp);
\end{verbatim}
	in order to register the {\tt J4XXPHitKeeper} to {\tt J4TrackingAction} and to
	{\tt J4HistoryKeeper}.
\item In {\tt J4XXSD}'s {\tt ProcessHits($\cdots$)} method, do 
\begin{verbatim}
	  if (J4XXPHitKeeper::GetInstance()->IsNext()) {
		  // create and store a J4XXPHit object 
	  }
\end{verbatim}
\item In {\tt ProcessHits($\cdots$)} of each calorimeter sensitive detector, which
	usually corresponds to a single calorimeter cell, 
	store the centers of gravity and energy deposits of particles
	from different {\tt PHit}s as different calorimeter hits even in the same cell and
	mark them with the current {\tt PHit} ID obtainable from an
	appropriate {\tt J4PHitKeeper} object. 
\end{itemize}
This ensures the history keeping to be continued until any one of
the pre-registered boundaries is hit and beyond which the calorimeter 
hits are marked with the {\tt PHit} ID put to the {\tt PHit} created on that
boundary.

The {\tt PHit} and {\tt BreakPoint} information can be used in SATELLITES
to perform the CPFA as well as to decompose the various factors contributing
to the jet energy resolution as we will see later.

%  ------------
%  Application to PFA
%  ------------
\section{Application to Studies of Fundamental Limits on the PFA Performance}

Various factors may affect the PFA performance.
The following is a list of possible contributors to the jet energy resolution:
\begin{itemize}
\item calorimeter resolution and acceptance,
\item tracker resolution and acceptance,
\item kink and V$^0$ influence,
\item effects of missing energies due to neutrinos,
\item effects of particle ID and mass assignment.
\end{itemize}
In this section we will investigate how significant these factors are,
making full use of the tool we described so far.

\subsection{Detector Model}

\begin{figure}[ht]
\begin{center}\begin{minipage}{\figurewidth}
\hfill
\begin{minipage}[t]{7cm}
\centerline{
\epsfig{file=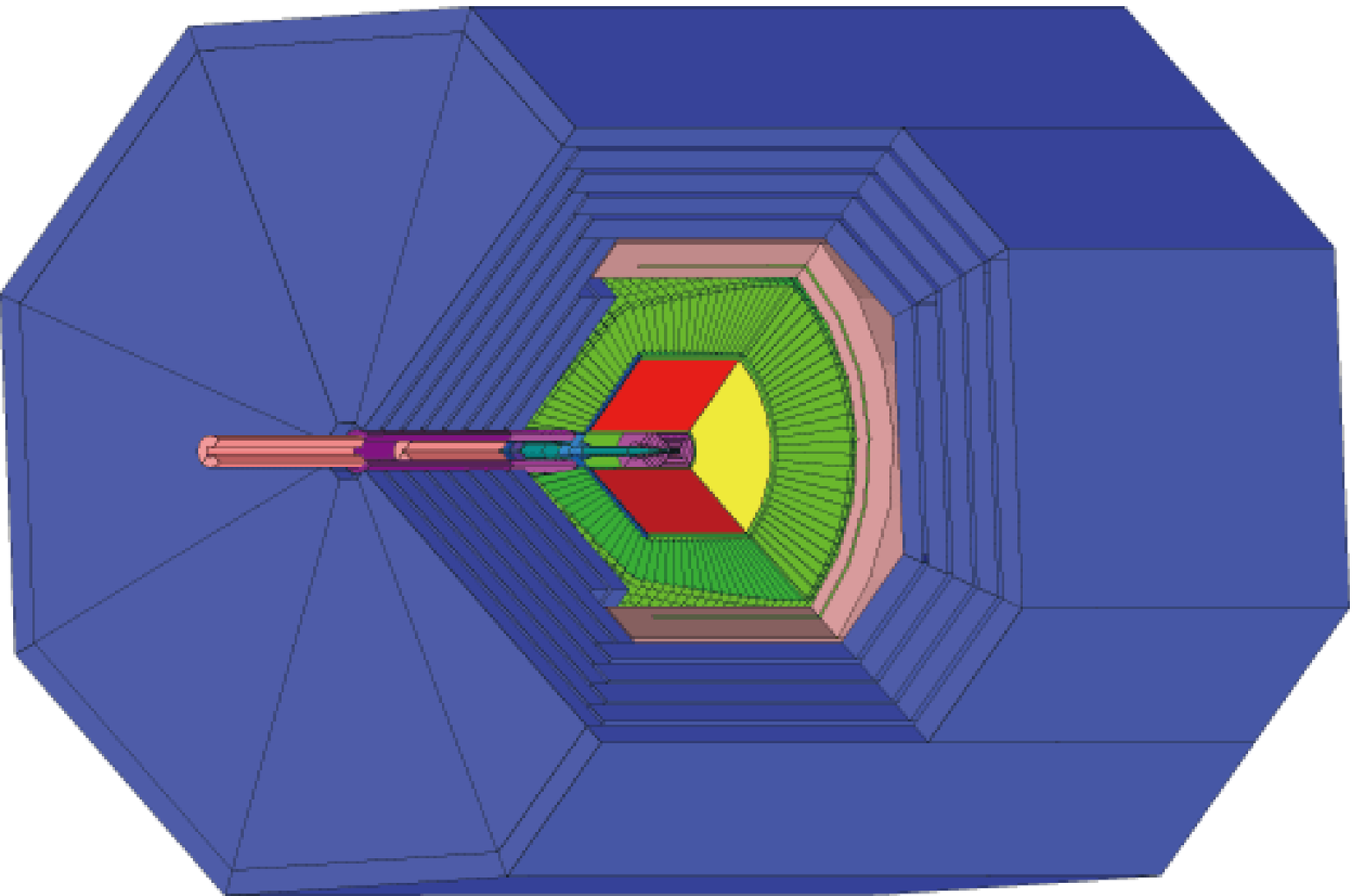,height=4.5cm}}
\end{minipage}
\hfill
\begin{minipage}[t]{7cm}
\centerline{
\epsfig{file=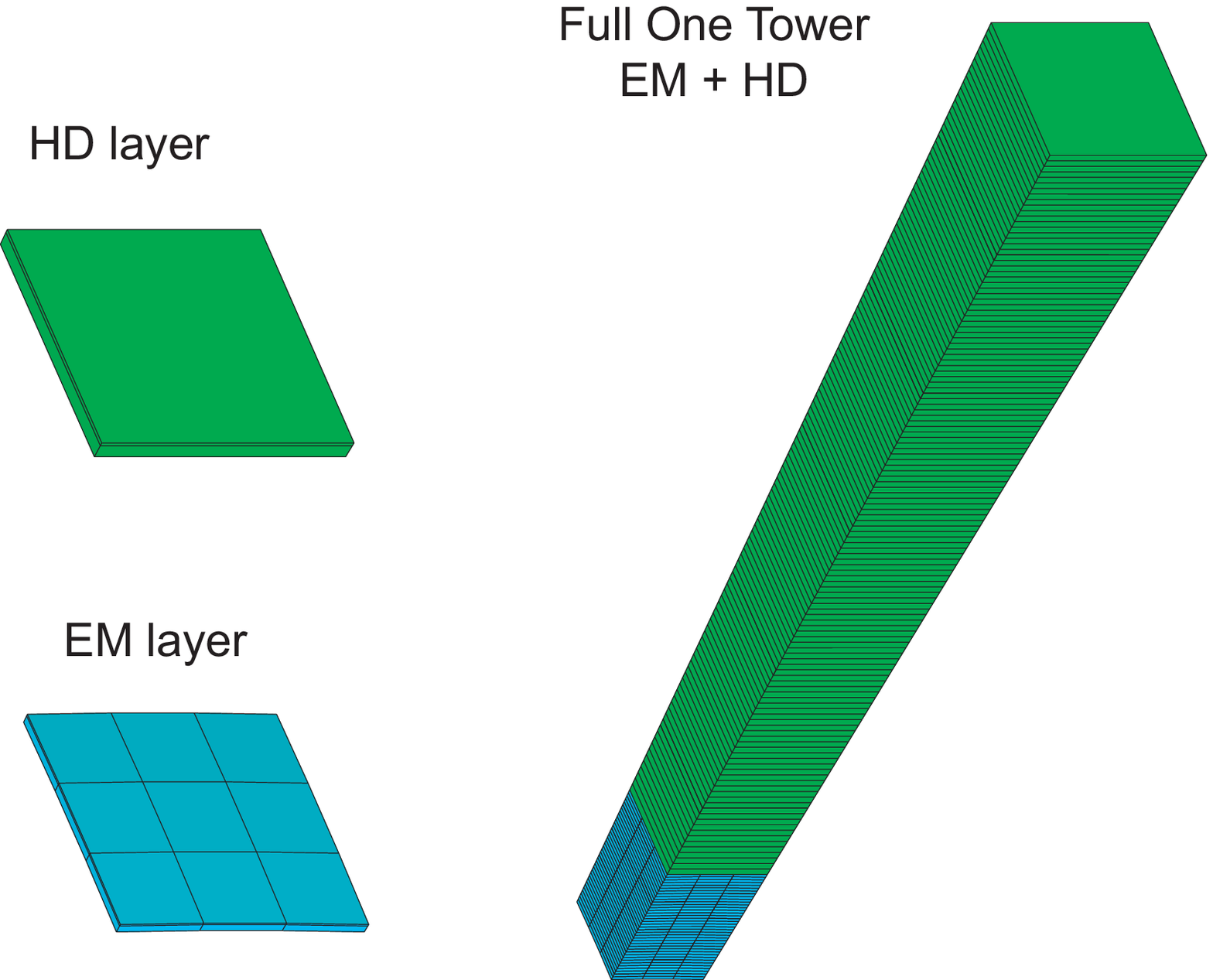,height=5cm}}
\end{minipage}
\hfill
\caption[Fig:gldtower]{\label{Fig:gldtower} \small \it
Detector model (left), a calorimeter tower (right) used in this study
}
\end{minipage}\end{center}
\end{figure}

As shown in Fig.\ref{Fig:gldtower} (left)
the detector model we use in this paper features a Time Projection Chamber (TPC)
as its central tracker and a lead-scintillator-sandwich-type calorimeter (CAL) with
a tower geometry pointing to the interaction point, both of them
installed inside a $3\,$T super-conducting solenoidal magnet (SOL)
with a bore radius of $3.75\,$m and a half length of $4.75\,$m.
The SOL is surrounded by a muon detector that also serves as a
return yoke for the magnetic field.
The detector model also incorporates a vertex detector (VTX)
consisting of six layers of silicon pixel detectors
and an inner tracker (IT) comprises
four cylindrical and seven end-cap layers at each end of silicon detectors.
%
% ------------------
%  Fig.2
% ------------------
%
The TPC has inner and outer radii of $45$ and $200\,{\rm cm}$, respectively,
and is $5.2\,{\rm m}$ long.
The tracking system provides a momentum resolution
of $\sigma_{p_T} / p_T = 1 \times 10^{-4} p_T \, [{\rm GeV/c}]$
with the TPC alone, which can be improved by a factor of about two
when combined with the IT and VTX.

The inner radius of the barrel part of CAL is $2.1\,{\rm m}$
while the front face of its endcap part is at $2.7\,{\rm m}$ away from the interaction point.
As shown in Fig.\ref{Fig:gldtower} (right)
each CAL tower has two sections:
electromagnetic (EM) calorimeter section made of
38 layers of $4\,$mm thick lead radiator plate and  $1\,$mm thick scintillator tile pairs
corresponding to $27.1$ radiation lengths,
and hadron (HD) calorimeter section made of
130 layers of $8\,$mm thick lead plate and $2\,$mm thick scintillator tile pairs
that follow the EM section.
Altogether the single tower has a tower height of $150\,$cm
corresponding to more than 6 interaction lengths.
%and a cross section of $4 \times 4\,$cm$^2$ at its front face.
The calorimeter provides energy resolutions (stochastic) of
$\sigma_E / E \simeq 0.15 / \sqrt{E \, [{\rm GeV}]}$
for electromagnetic showers
and
$\sigma_E / E \simeq 0.43 / \sqrt{E \, [{\rm GeV}]}$
%$\sigma_E / E \simeq 0.40 / \sqrt{E \, [{\rm GeV}]}$
for hadron showers.
The single tower has a cross section of about $12\,$cm$\times 12\,$cm
at the front face which is subdivided into $3 \times 3 = 9$ columns 
in the EM section. 
These finite cell sizes, however,
do not affect the CPFA results we show below,
since the CPFA effectively realizes infinitely fine segmentation.
The model detector system described here has been implemented 
in JUPITER with Geant4.8.2p01.

Notice that the calorimeter geometry we adopted here differs from
the GLD design\cite{Ref:gld} having a dodecagonal shape
and sampling layers parallel to the beam axis in the barrel part
and perpendicular to the beam axis in the end-cap parts.
This is to avoid the polar angle dependence of the calorimeter resolution
due to the variation of the effective sampling thickness, thereby enabling us to
access the best attainable PFA performance\footnote{
The calorimeter resolution depends on the layer configuration and materials
as well as the used calorimeter calibration method to convert the energy deposits 
in the sampling layers to the total energy of the incident particle. 
Their optimization is beyond the scope of this paper.
}.

\subsection{Monte Carlo Data Sample}

\begin{figure}[ht]
\begin{center}\begin{minipage}{\figurewidth}
\centerline{
\epsfig{file=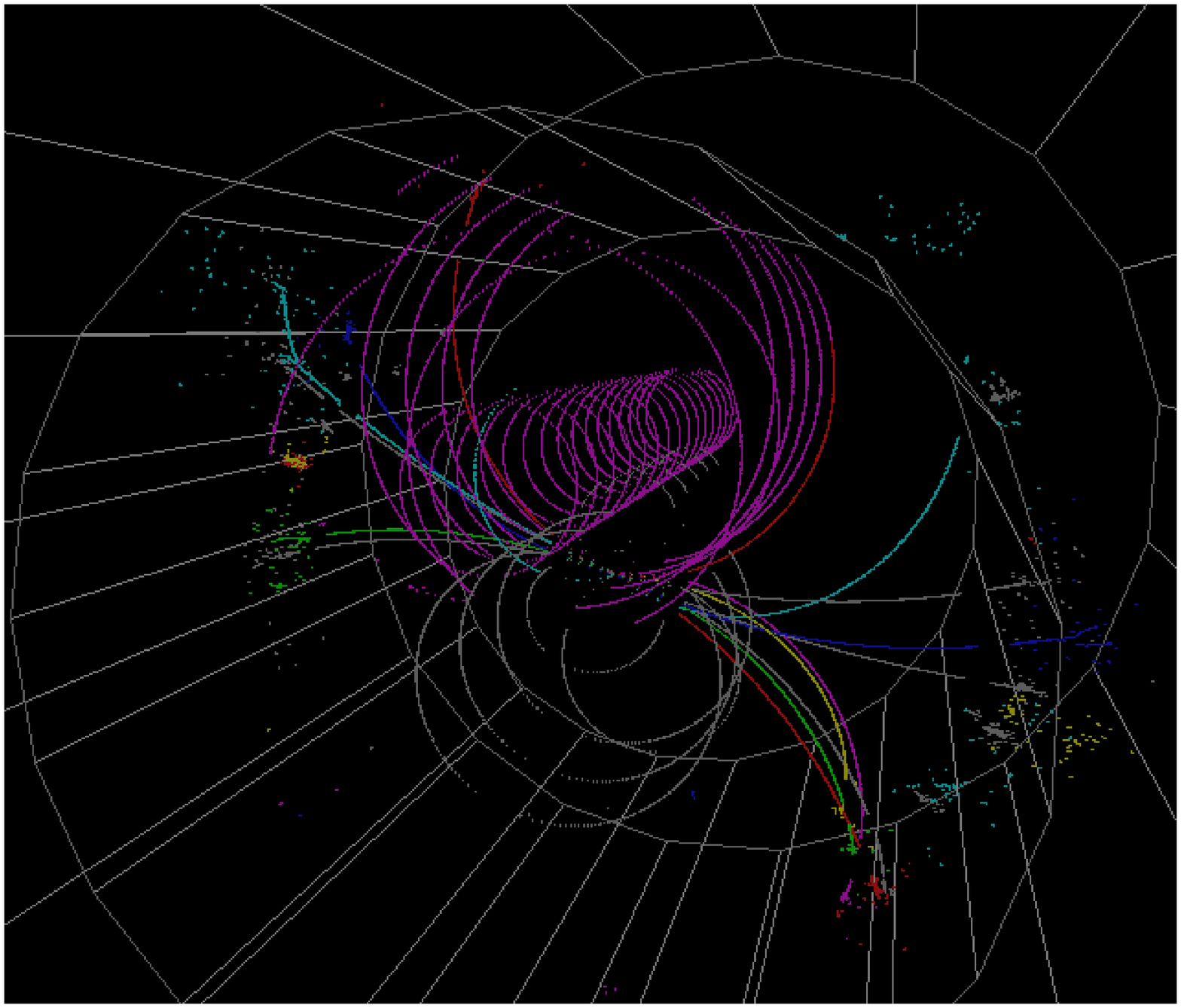,height=7cm}}
\caption[Fig:Z0evt]{\label{Fig:Z0evt} \small \it
A typical $Z^0 \to q \bar{q}$ event.
}
\end{minipage}\end{center}
\end{figure}

We have used PYTHIA version 6.319 to generate 4-momenta
of final-state particles for the $e^+e^- \to q \bar{q}$,
the $e^+e^- \to Z^0 Z^0$,
and the $e^+e^- \to Z^0 H^0$ processes.
For $e^+e^- \to q \bar{q}$ and $Z^0 Z^0$ events we restrict final-state quark flavors
to $u$, $d$, and $s$ to minimize the effects of neutrinos
and the initial state radiation (ISR) switched off to avoid the effect of
ISR photons escaping into the beam-pipe.
Switching off ISR is important in particular at higher energies,
since otherwise the radiative return to $Z^0$: $e^+e^- \to \gamma Z^0 \to \gamma + q \bar{q}$
would dominate the cross section.
For the $e^+e^- \to Z^0 Z^0$ events we have set the $Z^0$ natural width to zero
and one $Z^0$ is forced to decay into $\nu \bar{\nu}$
so that the apparent resolution of the reconstructed $Z^0$ mass is
affected neither by the natural width nor by confusions in jet clustering.
At a given energy point we have generated 10k events of each process
and the 4-momenta of the final-state particles have been fed into JUPITER in the HEPEVT format
and processed through the detector model described above
using a Geant4 physics list that implements various interactions in the
detector materials such as  multiple scattering, energy loss, electromagnetic showering,
hadronic interactions, etc..
Fig.\ref{Fig:Z0evt} shows a typical $Z^o$ pole event simulated this way.
%
% ------------------
%  Fig.3
% ------------------
%
In the figure, calorimeter hits belonging to the same shower cluster
due to the same parent particle are drawn in the same color
using {\tt PHit} information.
We can see a clear correspondence between a cluster and its
parent charged track for a charged PFO.\\

As basic selection cuts we require, in the following, the number of PFOs in a jet to be
5 or more and the absolute value of the cosine of its polar angle to be less than 0.8,
unless otherwise stated.

\subsection{Analysis Methods}

In what follows
we will use two analysis methods to estimate the size of contributions
from various factors to the PFA performance.
The first method directly compares the measured energy and corresponding
MC truth on an event-by-event basis and hence is very clear cut
but cannot be applied to the jet invariant mass resolution.
The second method relies on an assumption about the statistical independence 
of various factors contributing to the jet energy or jet invariant mass resolutions.
It is indirect but allows us to decompose various contributions not only
to the jet energy resolution but also to the jet invariant mass resolution.

\subsubsection{Direct Method}

The first method starts from the following energy sum rule
that holds on an event-by-event basis:
\begin{eqnarray}
E^{\rm true}_{\rm CM} 
& = & E^{\rm true}_{\rm tk} + E^{\rm true}_{\rm EM} + E^{\rm true}_{\rm HD} 
		+ E^{\rm true}_{\nu \,\&\, {\rm AH}}
\cr \rule{0in}{4ex}
& = & (E^{\rm meas}_{\rm tk} - \Delta E_{\rm tk})  
	+ (E^{\rm meas}_{\rm EM} - \Delta E_{\rm EM})
	+ (E^{\rm meas}_{\rm HD} - \Delta E_{\rm HD}) 
	+ E^{\rm true}_{\nu \,\&\, {\rm AH}} 
\cr \rule{0in}{4ex}
& = & E^{\rm meas}_{\rm CM} 
	- \Delta E_{\rm tk} - \Delta E_{\rm EM} - \Delta E_{\rm HD}  + E^{\rm true}_{\nu \,\&\, {\rm AH}} ,
\nonumber
\end{eqnarray}
where $E^{\rm meas / true}_{\rm tk}$, $E^{\rm meas / true}_{\rm EM}$, and
$E^{\rm meas / true}_{\rm HD}$ are the measured/true energy sum of
charged tracks, photons, and neutral hadrons, respectively
and $E^{\rm true}_{\nu \,\&\, AH} $ is the undetected energy due to
neutrinos and the acceptance holes of the detector.
By looking at the distribution of the measurement errors for each component
and fitting a Gaussian to it, we can estimate the contribution from that component
to the jet energy resolution.
For the Gaussian fitting we iteratively adjust the fit range 
so that the fit range would correspond to 2 $\sigma$s.

\subsubsection{Indirect Method}

The first method implies that
\begin{eqnarray}
\sigma_{E_{\rm CM}}^2
& = & \left<  (E^{\rm meas}_{\rm CM} - E^{\rm true}_{\rm CM})^2 \right>
\cr \rule{0in}{4ex}
& = & \left< (\Delta E_{\rm tk})^2 \right>
	+ \left< (\Delta E_{\rm EM})^2 \right>
	+ \left< (\Delta E_{\rm HD})^2 \right>
	+ \left< (E^{\rm true}_{\nu \,\&\, {\rm AH}})^2 \right>
\cr \rule{0in}{4ex}
& = &  \sigma_{E, \, \rm tk}^2 + \sigma_{E, \, \rm HD}^2 + \sigma_{E, \, \rm others}^2
	 + \sigma_{E, \,\rm others}^2 ,
\nonumber
\end{eqnarray}
provided that the measurement errors as well as the undetected energy are
mutually statistically independent\footnote{
The assumed statistical independence breaks down for instance when
there is energy double counting.
We will discuss such a case later when necessary.
}.
In the second method, we assume that this holds generically for both the jet energy and the
jet invariant mass resolutions:
\begin{eqnarray}
\sigma^2 & = &  \sigma_{\rm tk}^2 + \sigma_{\rm EM}^2 + \sigma_{\rm HD}^2 + \sigma_{\rm others}^2 ,
\nonumber
\end{eqnarray}
where $\sigma_{\rm tk}^2$, $\sigma_{\rm EM}^2$, $\sigma_{\rm HD}^2$, 
and $\sigma_{\rm others}^2$ are the contributions from the detector
resolutions for charged tracks, photons, neutral hadrons, and various other effects, respectively.
If we want to estimate $\sigma_{\rm EM}^2$, for instance, 
we replace the measured photon energies with their corresponding true values
obtained from the history keeper.
Then the resultant resolution will be
\begin{eqnarray}
\sigma_{\rm EM=exact}^2 
& = &  \sigma_{\rm tk}^2 + \sigma_{\rm HD}^2 + \sigma_{\rm others}^2 ,
\nonumber
\end{eqnarray}
since the contribution to the resolution from the measurement errors for photons
($\sigma_{\rm EM}^2$) should vanish then.
We can hence obtain $\sigma_{\rm EM}^2$ as
\begin{eqnarray}
\sigma_{\rm EM}^2 & = &
\sigma^2 - \sigma_{\rm EM=exact}^2 .
\nonumber
\end{eqnarray}
We will use this method repeatedly to decompose the jet invariant mass resolution
into various factors.

\subsection{$e^+e^- \to q \bar{q} $ Events}
\subsubsection{Performance on the $Z^0$ pole}

\begin{figure}[ht]
\begin{center}\begin{minipage}{\figurewidth}
\begin{minipage}[t]{5.2cm}
\centerline{
\epsfig{file=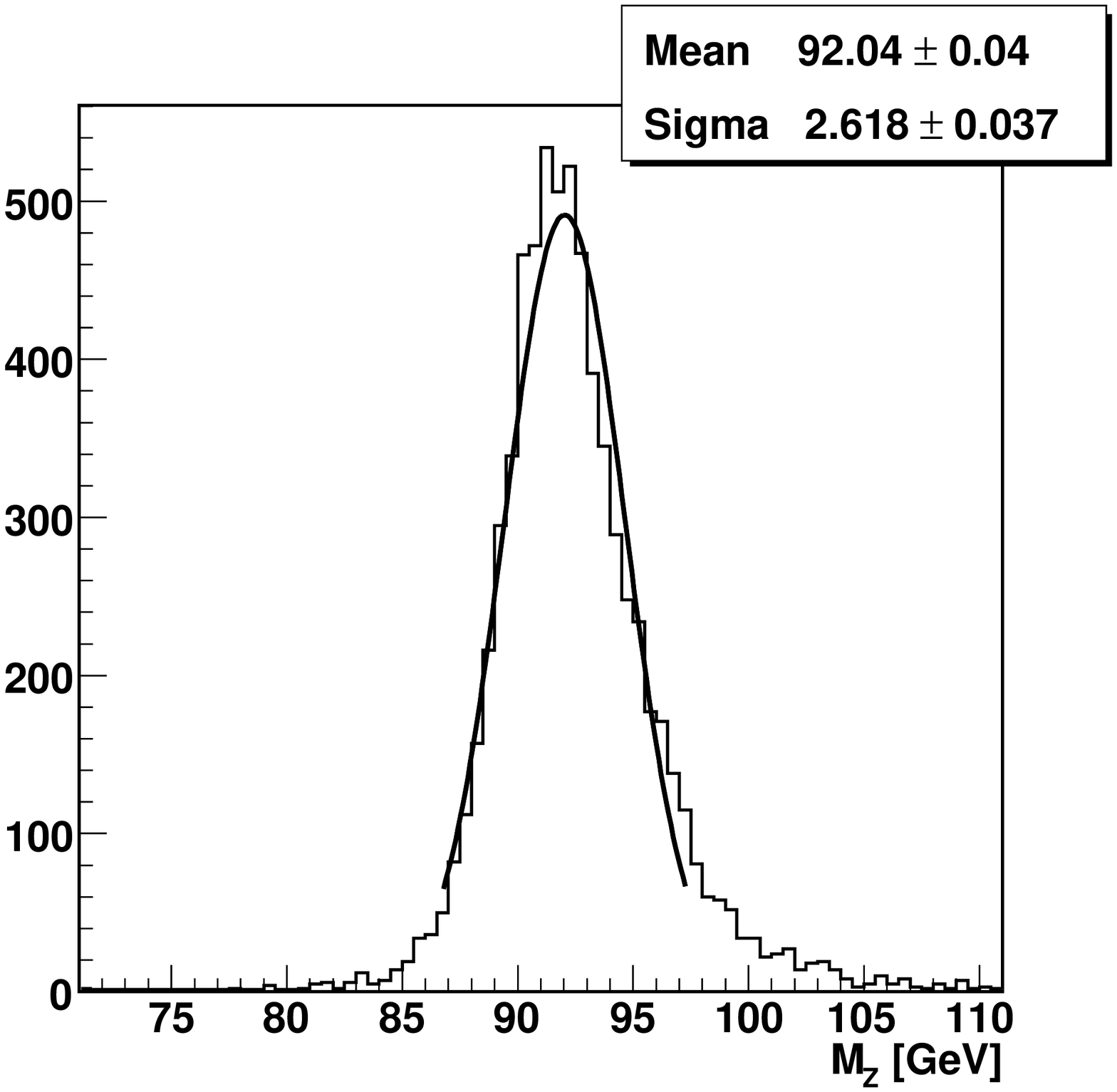,height=5cm}}
\end{minipage}
\hfill
\begin{minipage}[t]{5.2cm}
\centerline{
\epsfig{file=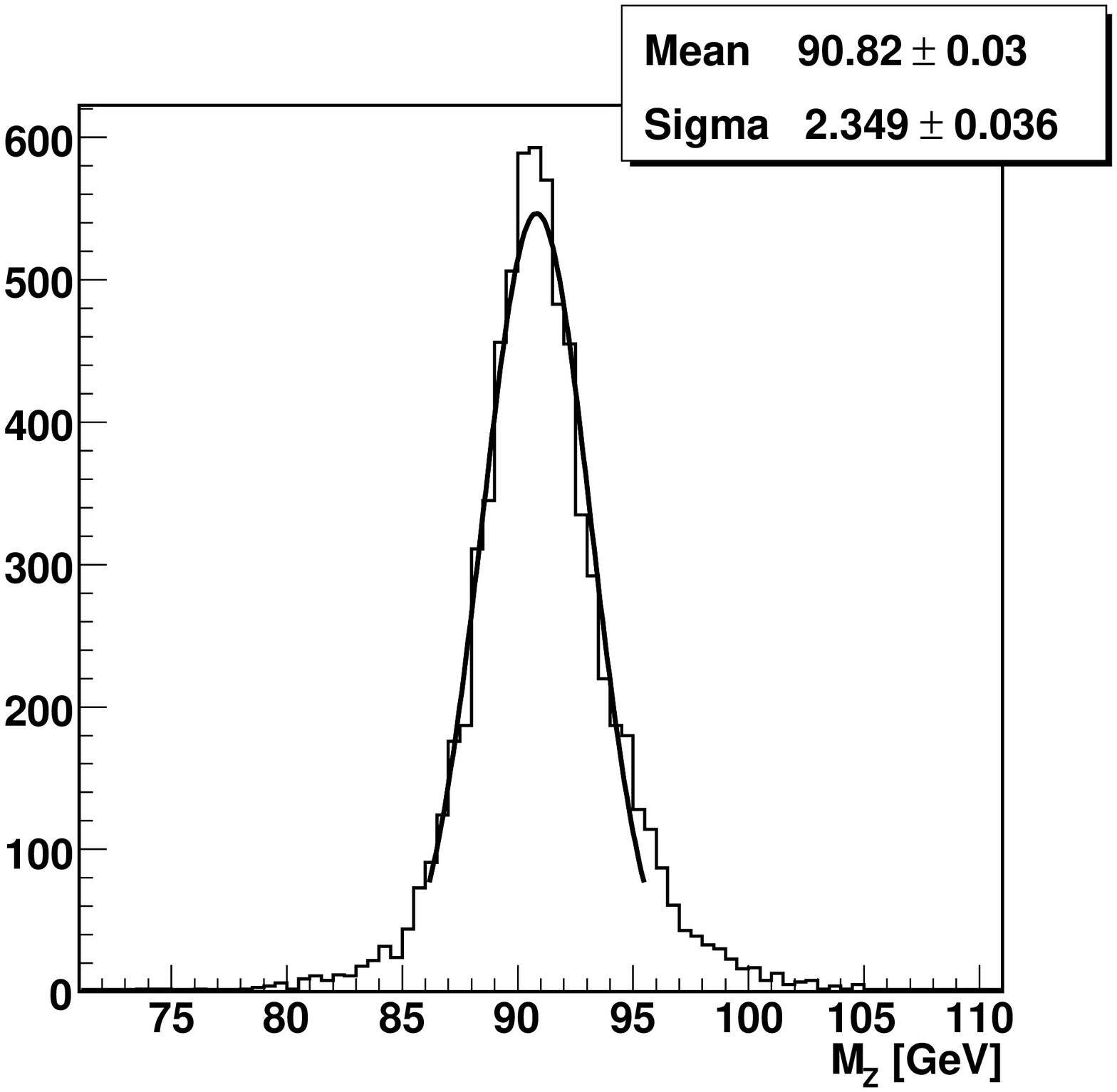,height=5cm}}
\end{minipage}
\hfill
\begin{minipage}[t]{5.2cm}
\centerline{
\epsfig{file=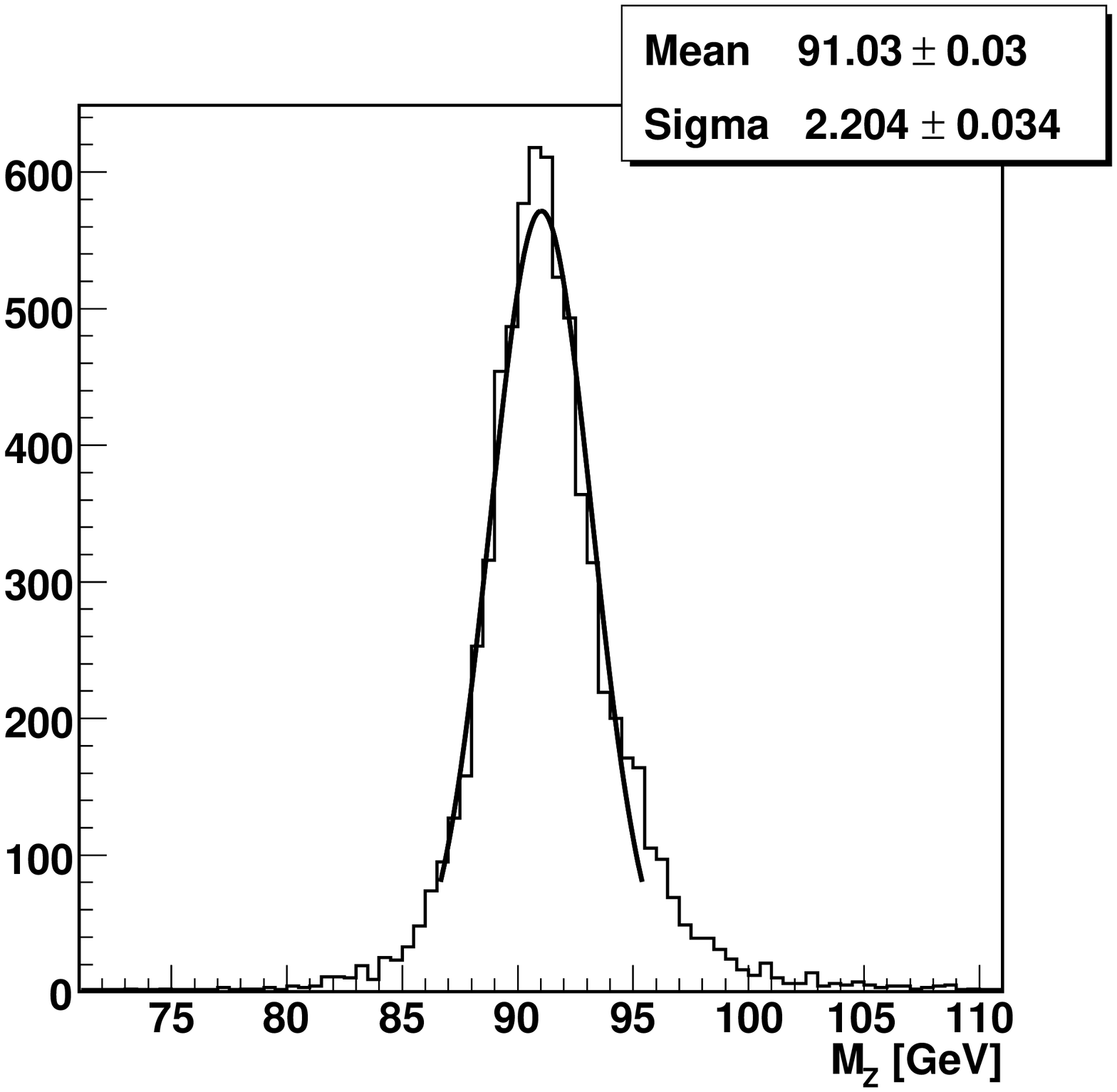,height=5cm}}
\end{minipage}
\caption[Fig:kinkeffect]{\label{Fig:kinkeffect} \small \it
Comparison of reconstructed $Z^0$ mass distributions in the
three schemes for kink treatment:
a) no kink treatment, b) kink daughter scheme, and c) kink mother scheme.
}
\end{minipage}\end{center}
\end{figure}

We start from the treatment of kink particles such as $K^{\pm}$s and $\pi^{\pm}$s
that decay in the tracking system.
A kink has a mother track and its charged and neutral daughters in general.
For instance a kink from the $K^{\pm} \to \pi^{\pm} \, \pi^{0}$ decay may yield
a daughter $\pi^{\pm}$ track and two neutral clusters from the $\pi^{0} \to \gamma \gamma$
decay.
In this case, we have one charged PFO for the parent $K^{\pm}$ track, 
one charged PFO for the daughter $\pi^{\pm}$ track, and
two neutral PFOs for the daughter $\gamma$s from the $\pi^{0}$ decay.
In practice there are three approaches we can take here:
\begin{itemize}
\item {No kink treatment: }
If we don't care about the kink all of these PFOs will be used in the jet reconstruction
and the energy will be double counted.
\item {Kink daughter scheme: }
To avoid the double counting, we may use just kink daughters throwing away
the kink mother.
\item {Kink mother scheme: }
We use the kink mother throwing away the charged kink daughter.
The neutral kink daughters will be be double counted in this case.
\end{itemize}
The kink mother scheme turns out to give the best resolution 
(30\% improvement as compared to the no kink treatment case)
%(29\% improvement as compared to the no kink treatment case)
as shown in Figs.\ref{Fig:kinkeffect} a) through c)
%
% ------------------
%  Fig.4
% ------------------
%
in spite of the risk of energy double counting.
This is because of the dominance of $K^{\pm} \to \mu^{\pm} \nu$ decay
where the neutral daughter escapes detection.
In what follows we will use the kink mother scheme.\\

If incorrectly treated, 
the V$^0$s such as $\gamma$s converted into $e^+e^-$ pairs,
$K^0_S$s, and $\Lambda^0$s
might also affect the jet invariant mass measurements,
since their momenta would be wrongly reconstructed.
In order to investigate this effect, 
let us take a look at the $Z^0$ mass distributions with and without
a V$^0$ finder relying on Monte Carlo truth from the history keeper:
when a pair of charged tracks in the tracking system is found to be
coming from a V$^0$, we re-evaluate the reconstructed daughter momenta 
at its decay vertex to reconstruct the V$^0$ momentum.

\begin{figure}[ht]
\begin{center}\begin{minipage}{\figurewidth}
\centerline{
\hfill
\begin{minipage}[t]{5.2cm}
\centerline{
\epsfig{file=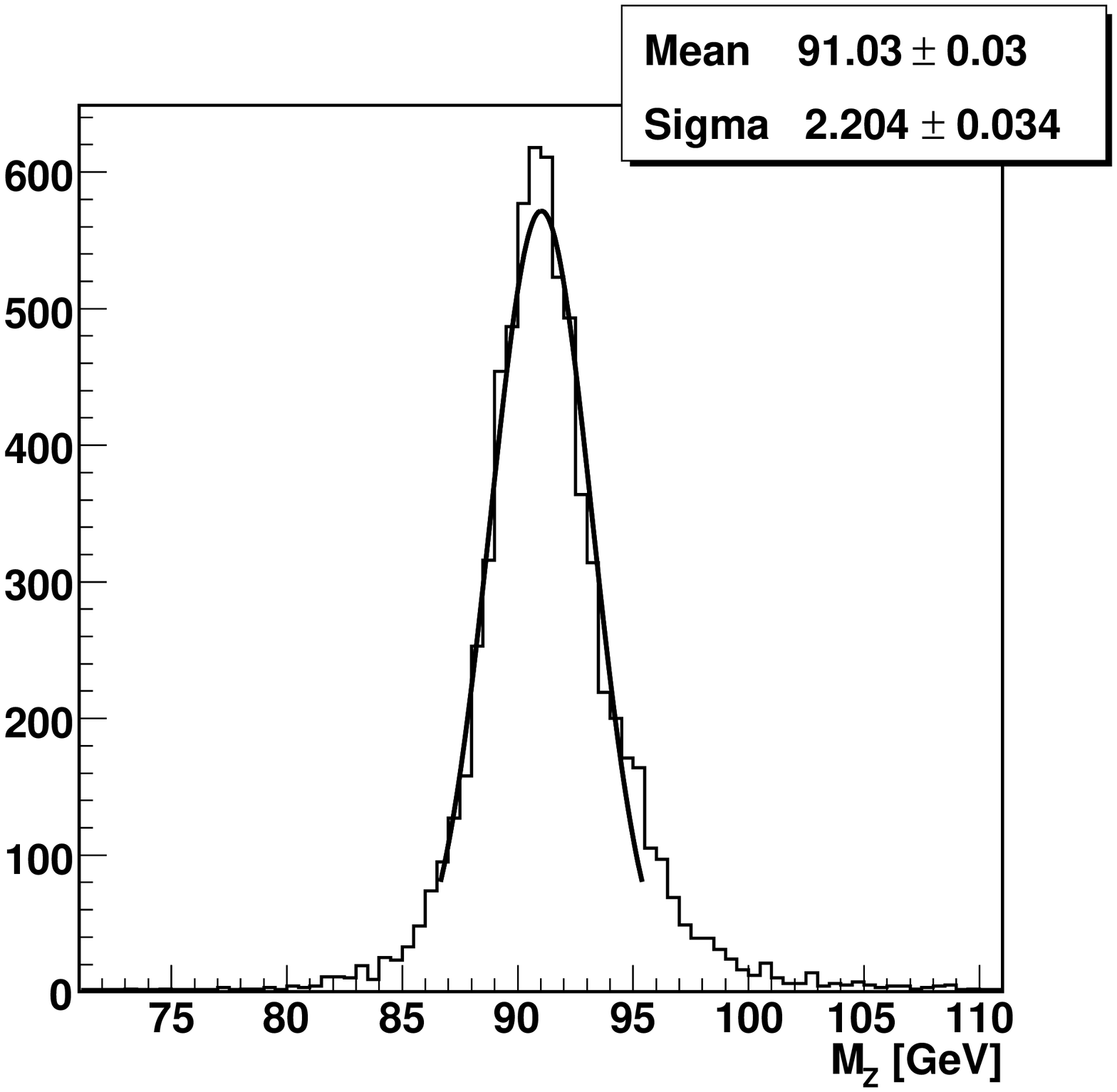,height=5cm}}
\end{minipage}
\hfill
\begin{minipage}[t]{5.2cm}
\centerline{
\epsfig{file=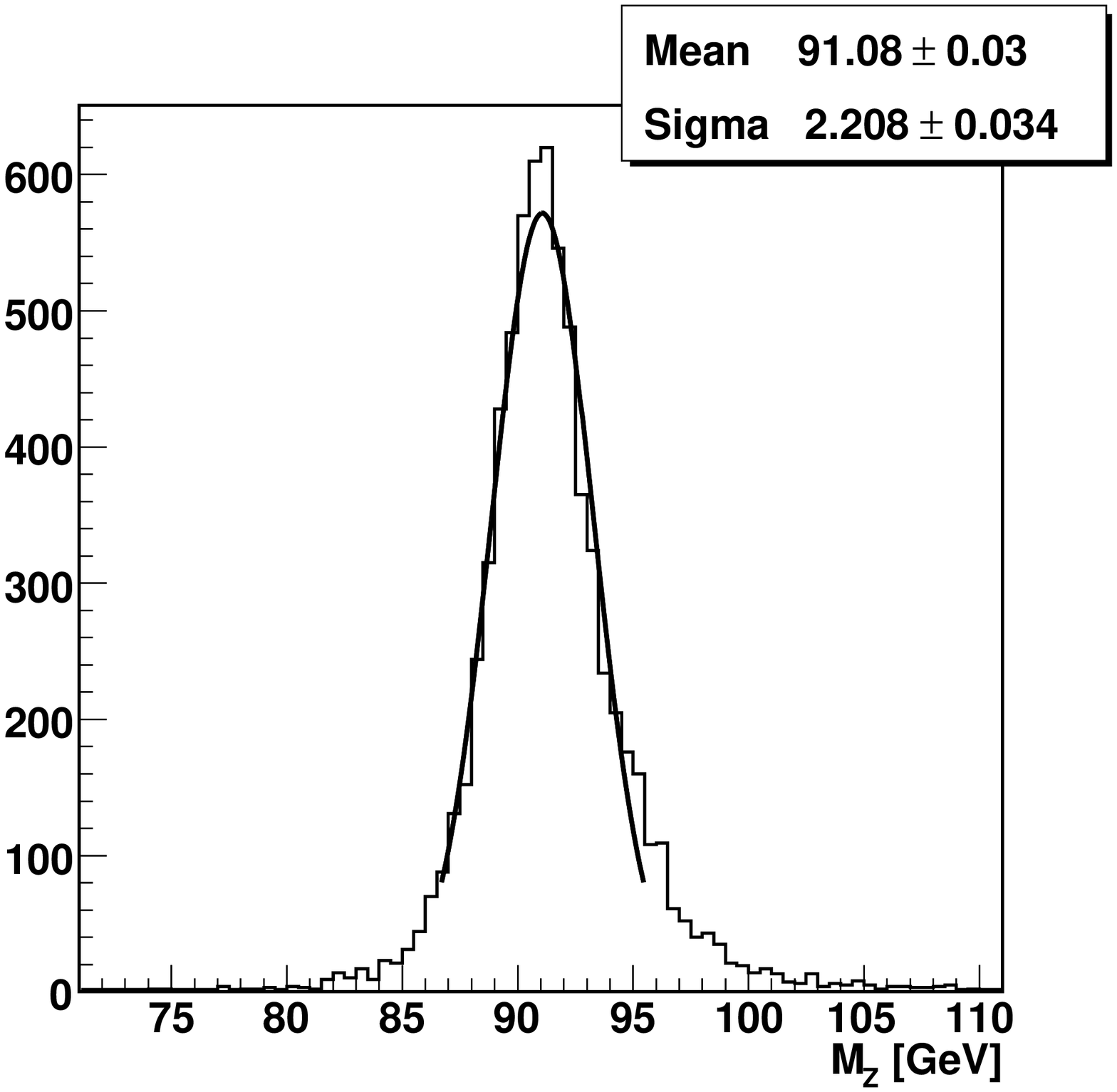,height=5cm}}
\end{minipage}
\hfill
}
\caption[Fig:v0effect]{\label{Fig:v0effect} \small \it
Comparison of reconstructed $Z^0$ mass distributions (a) with
and (b) without the V$^0$ finder explained in the text.
}
\end{minipage}\end{center}
\end{figure}

The effect of the V$^0$ treatment on the $Z^0$ mass resolution, however, 
turned out to be less significant than the effect of kink treatments
as shown in Figs.$\,$\ref{Fig:v0effect} a) and b).
%
% ------------------
%  Fig.5
% ------------------
%
This can be attributed to the fact that
for a high momentum V$^0$
the relative error in the sum of the reconstructed momenta of the V$^0$ daughters
is expected to be small since their opening angle becomes large
only when one of them has a negligible momentum.
Besides, the V$^0$ treatment does not affect
the total energy of the system unlike the kink treatments.
Although the effect is small we apply the V$^0$ treatment in what follows
for completeness.\\

\begin{figure}[ht]
\begin{center}\begin{minipage}{\figurewidth}
\hfill
\begin{minipage}[t]{5.2cm}
\centerline{
\epsfig{file=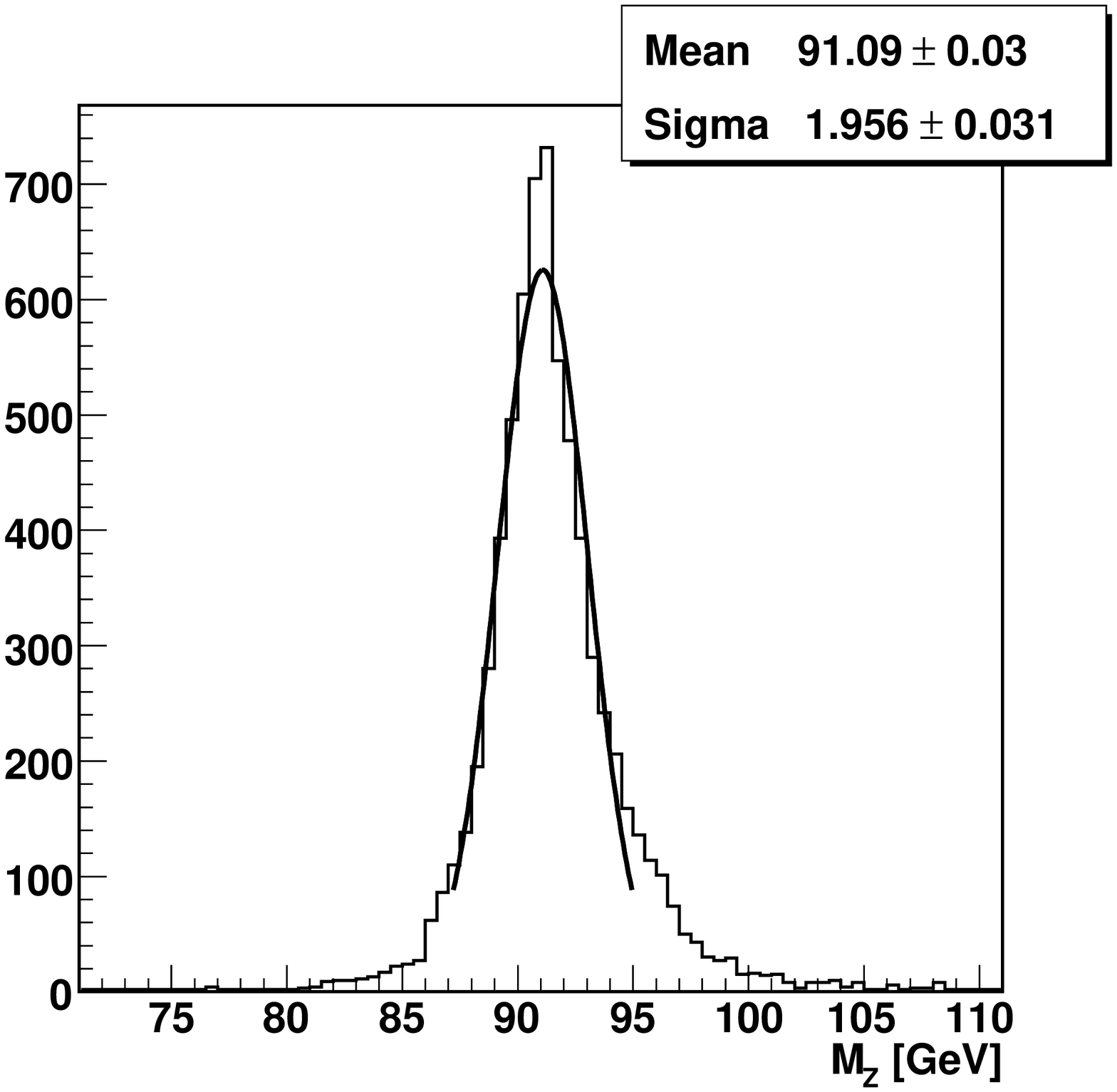,height=5cm}}
\end{minipage}
\hfill
\begin{minipage}[t]{5.2cm}
\centerline{
\epsfig{file=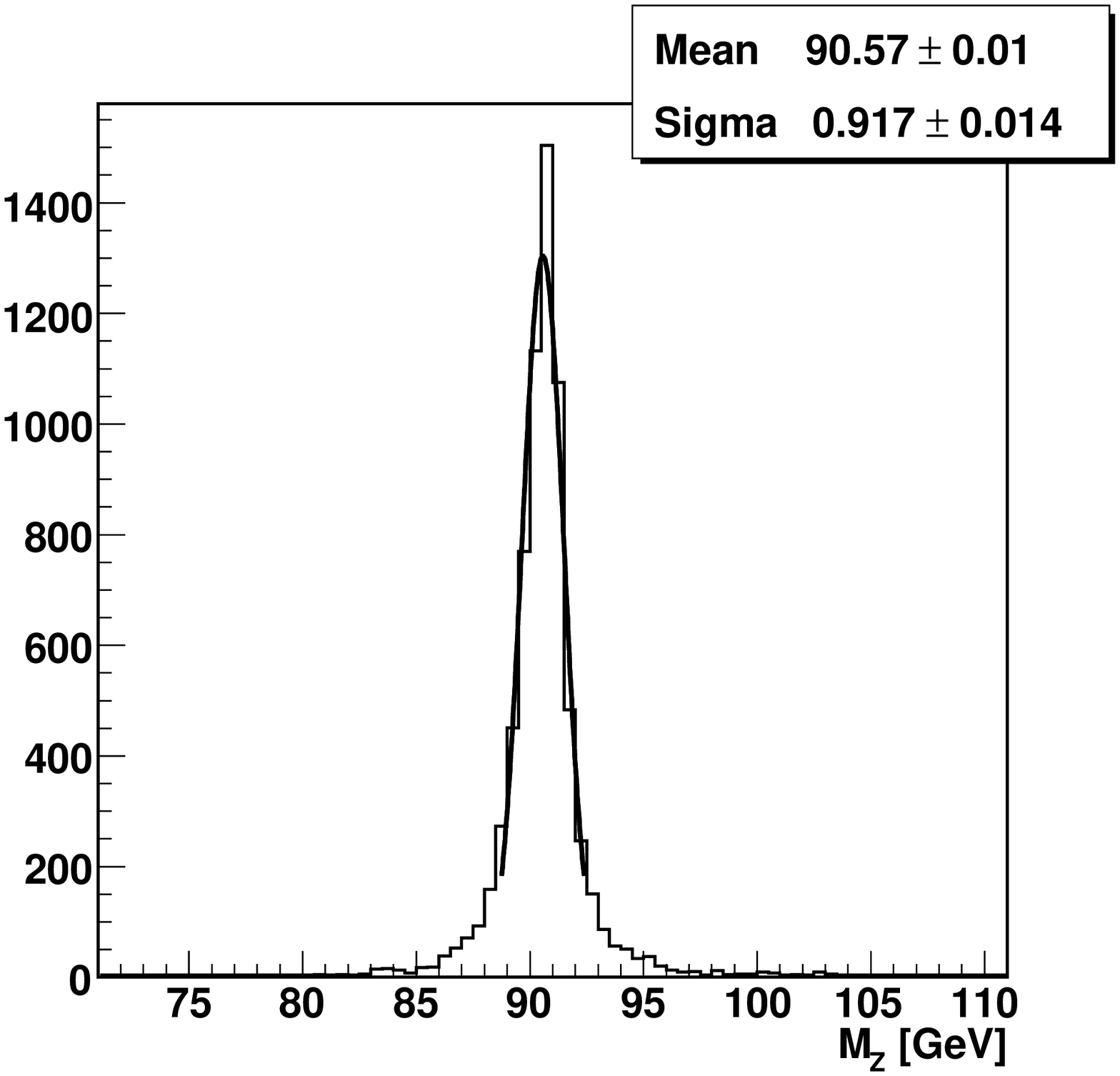,height=5cm}}
\end{minipage}
\hfill
\begin{minipage}[t]{5.2cm}
\centerline{
\epsfig{file=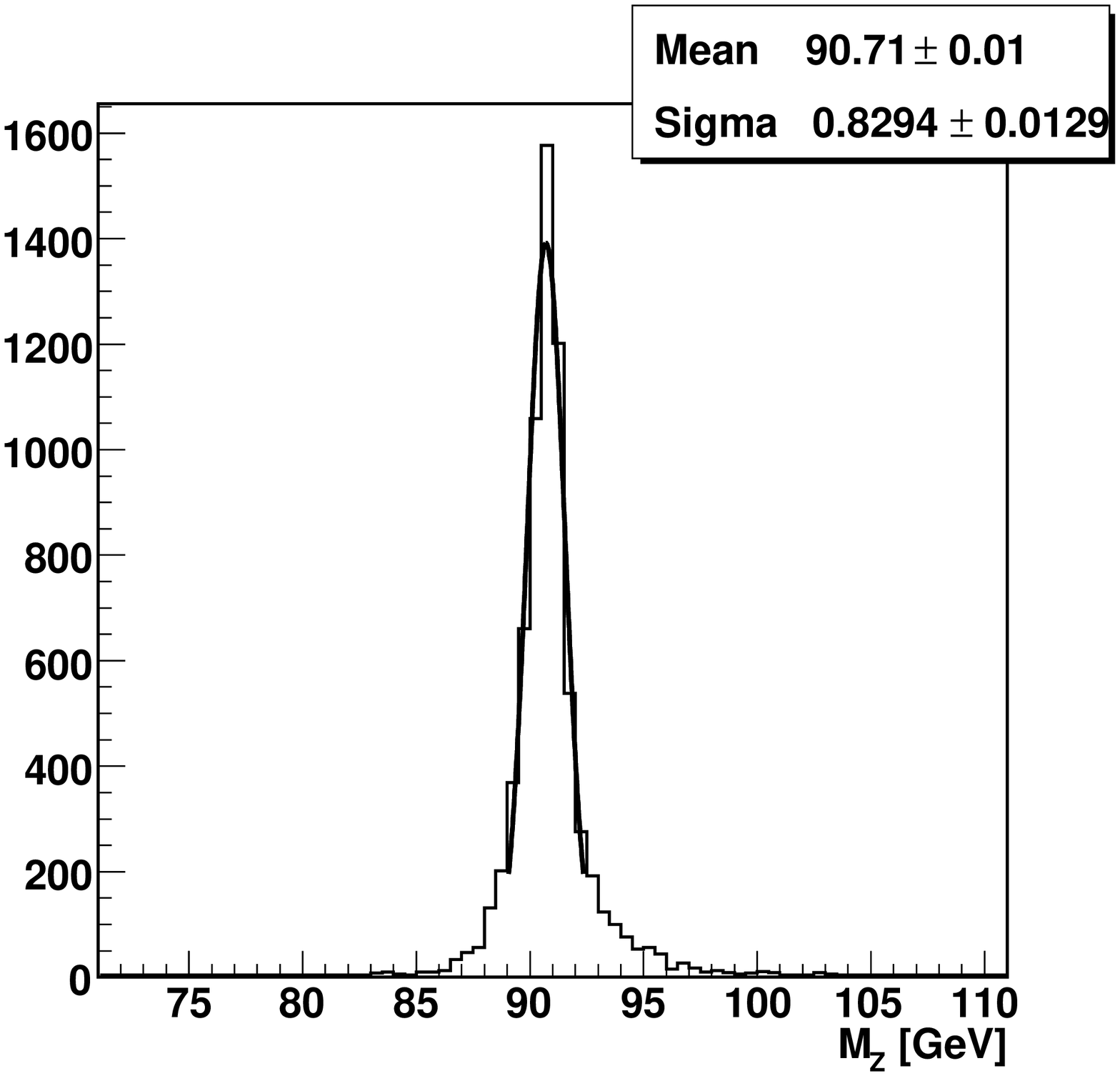,height=5cm}}
\end{minipage}
\hfill
\caption[Fig:mresol91]{\label{Fig:mresol91} \small \it
Comparison of reconstructed $Z^0$ mass distributions with
the 4-momenta of PFOs replaced by MC truths for
(a) electromagnetic showers with no associated tracks, 
(b) both electromagnetic and hadron showers with no associated tracks, and
(c) all of them.
}
\end{minipage}\end{center}
\end{figure}

With the kinks and V$^0$s treated we can now investigate the effects
of the detector resolutions on the jet invariant mass measurements.
Fig.$\,$\ref{Fig:mresol91} a) shows the $Z^0$ mass distribution
with the 4-momenta of PFOs corresponding to neutral
electromagnetic PFOs (photons, electrons, or positrons 
with no associated tracks) replaced by the MC truths.
%
% ------------------
%  Fig.6
% ------------------
%
The resolution difference between this and Fig.$\,$\ref{Fig:v0effect} a)
hence represents the contribution of the calorimeter resolution
for electromagnetic showers due mostly to photons
from $\pi^0$s: $\sqrt{2.20^2 - 1.96^2}=1.02\,$GeV ((1.02/2.20)$^2$=21\%). %w/o albedo
% from $\pi^0$s: $\sqrt{2.22^2 - 2.04^2}=0.88\,$GeV ((0.88/2.22)$^2$=16\%). % hdc
% $\sqrt{1.31^2 - 0.978^2} = 0.87\,$GeV from $(hdexact)^2 - (emhdexact)^2$

Further replacing the 4-momenta of PFOs corresponding to hadrons with
no associated tracks by the MC truths\footnote{
For neutrons kicked out from detector materials, we subtract the
neutron mass from each of their true energies
so as not to overestimate the parent jet energy. 
} from the history keeper, we have Fig.$\,$\ref{Fig:mresol91} b).
The difference between Figs.$\,$\ref{Fig:mresol91} a) and b)
gives the contribution from the calorimeter resolution
for hadron showers:
$\sqrt{1.96^2 - 0.917^2}=1.73\,$GeV (62\%). % w/o albedo
% $\sqrt{2.04^2 - 0.978^2}=1.79\,$GeV (65\%). % hdc
% $\sqrt{2.22^2 - 1.31^2} = 1.79\,$GeV from $(std)^2 - (hdexact)^2$

In order to estimate the contribution from the tracker resolution,
we switch off the tracker resolution in Fig.$\,$\ref{Fig:mresol91} c)
by replacing the 4-momenta of the charged PFOs with the 
corresponding MC truths from the history keeper.
The resultant difference between Figs.$\,$\ref{Fig:mresol91} b) and c)
yields the contribution from the tracker resolution:
$\sqrt{0.917^2 - 0.829^2}=0.39\,$GeV (3\%). % w/o albedo
%$\sqrt{0.978^2 - 0.829^2}=0.52\,$GeV (3\%).
\\

\begin{figure}[ht]
\begin{center}\begin{minipage}{\figurewidth}
\hfill
\begin{minipage}[t]{5.2cm}
\centerline{
\epsfig{file=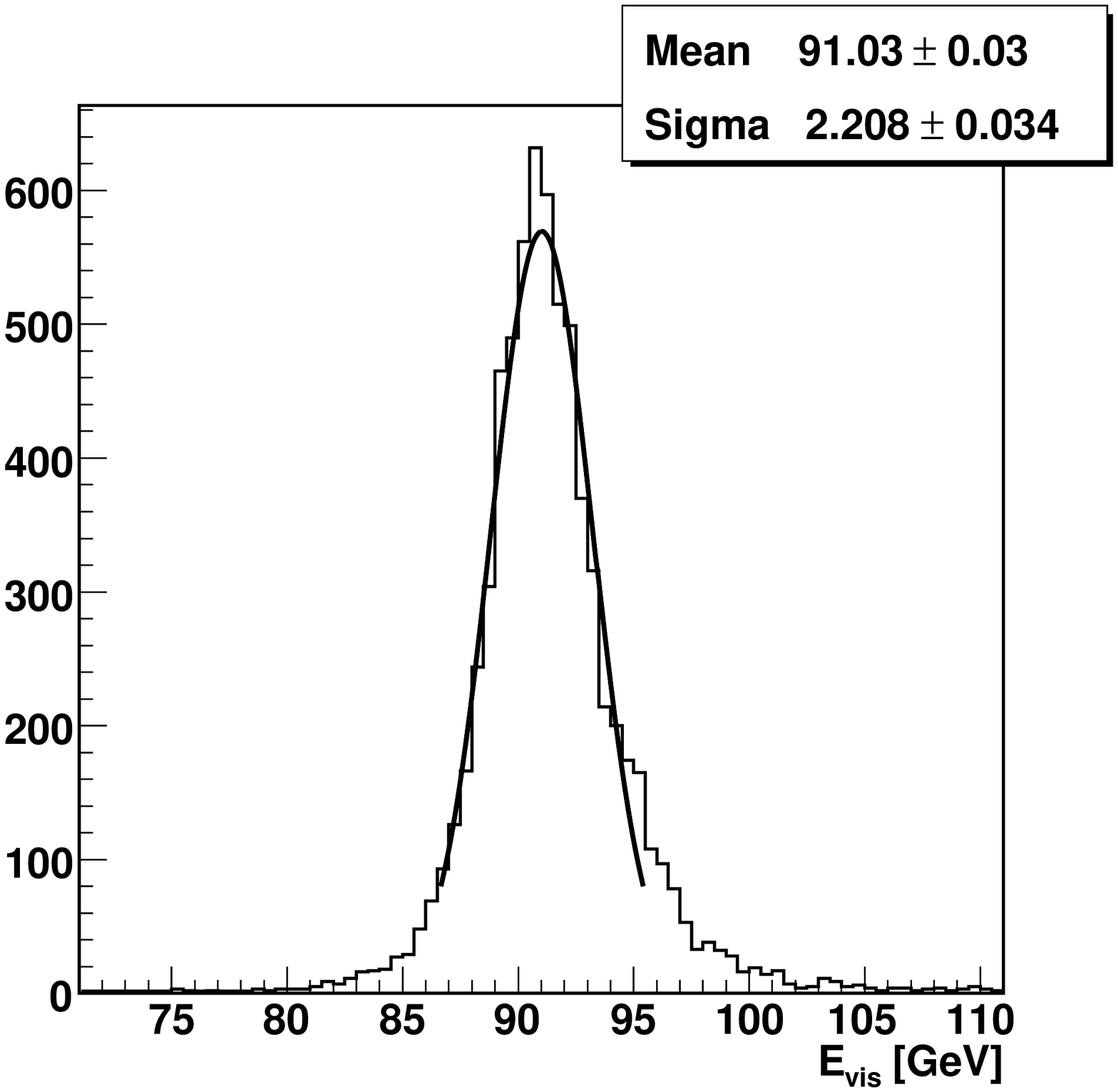,height=5cm}}
\end{minipage}
\hfill
\begin{minipage}[t]{5.2cm}
\centerline{
\epsfig{file=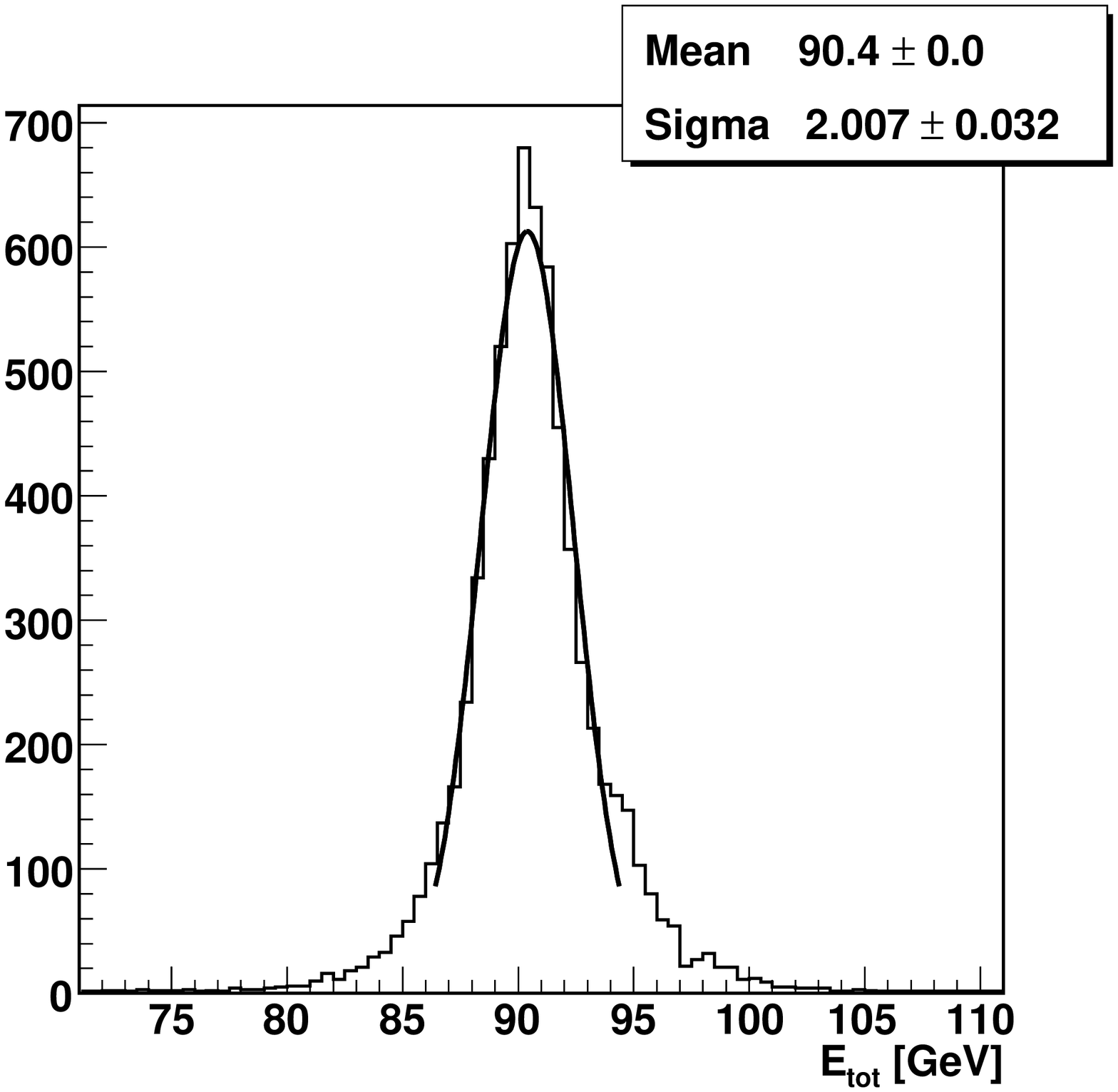,height=5cm}}
\end{minipage}
\hfill
\begin{minipage}[t]{5.2cm}
\centerline{
\epsfig{file=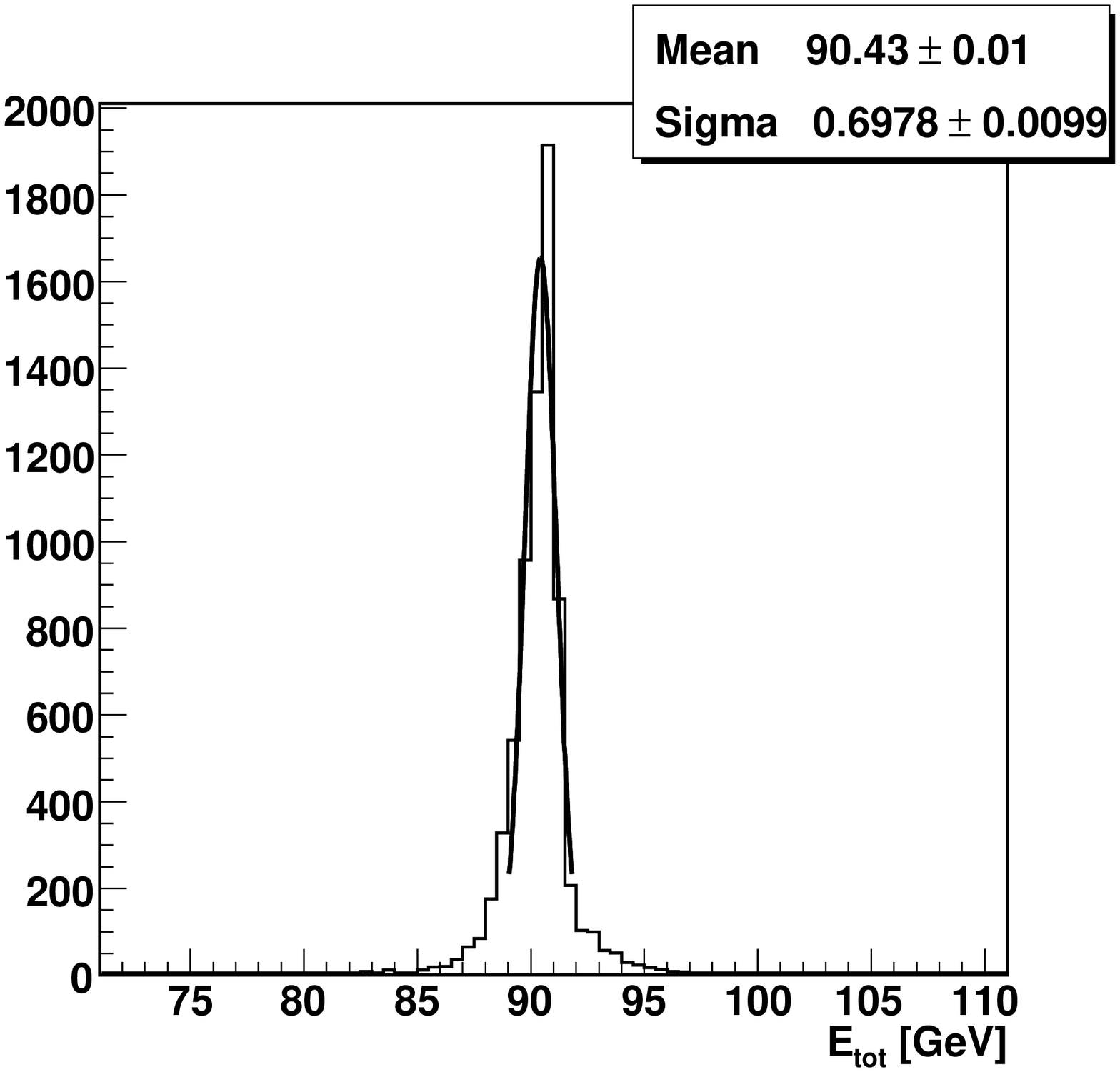,height=5cm}}
\end{minipage}
\hfill
\caption[Fig:evis91]{\label{Fig:evis91} \small \it
Visible energy distributions:  
(a) that corresponding to the mass distribution in Fig.$\,$\ref{Fig:v0effect} a), 
(b) that with the double counted daughter energies avoided by using the history keeper, and
(c) the same as (b) but the 4-momenta of PFOs replaced by the MC truths as
in Fig.$\,$\ref{Fig:mresol91} c).
}
\end{minipage}\end{center}
\end{figure}

The remaining $0.829\,$GeV (14\%) must be coming from
the effects of undetected particles due to acceptance holes or neutrinos, 
energy double counting, 
particle misidentification, etc..
Since the undetected particles only make a tail on the
lower mass side of the peak,
the tail on the higher mass side suggests
some double counting of energy due for instance to 
the effect of neutral kink daughters discussed above
or the effect of incorrectly included mass energies of 
particles kicked out from detector materials\footnote{
As mentioned above the energies of the neutrons
kicked out from the detector materials should be corrected for
their mass energies to avoid overestimation.
This also applies to protons or electrons kicked out from
the detector materials, though they are not corrected for
in the figures.
}.
To study these effects, let us now switch to the direct method.
Fig.$\,$\ref{Fig:evis91} a) shows the visible energy distribution
for the same sample of PFOs as used for
the mass distribution in Fig.$\,$\ref{Fig:v0effect} a).
%
% ------------------
%  Fig.7
% ------------------
%
The visible energy distribution is almost identical to
the mass distribution as expected.
On the other hand, Fig.$\,$\ref{Fig:evis91} b) is the same distribution
with the double counted daughter energies 
eliminated by using the history keeper.
The difference of the two gives the contribution from
the double counting:
$\sqrt{2.20^2 - 2.01^2}=0.92\,$GeV (17\%).
% $\sqrt{2.22^2 - 2.01^2}=0.94\,$GeV (18\%).
% $\sqrt{2.22^2 - 2.09^2}=0.74\,$GeV (11\%).
Notice that this includes the contribution from the finite detector resolutions
for the double counted energies.
The contribution from the double counted energies alone can be obtained from
the difference between Figs.$\,$\ref{Fig:mresol91} c) and \ref{Fig:evis91} c):
$\sqrt{0.829^2 - 0.698^2}=0.45\,$GeV (4\%).
As seen in \ref{Fig:evis91} c), the distribution still has a higher tail
indicating that there still remains some contribution from energy overestimation
due to the mass energies of particles kicked out from the detector materials.\\

\begin{figure}[ht]
\begin{center}\begin{minipage}{\figurewidth}
\hfill
\begin{minipage}[t]{5.2cm}
\centerline{
\epsfig{file=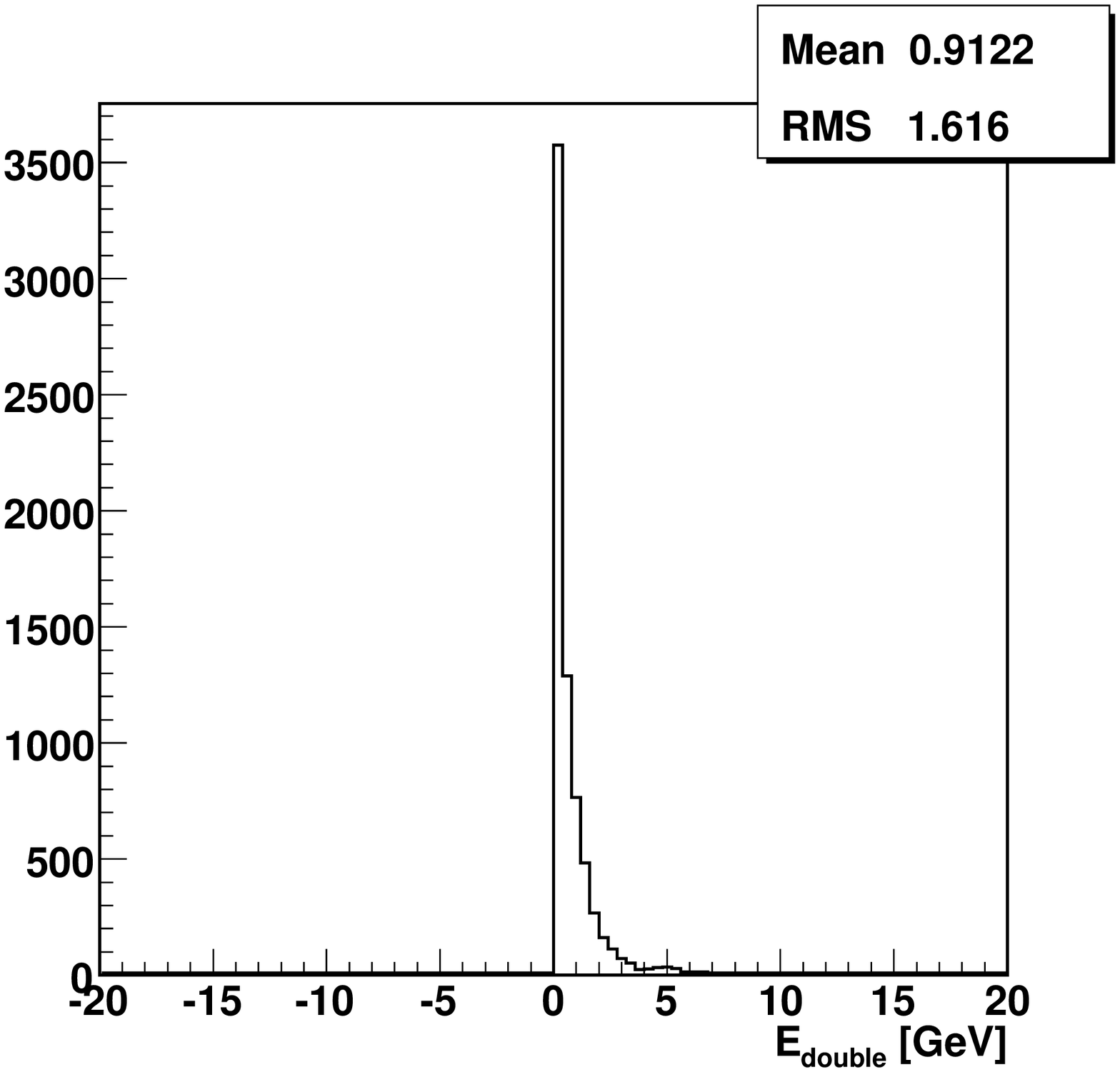,height=5cm}}
\end{minipage}
\hfill
\begin{minipage}[t]{5.2cm}
\centerline{
\epsfig{file=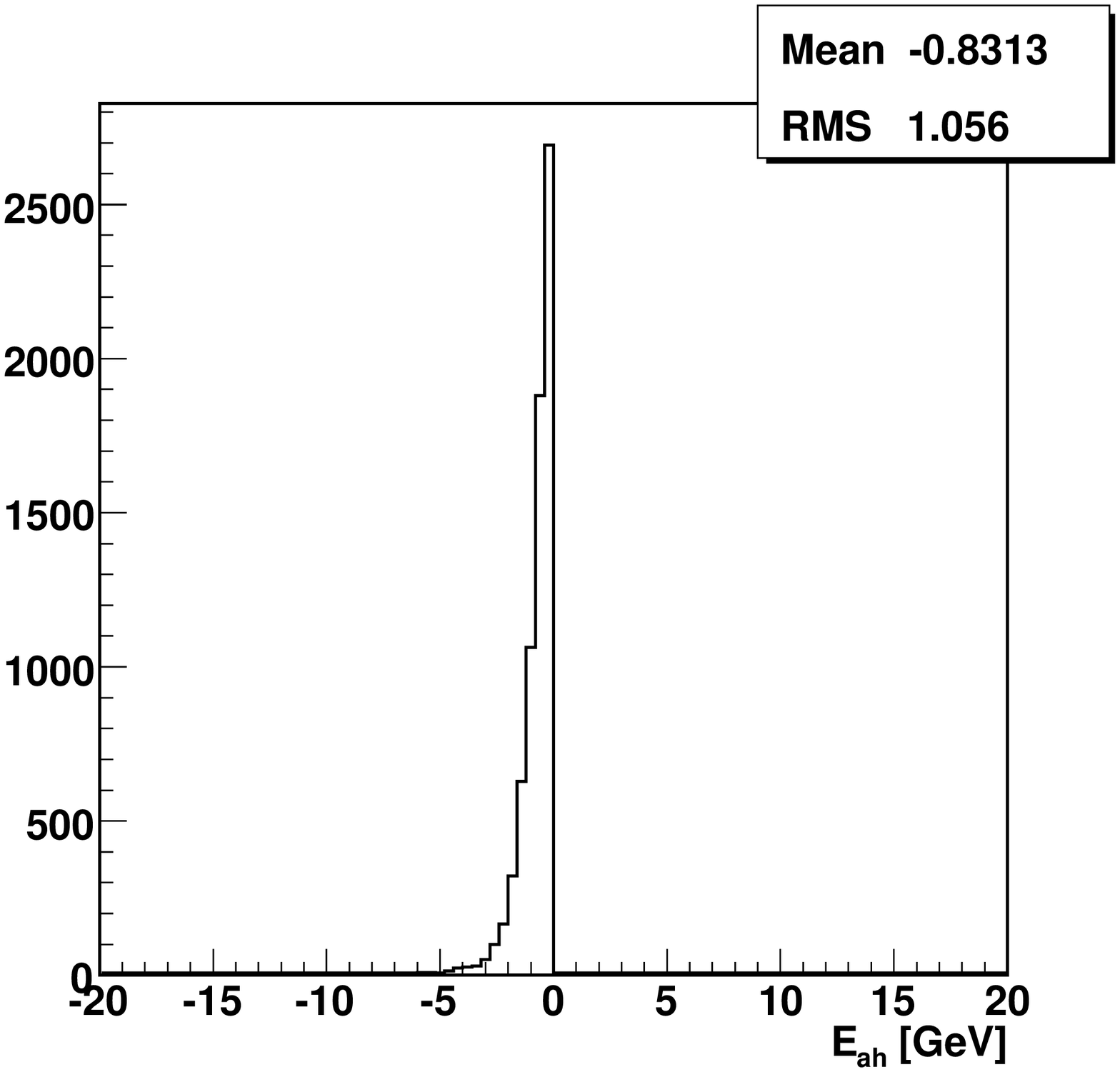,height=5cm}}
\end{minipage}
\hfill
\begin{minipage}[t]{5.2cm}
\centerline{
\epsfig{file=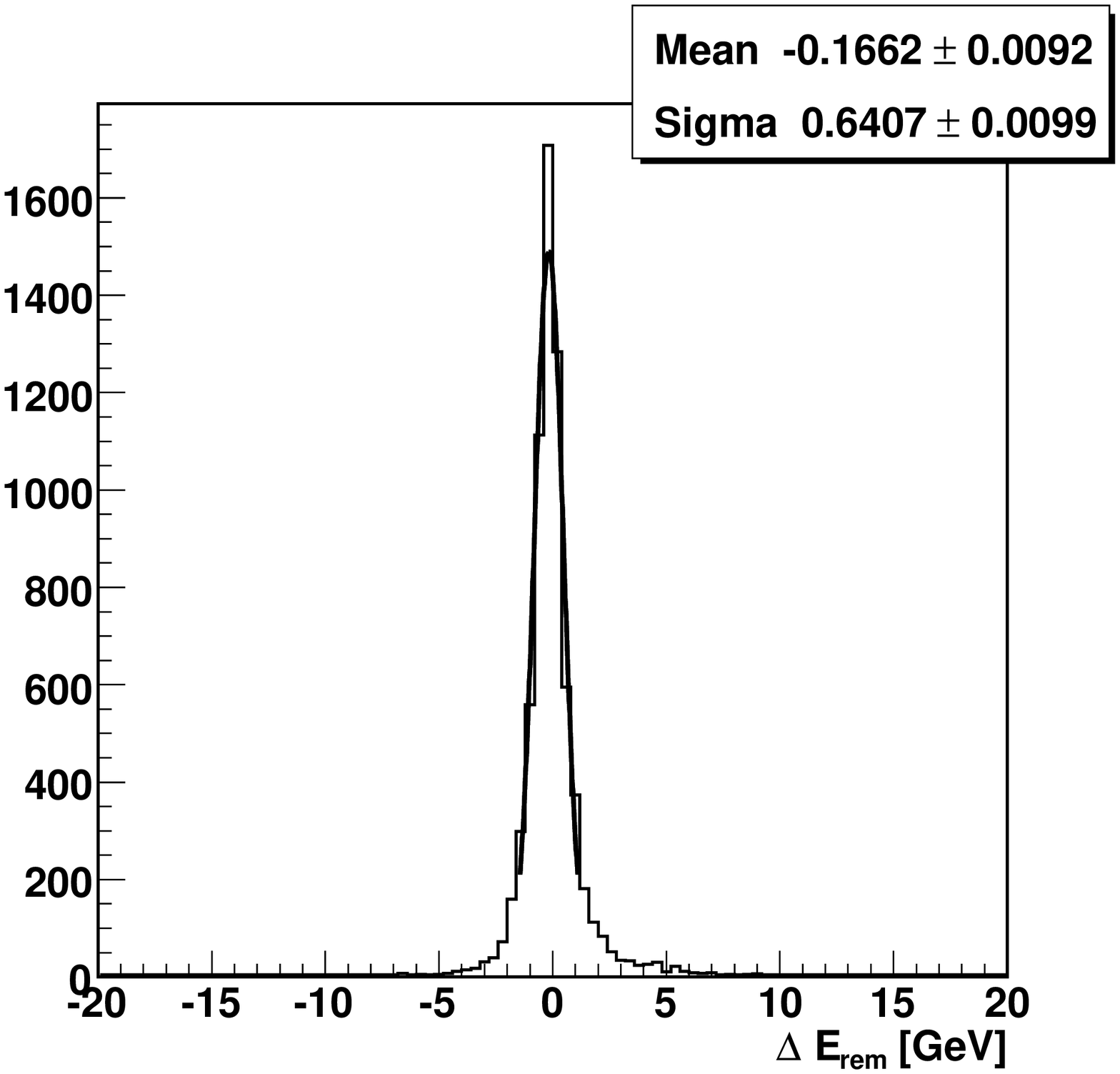,height=5cm}}
\end{minipage}
\hfill
\caption[Fig:edouble91]{\label{Fig:edouble91} \small \it
Distributions of
(a) the true event-by-event sum of double counted energies, 
(b) the true event-by-event sum of undetected energies with an overall minus sign, and
(c) the event-by-event sum of (a) and (b).
}
\end{minipage}\end{center}
\end{figure}

\begin{figure}[ht]
\begin{center}\begin{minipage}{\figurewidth}
\centerline{
\begin{minipage}[t]{5.2cm}
\centerline{
\epsfig{file=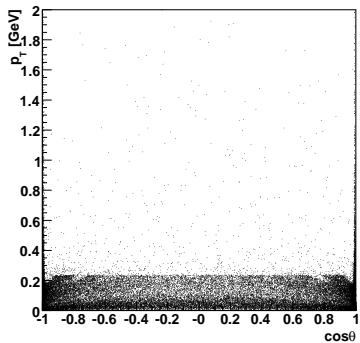,height=5cm}}
\end{minipage}
}
\caption[Fig:ah]{\label{Fig:ah} \small \it
Distribution of $(\cos\theta, p_T)$ of undetected primary
particles on the $Z^0$ pole.
}
\end{minipage}\end{center}
\end{figure}

To see their contributions more clearly we plot 
the exact values of the double counted energies 
and the exact values of undetected energies (with an overall minus sign)
in Figs.$\,$\ref{Fig:edouble91} a) and b), respectively.
%
% ------------------
%  Fig.8
% ------------------
%
As expected, the double counted energies have only a higher tail
while the undetected energies have only a lower tail, thereby
resulting in a more or less symmetric distribution for their sum
as shown in Fig.$\,$\ref{Fig:edouble91} c).
Since we restrict the $Z^0$ decays only to $u$, $d$, and $s$ quarks,
the effect of neutrinos is negligible and the undetected energy
in Fig.$\,$\ref{Fig:edouble91} b) is mostly due to
acceptance holes: $p_{T}$ cutoff of about $0.23\,$GeV\footnote{
The $p_{T}$ cutoff at $0.23\,$GeV for charged particles from the IP
is due to the TPC acceptance.
If we can efficiently perform self-tracking with the IT and VTX only,
we may lower the cutoff to a negligible level.
} for
charged particles from the IP and the forward and the backward
regions near the beam pipe
(see Fig.$\,$\ref{Fig:ah}).
%
% ------------------
%  Fig.9
% ------------------
%

\begin{figure}[ht]
\begin{center}\begin{minipage}{\figurewidth}
\centerline{
\hfill
\begin{minipage}[t]{5.2cm}
\centerline{
\epsfig{file=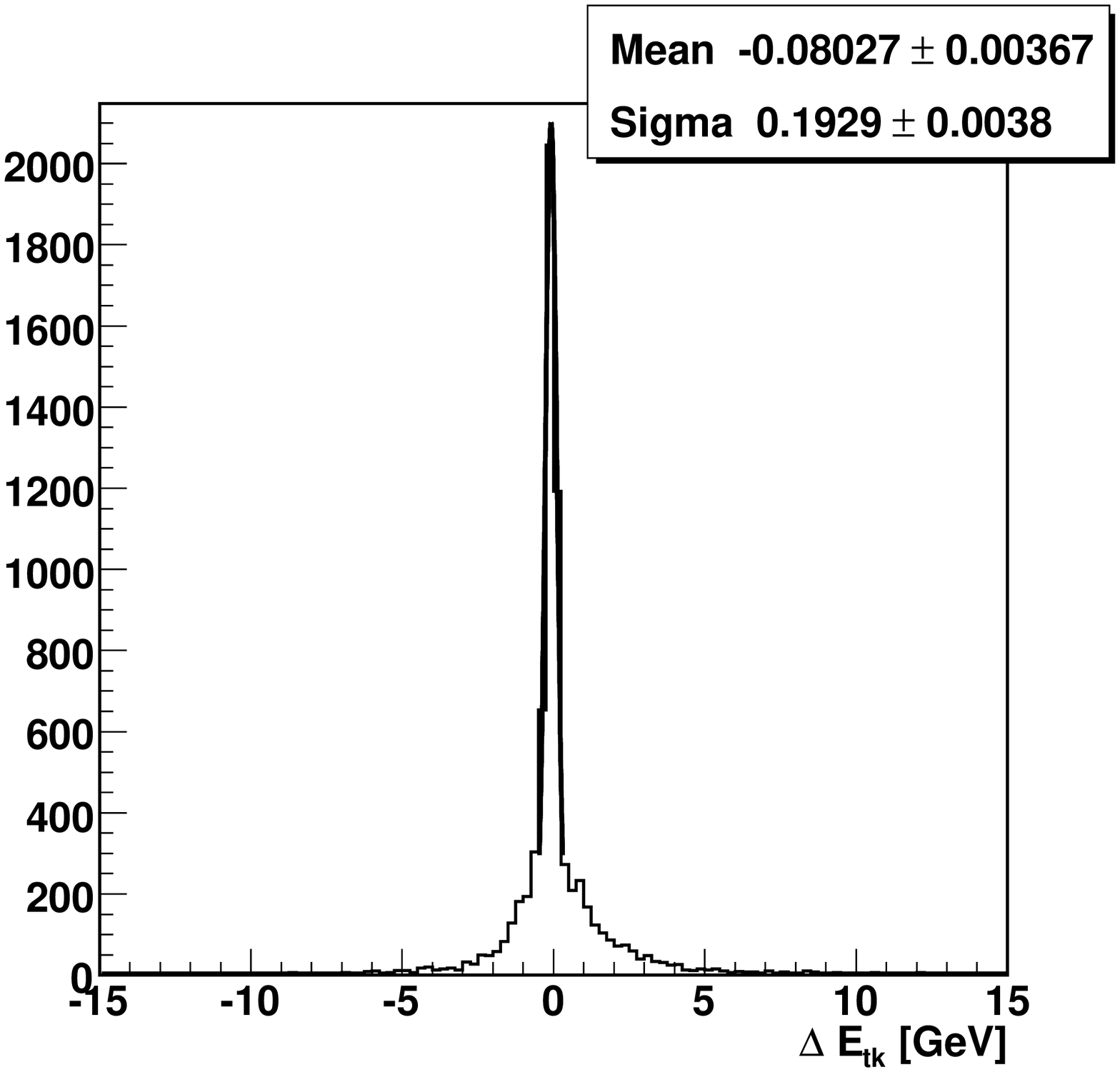,height=5cm}}
\end{minipage}
\hfill
\begin{minipage}[t]{5.2cm}
\centerline{
\epsfig{file=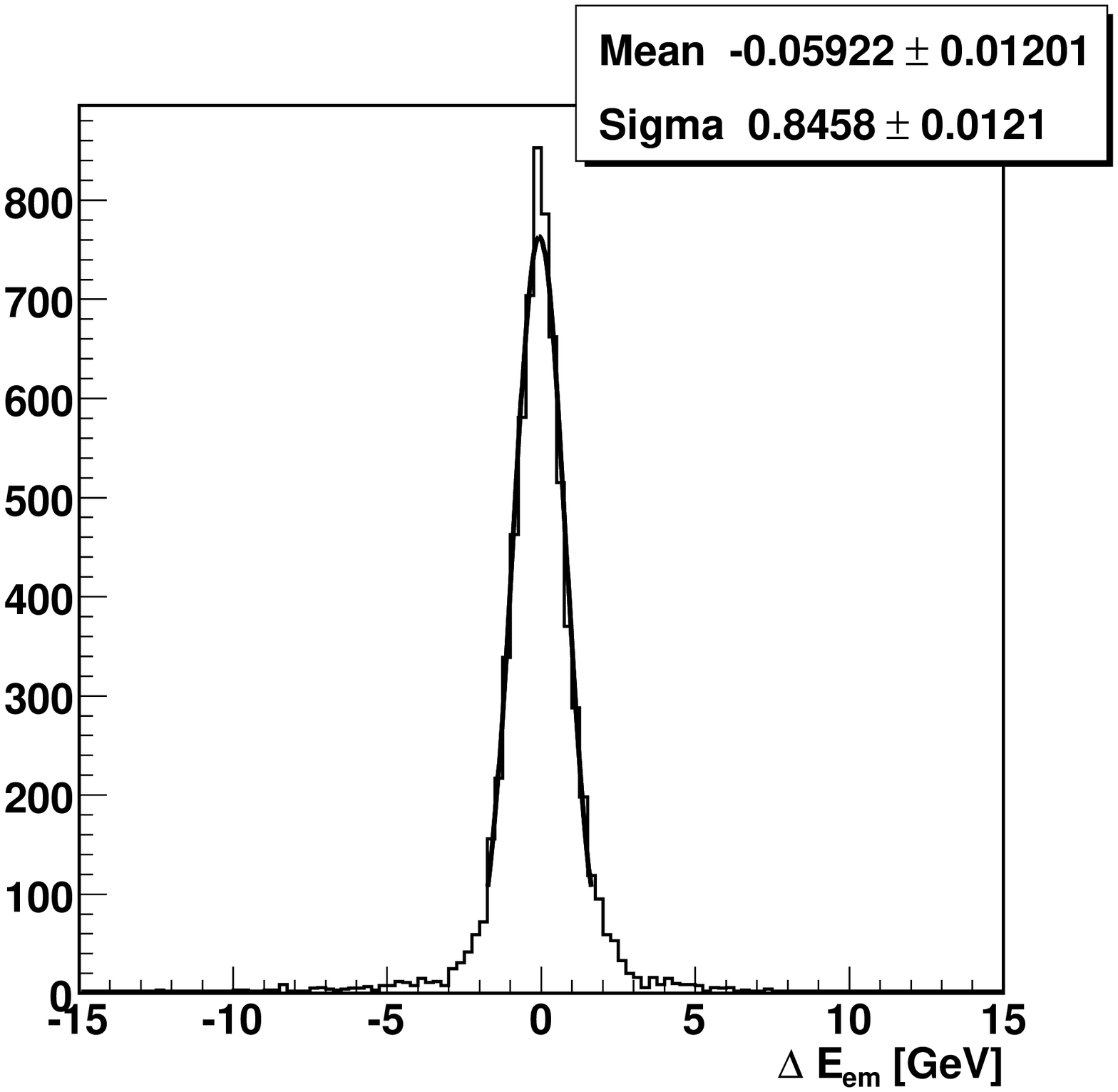,height=5cm}}
\end{minipage}
\hfill
\begin{minipage}[t]{5.2cm}
\centerline{
\epsfig{file=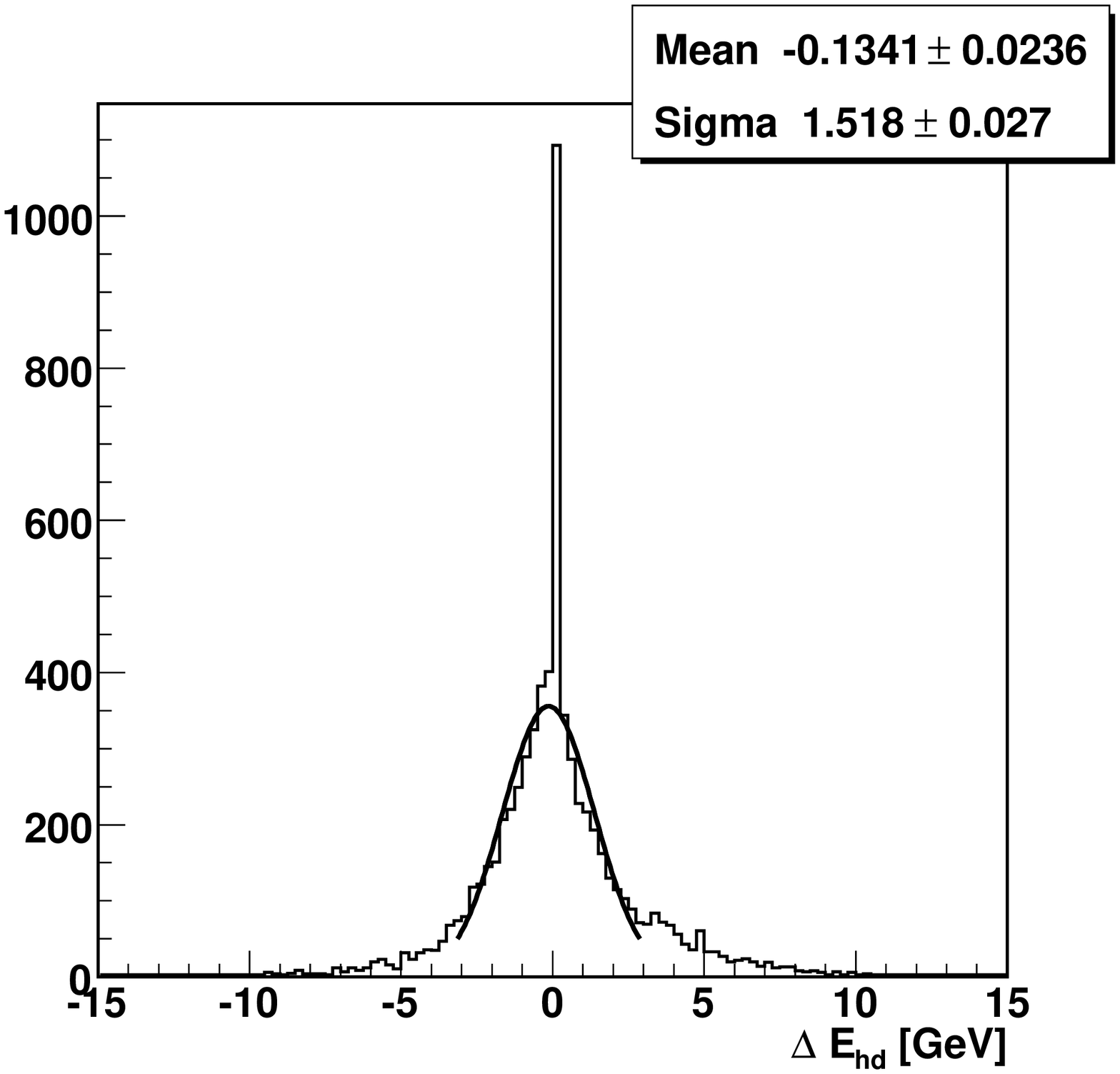,height=5cm}}
\end{minipage}
\hfill
}
\caption[Fig:deresol91]{\label{Fig:deresol91} \small \it
Distributions of the difference of the total reconstructed energies
from the corresponding MC truths for  
(a) charged PFOs, 
(b) neutral PFOs with electromagnetic showers, and
(c) neutral PFOs with hadron showers.
}
\end{minipage}\end{center}
\end{figure}

With the effect of the double counting eliminated, we can now
break up the contributions to the jet energy resolution
from detector resolutions on an event-by-event basis using the history keeper.
Figs.$\,$\ref{Fig:deresol91} a) through c) plot the
differences of the reconstructed energies from their
corresponding MC truths for the charged PFOs, 
the neutral electromagnetic PFOs,
and the neutral hadronic PFOs, respectively. 
Notice that as the MC truths we use the energies of primary
particles to ensure that their sum to have a $\delta$-function-like
distribution centered at $E_{\rm CM} = 91.18\,$GeV.
%
% ------------------
%  Fig.10
% ------------------
%
The measurement error distribution for the charged PFOs has
a sharp peak with a rather broad component beneath it.
Since we are comparing the measured charged track energies with those of their
corresponding primary particles,
the reconstructed energies can be significantly lower than those expected
from the detector resolution when some secondary particles are left undetected
as in the case of kink tracks or when the $dE/dx$ contribution is non-negligible.
On the other hand, the energies of some secondary charged particles 
kicked out from the detector materials, protons in particular, 
can be overestimated if the mass energies are included.
These two effects explain the broad component.
By the same token, the apparent resolutions are worse than those expected from
the detector performance also for the neutral PFOs.
One can also notice a sharp peak at zero in Fig.$\,$\ref{Fig:deresol91} c).
This is due to events in which there is no detected neutral hadrons.

Figs.$\,$\ref{Fig:deresol91} a) through c) tell us the contributions from 
(a) the tracker resolution ($\sigma_{\rm tk} = 0.2\,$GeV: narrow component only), 
(b) the calorimeter resolution for electromagnetic showers ($\sigma_{\rm EM} = 0.85\,$GeV),
and (c) the calorimeter resolution for hadronic showers ($\sigma_{\rm HD} = 1.52\,$GeV).
%and (c) the calorimeter resolution for hadronic showers ($\sigma_{\rm HD} = 1.60\,$GeV).
These values can be compared with the previous values obtained with
the indirect method:
$\sigma_{\rm tk} = 0.39\,$GeV,
$\sigma_{\rm EM} = 1.02\,$GeV, and
$\sigma_{\rm HD} = 1.73\,$GeV.
% $\sigma_{\rm tk} = 0.5\,$GeV,
% $\sigma_{\rm EM} = 0.88\,$GeV, and
% $\sigma_{\rm HD} = 1.79\,$GeV.
Considering the energy double counting in the previous values,
the agreement is reasonable.\\

\begin{figure}[ht]
\begin{center}\begin{minipage}{\figurewidth}
\centerline{
\hfill
\begin{minipage}[t]{5.2cm}
\centerline{
\epsfig{file=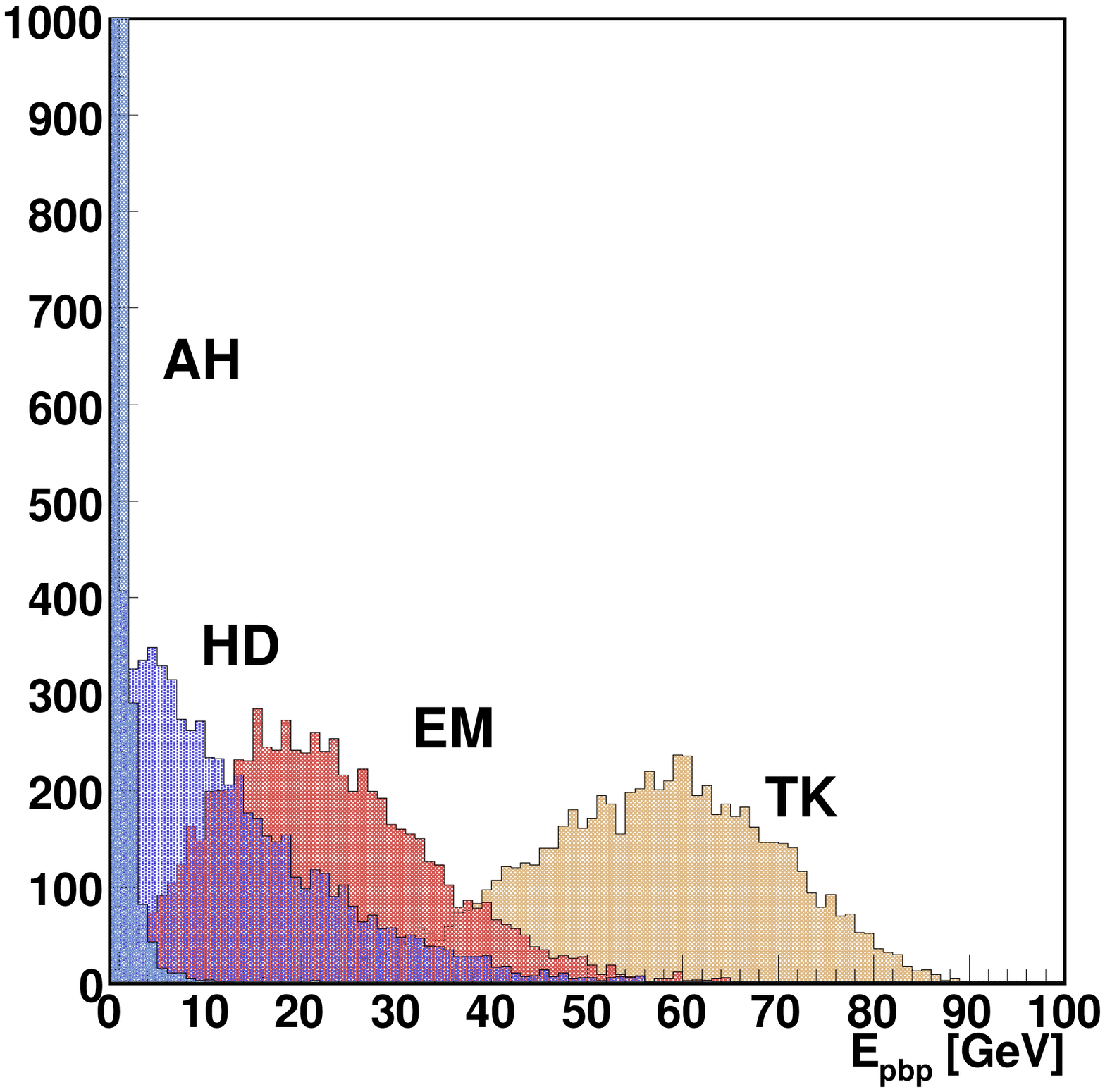,height=5cm}}
\end{minipage}
\hfill
\begin{minipage}[t]{5.2cm}
\centerline{
\epsfig{file=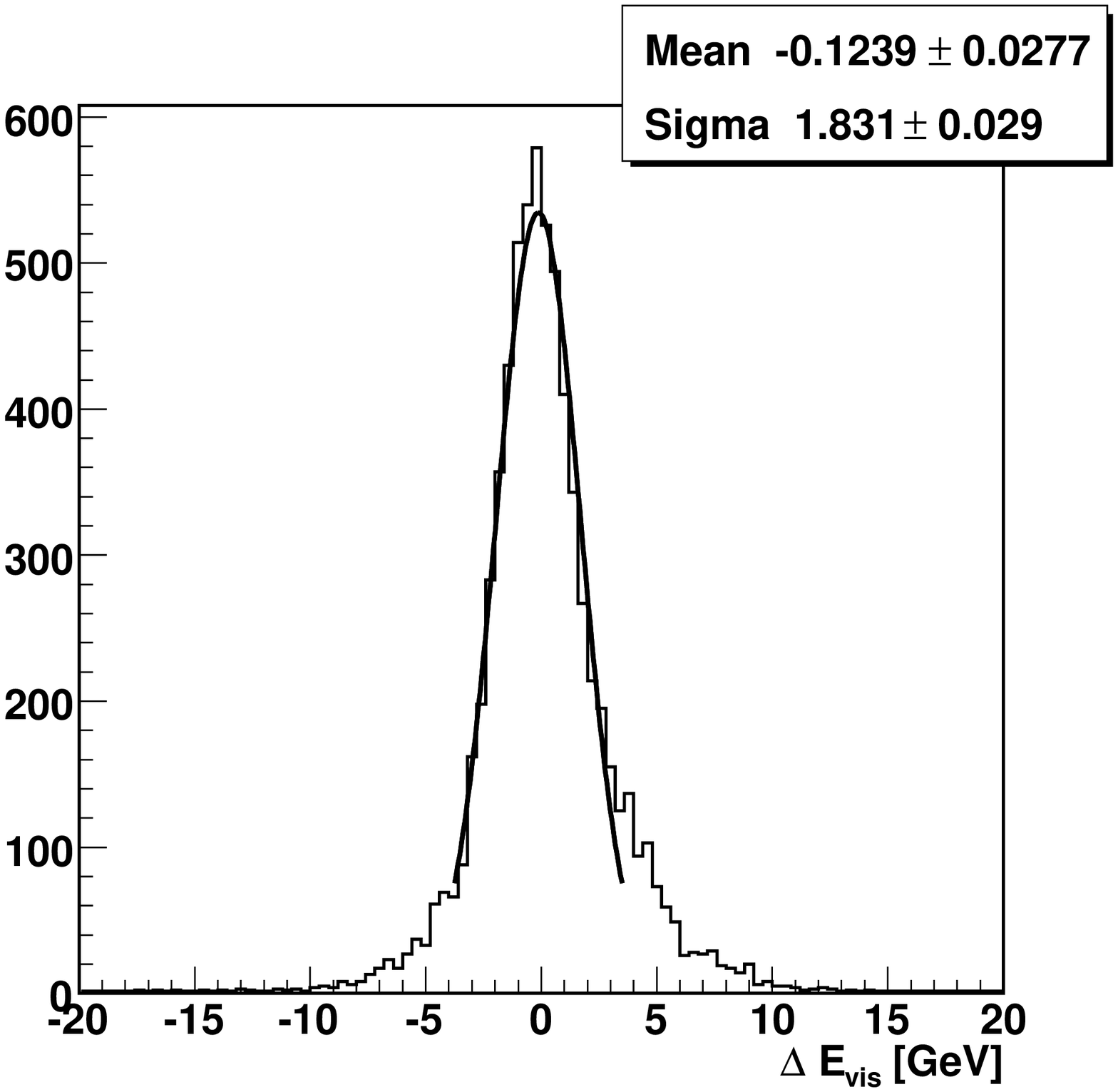,height=5cm}}
\end{minipage}
\hfill
}
\caption[Fig:devis91]{\label{Fig:devis91} \small \it
%(a) MC truths for the energy sum distributions for charged (gold), neutral electromagnetic (brown),
%and neutral hadronic (blue) PFOs corresponding to Figs.$\,$\ref{Fig:deresol91} a) through c), and
%undetected primary particles (dark blue); 
(a) MC truths for the energy sum distributions for charged (TK), neutral electromagnetic (EM),
and neutral hadronic (HD) PFOs corresponding to Figs.$\,$\ref{Fig:deresol91} a) through c), and
%undetected primary particles due to acceptance holes and neutrinos (AH); 
undetected primary particles due to acceptance holes (AH); 
(b) the measurement error in the total visible energy corresponding to the sum of these three.
}
\end{minipage}\end{center}
\end{figure}

Fig.$\,$\ref{Fig:devis91} a) shows the decomposition of the true visible
energy into the three components corresponding to
those shown in Figs.$\,$\ref{Fig:deresol91} a) through c).
%
% ------------------
%  Fig.11
% ------------------
%
% Etk = 56.2 GeV (61.6%)
% Eem = 22.6 GeV (24.8%)
% Ehd = 11.6 GeV (12.7%)
% Eah = 0.8 GeV (0.9%)
%
%  \sigma_{E}^2 & = & \frac{1}{N} \, \sum_{i~1}^{N} \, \frac{1}{n_{{\rm tk}, i}} \, 
%  					\sum_{j \in {\rm tk}}   \, \sigma_j^2
%                                    + \frac{1}{N} \, \sum_{i~1}^{N} \, \frac{1}{n_{{\rm tk}, i}} \, 
%					\sum_{j \in {\rm EM}}}  \, \sigma_j^2 
%                                    + \frac{1}{N} \, \sum_{i~1}^{N} \, \frac{1}{n_{{\rm tk}, i}} \, 
%					\sum_{j \in {\rm Hd}}}  \, \sigma_j^2 
% \cr\rule{0in}{4ex}
%  				 & = & \frac{1}{N} \, \sum_{i~1}^{N} \, \frac{1}{n_{{\rm tk}, i}} \, 
%  					\sum_{j \in {\rm tk}}   \, \sigma_j^2
%                                    + \frac{1}{N} \, \sum_{i~1}^{N} \, \frac{1}{n_{{\rm tk}, i}} \, 
%					\sum_{j \in {\rm EM}}}  \, (0.15 \sqrt{E_j})^2 
%                                    + \frac{1}{N} \, \sum_{i~1}^{N} \, \frac{1}{n_{{\rm tk}, i}} \, 
%					\sum_{j \in {\rm Hd}}}  \, (0.40 \sqrt{E_j})^2 
% \cr\rule{0in}{4ex}
%  				& \simeq & \sigma_{\rm tk}^2
%                                    + 0.15 \times \left< E_{\rm EM} \right>
%                                    + 0.40 \times \left< E_{\rm HD} \right>
%
On the average, the charged, the neutral electromagnetic,
and the neutral hadronic PFOs share
$56.2$, $22.6$, and $11.6\,$GeV of the center of mass energy on the $Z^0$ pole.
Ignoring the constant terms in the calorimeter resolutions, we can relate
these average energies to the resolution contribution:
\begin{eqnarray}
\sigma_{E}^2 & = & \frac{1}{N_{\rm evt}} \, \sum_{i=1}^{N_{\rm evt}} \, \frac{1}{N_{{\rm tk}, i}} \, 
  					\sum_{j \in {\rm tk}}   \, \sigma_j^2
                                    + \frac{1}{N_{\rm evt}} \, \sum_{i=1}^{N_{\rm evt}} \, \frac{1}{N_{{\rm EM}, i}} \, 
					\sum_{j \in {\rm EM}}  \, \sigma_j^2 
                                    + \frac{1}{N_{\rm evt}} \, \sum_{i=1}^{N_{\rm evt}} \, \frac{1}{N_{{\rm HD}, i}} \, 
					\sum_{j \in {\rm HD}}  \, \sigma_j^2 
\cr\rule{0in}{5ex}
  			   & \simeq & \sigma_{\rm tk}^2
                                    + \frac{1}{N_{\rm evt}} \, \sum_{i~1}^{N_{\rm evt}} \, \frac{1}{N_{{\rm EM}, i}} \, 
					\sum_{j \in {\rm EM}}  \, (0.15 \times \sqrt{E_j})^2 
                                    + \frac{1}{N_{\rm evt}} \, \sum_{i~1}^{N_{\rm evt}} \, \frac{1}{N_{{\rm HD}, i}} \, 
					\sum_{j \in {\rm HD}}  \, (0.43 \times \sqrt{E_j})^2 
%					\sum_{j \in {\rm HD}}  \, (0.40 \times \sqrt{E_j})^2 
\cr\rule{0in}{4ex}
  				& = & \sigma_{\rm tk}^2
                                    + (0.15)^2 \times \left< E_{\rm EM} \right>
                                    + (0.43)^2 \times \left< E_{\rm HD} \right>  
%                                    + (0.40)^2 \times \left< E_{\rm HD} \right>  
~ = ~ \sigma_{\rm tk}^2  +  \sigma_{\rm EM}^2 + \sigma_{\rm HD}^2  .
\nonumber
\end{eqnarray}
This gives the following estimates for the contributions from the detector resolutions:
$\sigma_{\rm EM} = 0.15 \times \sqrt{22.6} = 0.71\,$GeV and 
$\sigma_{\rm HD} = 0.43 \times \sqrt{11.6} = 1.46\,$GeV.
%$\sigma_{\rm HD} = 0.40 \times \sqrt{11.6} = 1.36\,$GeV.
These values are significantly smaller than those obtained from
Figs.$\,$\ref{Fig:deresol91} b) and c).
As pointed out above, this is because the measurement errors are defined as
the differences between the measured PFO energies and those of the corresponding
primary particles, which sometimes decay before making those PFOs.
In order to confirm this we have looked at the difference between 
the measured PFO energies and the MC truths 
for the directly corresponding particles and found:
$\sigma_{\rm EM} = 0.70\,$GeV and 
$\sigma_{\rm HD} = 1.41\,$GeV, 
%$\sigma_{\rm HD} = 1.38\,$GeV, 
being in good agreement with the above estimates.
%
% Figs: Resolution contributions with BPs instead of PBPs.
% lower tail in dE_tk distribution from dE/dx.
% If mass energies subtracted for the secondary particles kicked out from detector materials
% there will be a higher tail.
%
Fig.$\,$\ref{Fig:devis91} b) shows the distribution of the sum of the contributions from the
three components.
The total detector resolution contribution of $\sigma_{\rm det} = 1.83\,$GeV is
%The total detector resolution contribution of $\sigma_{\rm det} = 1.90\,$GeV is
consistent with the quardratic sum:
$\sigma_{\rm det} = \sqrt{0.39^2 ({\rm tk})+ 0.85^2 ({\rm EM}) + 1.52^2 ({\rm HD})} = 1.78\,$GeV,
%$\sigma_{\rm det} = \sqrt{0.52^2 ({\rm tk})+ 0.85^2 ({\rm EM}) + 1.5^2 ({\rm HD})} = 1.82\,$GeV,
%$\sigma_{\rm det} = \sqrt{0.52^2 ({\rm tk})+ 0.85^2 ({\rm EM}) + 1.6^2 ({\rm HD})} = 1.88\,$GeV,
when we assume $\sigma_{\rm tk} = 0.39\,$GeV instead of $0.20\,$GeV
taken into account the broad component.
This confirms the statistical independence of 
the resolution contributions from the different detectors as expected.

\subsubsection{Energy Dependence}

\begin{figure}[h]
\begin{center}\begin{minipage}{\figurewidth}
\centerline{
\begin{minipage}[t]{5.2cm}
\centerline{
\epsfig{file=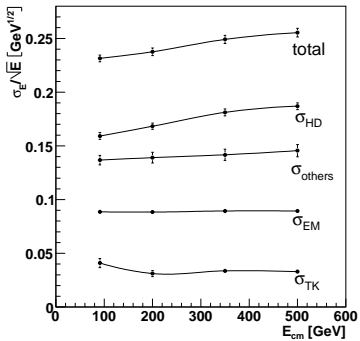,height=5cm}} % w/o albedo w/ e-dep calib
\end{minipage}
}
\caption[Fig:edep]{\label{Fig:edep} \small \it
Energy dependence of the energy resolutions normalized by
the square root of the center of mass energies.
The smooth curves in the figure are drawn just to guide eyes.
}
\end{minipage}\end{center}
\end{figure}

The next question is how the jet energy resolution depends on the energy.
Fig.$\,$\ref{Fig:edep} plots the energy resolutions normalized by
the square root of the center of mass energies for $E_{\rm CM} = $91.18, 200, 350, 500$\,$GeV.
%
% ------------------
%  Fig.12
% ------------------
%
From the bottom to the top,
the lines correspond to the contribution from the charged PFOs ($\sigma_{\rm tk}$),
that from the acceptance hole, neutrinos, and the energy double counting ($\sigma_{\rm others}$),
that from the neutral electromagnetic PFOs ($\sigma_{\rm EM}$),
that from the neutral hadronic PFOs ($\sigma_{\rm HD}$), and
the total jet energy resolution.
%
% Why tracking contribution has almost no E-dependence?
% Why neutral hadron contribution increases with E_{cm}?
%   --> How energy fractions depend on E_{cm}?
%          <E_tk>/Ecm = 0.62, 0.61, 0.61, 0.61
%          <E_em>/Ecm = 0.25, 0.25, 0.25, 0.25
%          <E_hd>/Ecm = 0.127, 0.128, 0.133, 0.138
%          <E_ah>/Ecm = 0.0092, 0.0055, 0.0038, 0.0029 
%          --> <Ehd> slightly increases with E_{cm} but not enough to explain the increase.
%   --> Looks like a_{hd} increases with E_{cm} --> shower leakage?
% What happens when normalized by E_{tk}, E_{EM}, and E_{HD}
%   --> a_tk = 0.026, 0.039, 0.034, 0.042 (sharp component)
%         a_em = 0.18, 0.18, 0.18, 0.18
%         a_hd = 0.47, 0.50, 0.52, 0.51
%      for Ecm = 91, 200, 350, 500 GeV
% What happens if we use the indirect method?
%   --> dEtk = 0.52, 0.66, 0.65, 0.94
%         dEem = 0.88, 1.60, 2.31, 2.59
%         dEhd = 1.79, 2.68, 3.75, 4.81
%         dEothers = 0.83, 1.01, 1.10, 1.18
%
The figure suggests that our model detector would have a jet energy resolution:
$\sigma_{E_{\rm jet}} \simeq 0.23 \times \sqrt{E_{\rm jet} \, \mbox{[GeV]}}$
with a slight increase with energy.
The increase can mostly be attributed to the contribution from the neutral hadronic PFOs.
The energy dependence of the contribution from the neutral hadronic PFOs is
largely due to our unsophisticated calorimeter calibration with a
single conversion factor for each of the electromagnetic and hadronic 
components.
It might hence be reduced by improving the calibration method.
As stated earlier, optimization of the calibration method as well as the calorimeter
configuration is, however, beyond the scope of this paper. 

Notice that the contributions from the calorimeter resolutions are estimated
by the direct method, while that from the tracker resolution is by the indirect method here.
The remaining contribution ($\sigma_{\rm others}$) hence includes the fluctuation
of the double counted energies due to the finite detector resolutions and hence
is significantly larger than the remnant contribution we discussed above
in examining Fig.$\,$\ref{Fig:mresol91} c). 
It is also remarkable that the tracker contribution ($\sigma_{\rm tk}$) too
seems to scale approximately
as $\sigma_{\rm tk} \propto \sqrt{E_{\rm CM}}$ 
at least in the energy range considered here.
This is because the tracker contribution is estimated by the indirect method
in order to take into account the broad component which might as well be
dominated by some stochastic processes.
As discussed previously, 
the broad component indeed comes from stochastic processes such as 
undetected daughter particles or wrongly included mass energies
for protons kicked out from the detector materials. 
This is in contrast to the sharp component that is solely determined by the tracker resolution
which roughly scales as 
$\sigma_{\rm tk, sharp} \simeq 5 \times 10^{-4} \left< E_{\rm tk} \right>$,
indicating the dominance of relatively low-momentum tracks
for which the effect of multiple scattering is significant.

\subsection{$e^+e^- \to Z^0 Z^0$ Events}

In the previous subsection we have examined how jet energy resolution
depends on jet energy.
It is practically more important and probably more interesting to examine how the invariant mass resolution
of a jet pair from a gauge boson changes with its momentum,
since it would determine the $W/Z$ separation capability 
with their jet invariant mass measurements.
In this subsection we will hence look at the invariant mass resolution for
the jet pair from a $Z^0$ boson decay in the $e^+e^- \to Z^0 Z^0$ process.

\begin{figure}[ht]
\begin{center}\begin{minipage}{\figurewidth}
\centerline{
\hfill
\begin{minipage}[t]{5.2cm}
\centerline{
\epsfig{file=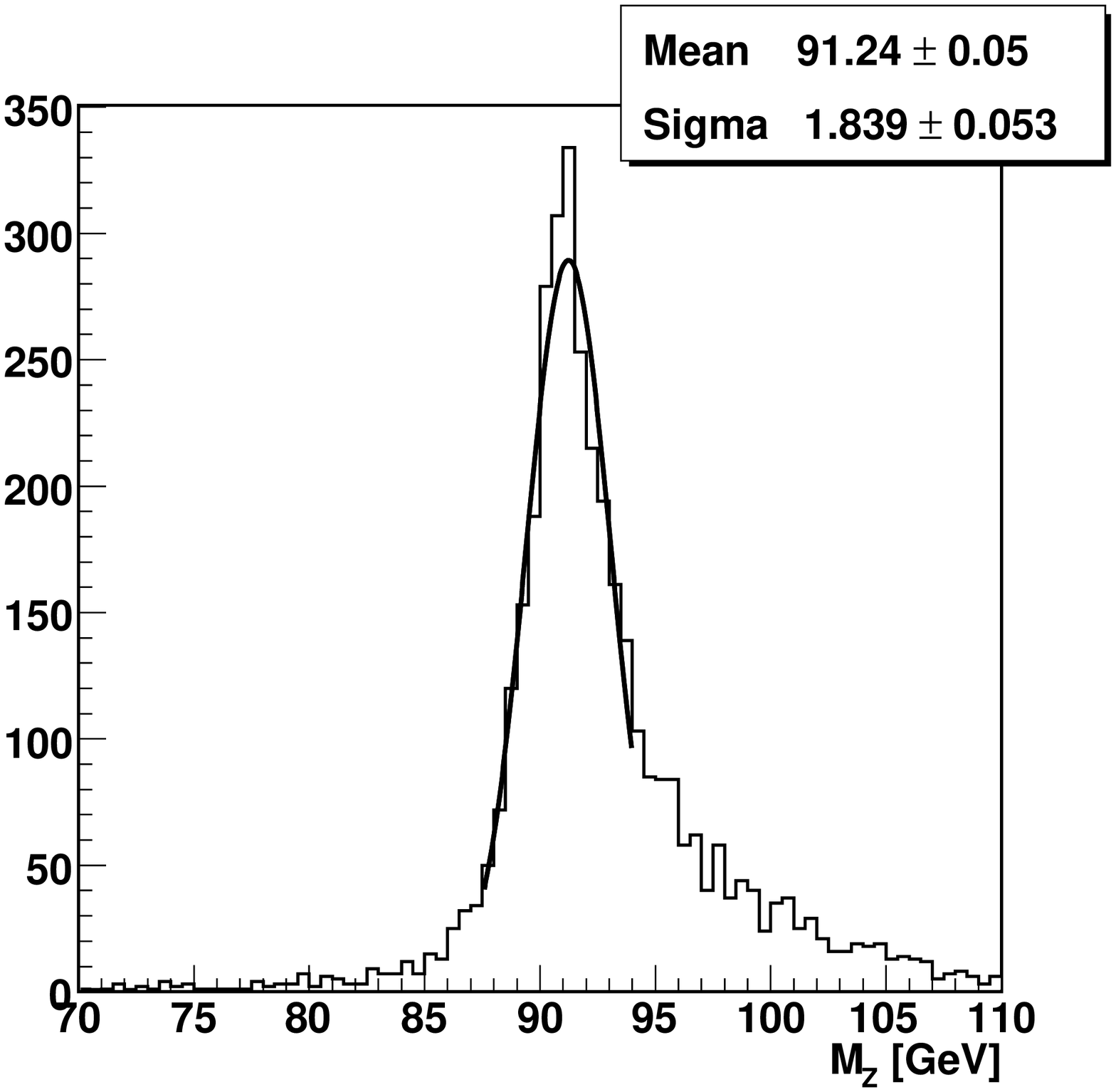,height=5cm}}
\end{minipage}
\hfill
\begin{minipage}[t]{5.2cm}
\centerline{
\epsfig{file=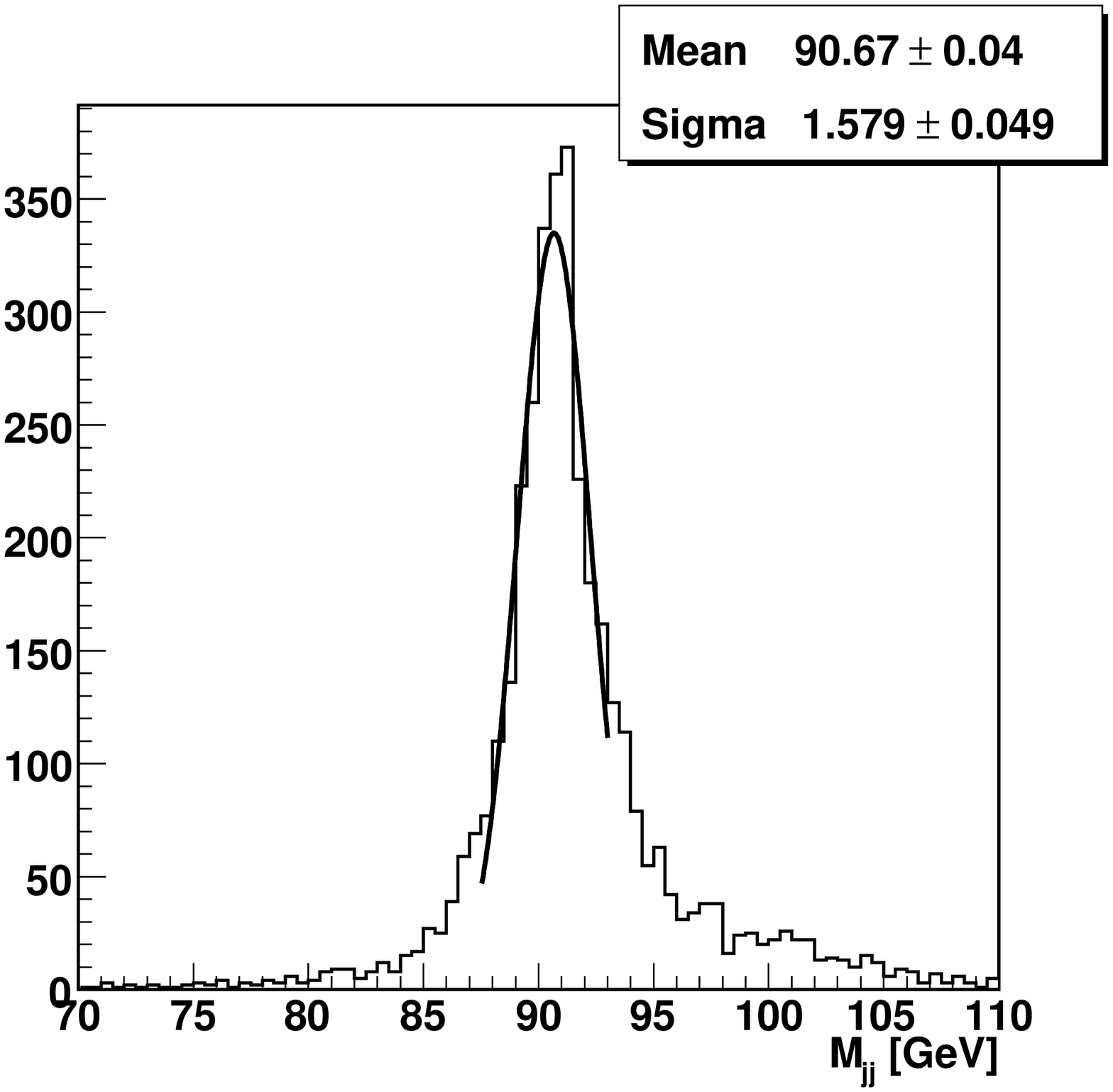,height=5cm}}
\end{minipage}
\hfill
\begin{minipage}[t]{5.2cm}
\centerline{
\epsfig{file=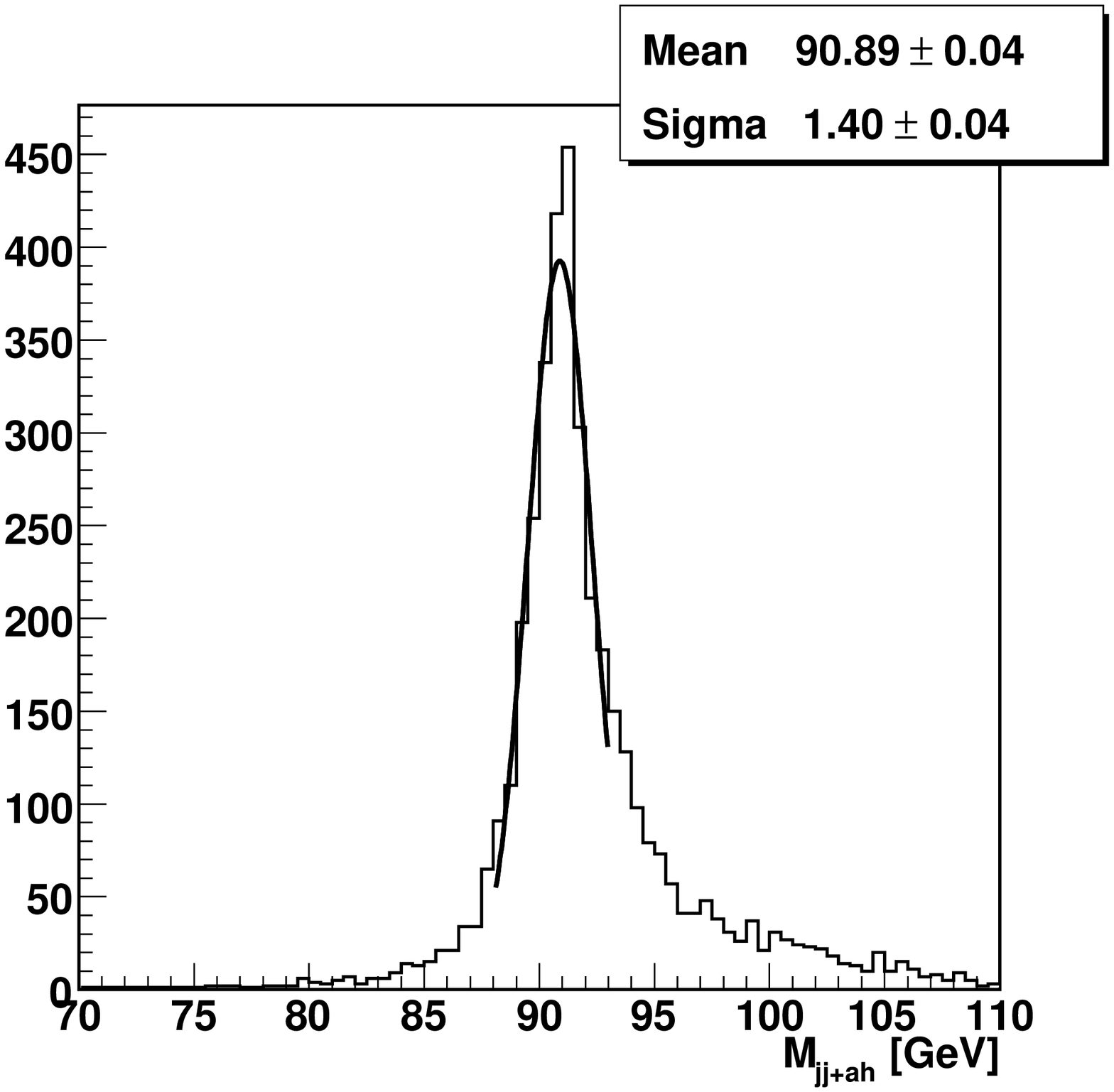,height=5cm}}
\end{minipage}
\hfill
}
\caption[Fig:mzz700]{\label{Fig:mzz700} \small \it
Jet invariant mass distributions for a jet pair from $e^+e^- \to Z^0 Z^0$ followed by
$Z^0 \to \nu \bar{\nu}$ and $Z^0 \to q \bar{q}$
(a) that with the standard treatment 
corresponding to the mass distribution in Fig.$\,$\ref{Fig:v0effect} a), 
(b) that with double counted daughter PFOs eliminated by using the history keeper, and
(c) the same as (b) but with the undetected primary particles artificially added in.
}
\end{minipage}\end{center}
\end{figure}

Fig.$\,$\ref{Fig:mzz700} a) shows the invariant mass distribution 
at $E_{\rm CM} = 700\,{\rm GeV}$ or, equivalently, a $Z^0$ energy of $350\,{\rm GeV}$.
Here we have relaxed the cut on  the absolute value of the cosine of the polar angle of each jet to 0.95 from its default 0.8,
since otherwise the acceptance becomes too small (less than $20\%$) due to the forward backward peaks 
of the $Z^0$ boson  production angle distribution.
Nevertheless the acceptance is significantly smaller (about $40\%$) at $E_{\rm CM} = 700\,{\rm GeV}$
than that for the on-pole $Z^0$ production.
%
% ------------------
%  Fig.13
% ------------------
%
The figure tells us that, in this highly boosted case, the core part of the distribution has a narrower width than 
that of  $Z^0$ bosons at rest, indicating the improvement of the relative error ($\sigma_{E}/E$) with energy,
while there is a significant tail towards the higher mass region.
The Gaussian fit range is changed here to $(-2, +1.5)$-$\sigma$s so as 
to avoid the higher tail affecting the fit.
By eliminating the double counted daughter PFOs using MC truths, 
we can suppress the higher tail to some extent as shown in Fig.$\,$\ref{Fig:mzz700} b)
but not completely.
In order to see the acceptance hole effect we plot the mass distribution
in Fig.$\,$\ref{Fig:mzz700} c) with the undetected primary particles artificially added in.
The width of the core part becomes significantly narrower, but the higher tail persists
as expected; their addition could only enhance the higher tail but not reduce it.

The origin of the higher tail has been traced to the over-counted mass energies for the charged PFOs
corresponding to secondary protons kicked out from the detector materials. 
As with the secondary neutrons kicked out from the detector materials, 
the mass energies of these secondary protons should be subtracted.
%
% ------------------
%  Fig.14
% ------------------
%
\begin{figure}[ht]
\begin{center}\begin{minipage}{\figurewidth}
\centerline{
\hfill
\begin{minipage}[t]{5.2cm}
\centerline{
\epsfig{file=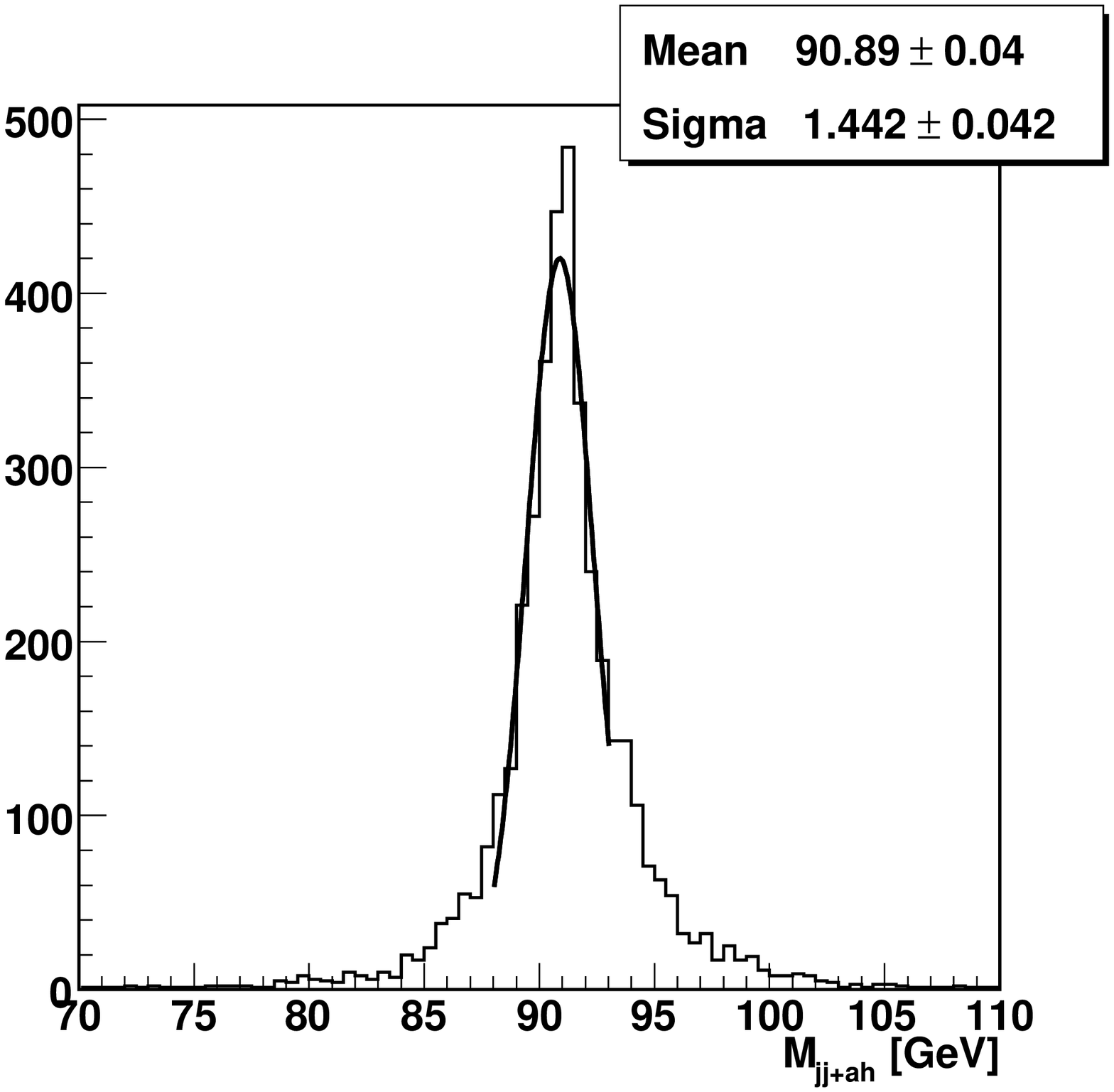,height=5cm}}
\end{minipage}
\hfill
\begin{minipage}[t]{5.2cm}
\centerline{
\epsfig{file=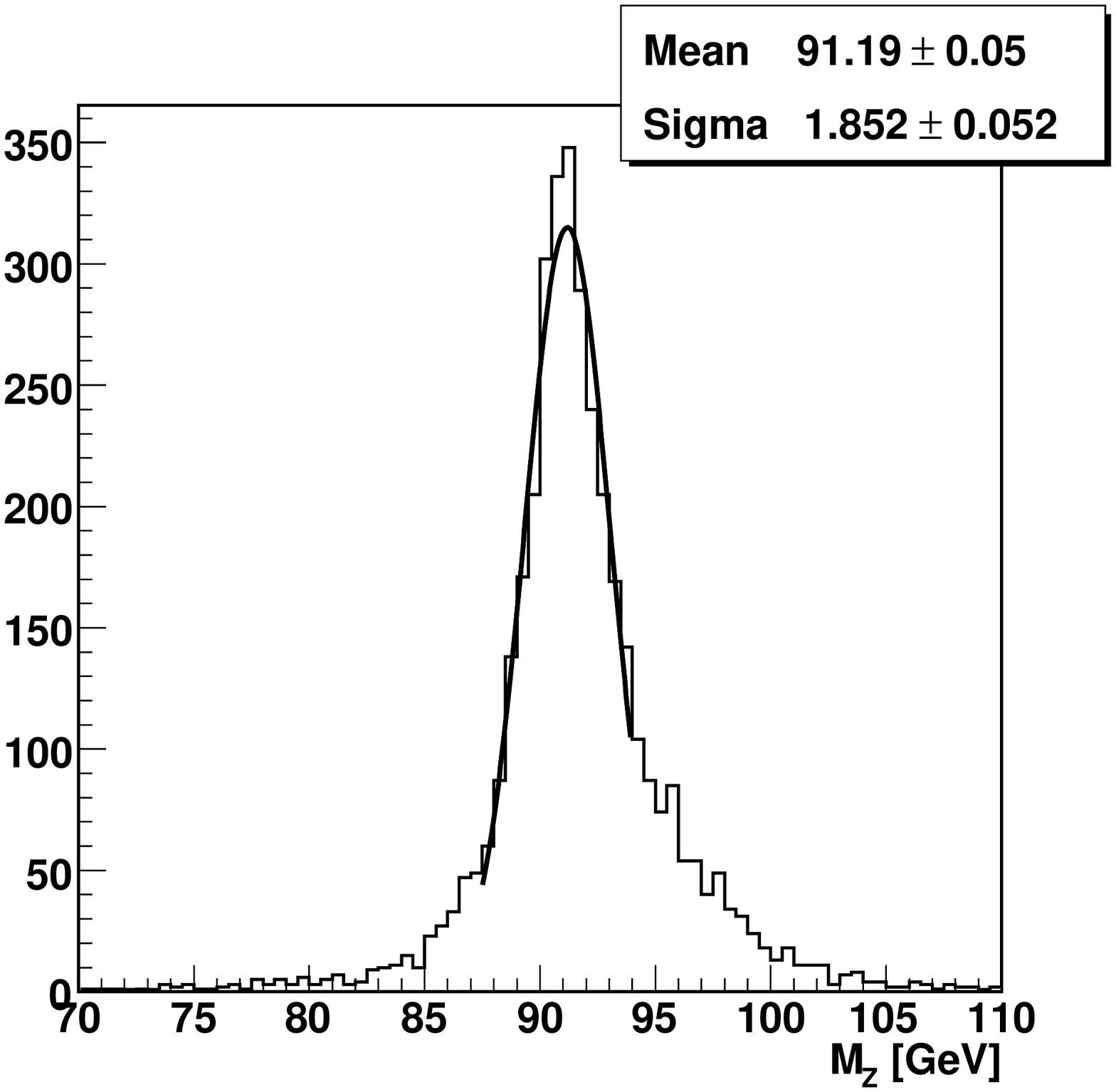,height=5cm}}
\end{minipage}
\hfill
\begin{minipage}[t]{5.2cm}
\centerline{
\epsfig{file=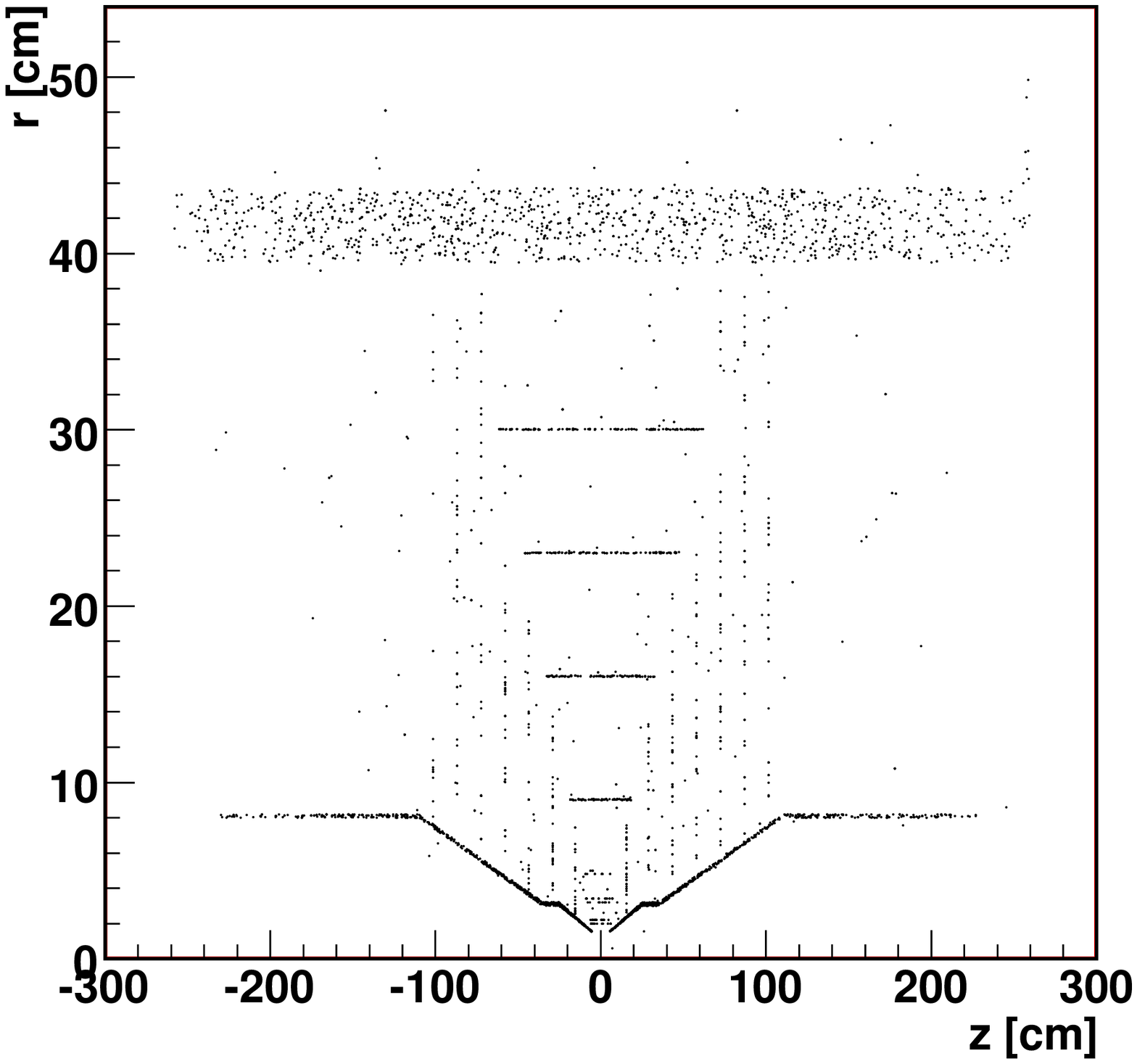,height=5cm}}
\end{minipage}
\hfill
}
\caption[Fig:mzz700mpc]{\label{Fig:mzz700mpc} \small \it
Jet invariant mass distributions for a jet pair from $e^+e^- \to Z^0 Z^0$ followed by
$Z^0 \to \nu \bar{\nu}$ and $Z^0 \to q \bar{q}$
(a) that corresponding to the mass distribution in Fig.$\,$\ref{Fig:mzz700} c), and
(b) that corresponding to the mass distribution in Fig.$\,$\ref{Fig:mzz700} a),
but with the mass energies subtracted for secondary protons kicked out from the detector materials. 
(c) The production points of secondary protons kicked out from the detector materials in the $r$-$z$ view.
}
\end{minipage}\end{center}
\end{figure}
Fig.$\,$\ref{Fig:mzz700} c) becomes Fig.$\,$\ref{Fig:mzz700mpc} a) after this mass energy subtraction.
Fig.$\,$\ref{Fig:mzz700mpc} c) plots the production points of the secondary protons kicked out from
the detector materials, which clearly reflects the material distributions for the beam pipe,
the VTX and IT layers and the inner wall of the TPC.
Since the production points are so well defined we might be able to separate significant fraction
of these secondary protons and subtract their mass energies provided that they are identified
as protons by the energy loss measurements in the TPC\footnote{
Notice that just blindly assigning the pion mass to all the charged PFOs cannot remove
the higher tail since this mass becomes negligible for high momentum protons.
It is essential that these secondary protons are identified as protons and
assigned the correct proton mass and then from their total energies the proton mass
should be subtracted so as to just count their kinetic energies. 
}.
Fig.$\,$\ref{Fig:mzz700mpc} b) is the invariant mass distribution assuming that 
such mass energy subtraction is practicable.
The higher tail is significantly reduced as compared to Fig.$\,$\ref{Fig:mzz700} a).
In what follows, however, we will not apply this mass energy correction.
\\

\begin{figure}[h]
\begin{center}\begin{minipage}{\figurewidth}
\centerline{
\begin{minipage}[t]{5.2cm}
\centerline{
\epsfig{file=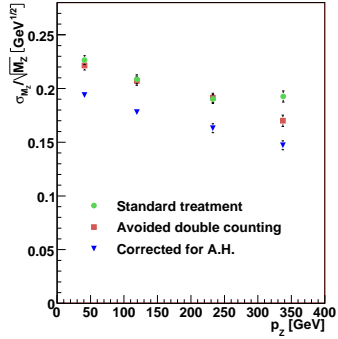,height=5cm}}
\end{minipage}
}
\caption[Fig:edepzz]{\label{Fig:edepzz} \small \it
Momentum dependence of the invariant mass resolutions for 2-jet systems from
$Z^0$ decays normalized by
the square root of the $Z^0$  mass.
}
\end{minipage}\end{center}
\end{figure}

Fig.$\,$\ref{Fig:edepzz} plots the invariant mass resolutions, as obtained by the
asymmetric Gaussian fits described above,
normalized by the square root of the $Z^0$ mass $M_{Z} = $91.18$\,$GeV
as a function of the $Z^0$  momentum.
%
% ------------------
%  Fig.15
% ------------------
%
The circles, squares, and reverse triangles are ones with the standard treatment,
with the double counting avoided, and
with the correction of undetected particle momenta, respectively.
The figure shows that the invariant mass resolution indeed improves with the $Z^0$ 
momentum, though it is not as quickly as $1/\sqrt{E_{Z}}$.
Denoting the two jet energies by $E_1$ and $E_2$ and ignoring the individual jet masses, we have
an approximate formula for
the relative error on the invariant mass of the jet pair:
\begin{eqnarray}
\frac{\sigma_{M}}{M} 
& \simeq & \frac{1}{2} \, \sqrt{\left( \frac{\sigma_{E_1}}{E_1} \right)^2 + \left( \frac{\sigma_{E_2}}{E_2} \right)^2}
~ \simeq ~ \frac{a}{2} \, \sqrt{\frac{1}{E_1} + \frac{1}{E_2}}, 
\nonumber
\end{eqnarray}
where $a \simeq 0.23 \sim 0.25$ as indicated by Fig.$\,$\ref{Fig:edep}.
This formula tells us that $\sigma_{M} / M$ would scale as $1/\sqrt{E_{Z}}$ if $E_1 = E_2$
but it corresponds to  the minimum expected value under the constraint that $E_1 + E_2 = E_{Z} = E_{\rm CM}/2$,
being consistent with the above observation.

\subsection{$e^+e^- \to Z^0 H^0$ Events}

As our last example, let us consider one of the most important processes
for the ILC experiment that is $e^+e^- \to Z^0 H^0$
and study the $H^0$ reconstruction in its hadronic decays.
Again we force $Z^0$ to decay into $\nu \bar{\nu}$ so as to
avoid the effect of jet clustering, but we impose no restriction 
on the $H^0$ decays.
We switch on the initial state radiation and beamstrahlung
to see possible complications we would encounter in reality.
The $Z^0$  and the $H^0$ widths are finite now but the
effect of their finiteness is negligible, 
since the $H^0$ width can be regarded as zero for the purpose of this analysis
and the $Z^0$ width only induces a small spread in the otherwise-monochromatic 
$H^0$ energy distribution.
Under these conditions we have generated 10k events 
at $E_{\rm CM} = 350\,{\rm GeV}$ with $M_H = 120\,{\rm GeV}$.

\begin{figure}[ht]
\begin{center}\begin{minipage}{\figurewidth}
\centerline{
\hfill
\begin{minipage}[t]{5.2cm}
\centerline{
\epsfig{file=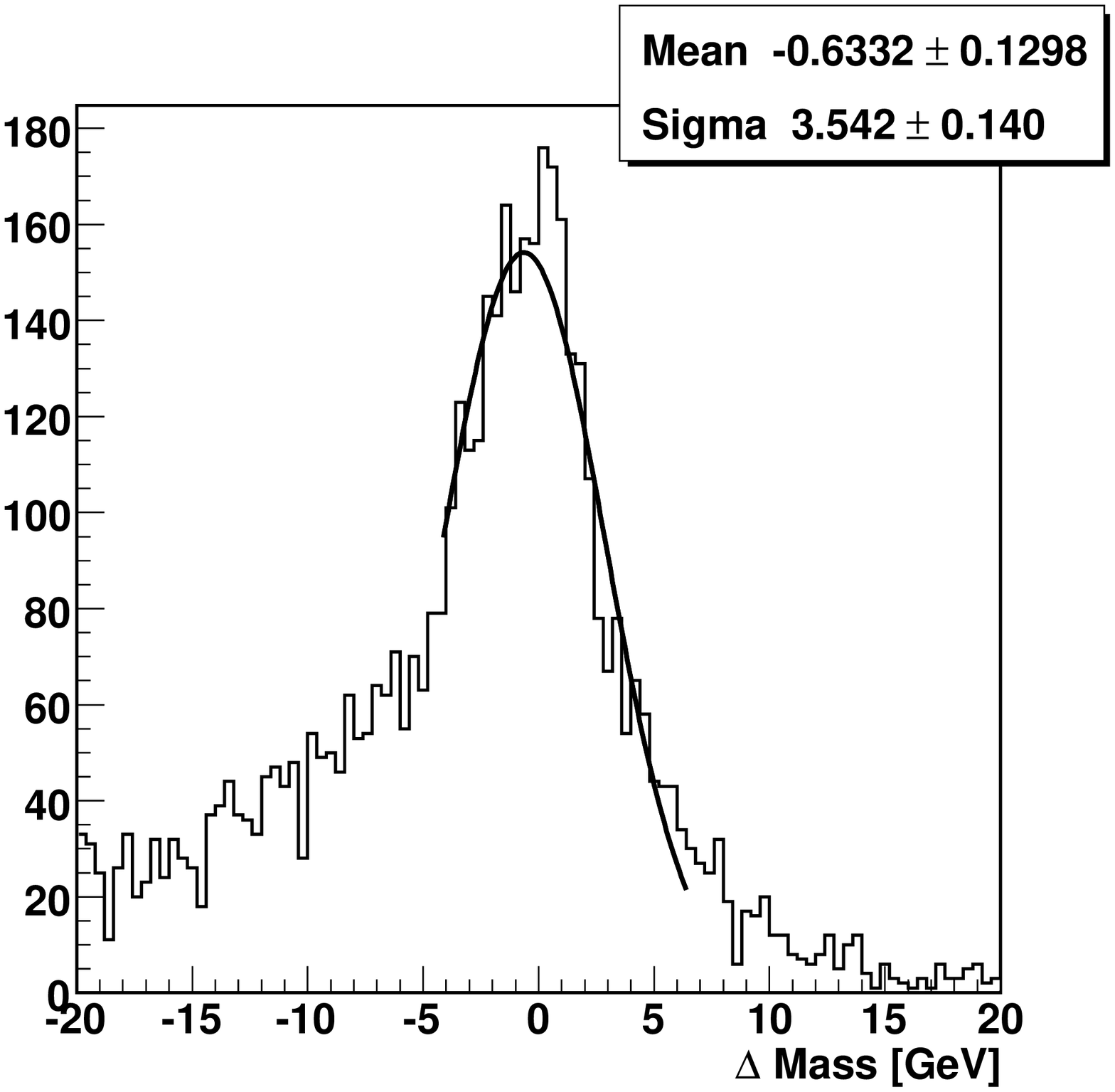,height=5cm}}
% <dmh> = -0.6322 +/- 0.1298
% sigma = 3.542 +/- 0.140
\end{minipage}
\hfill
\begin{minipage}[t]{5.2cm}
\centerline{
\epsfig{file=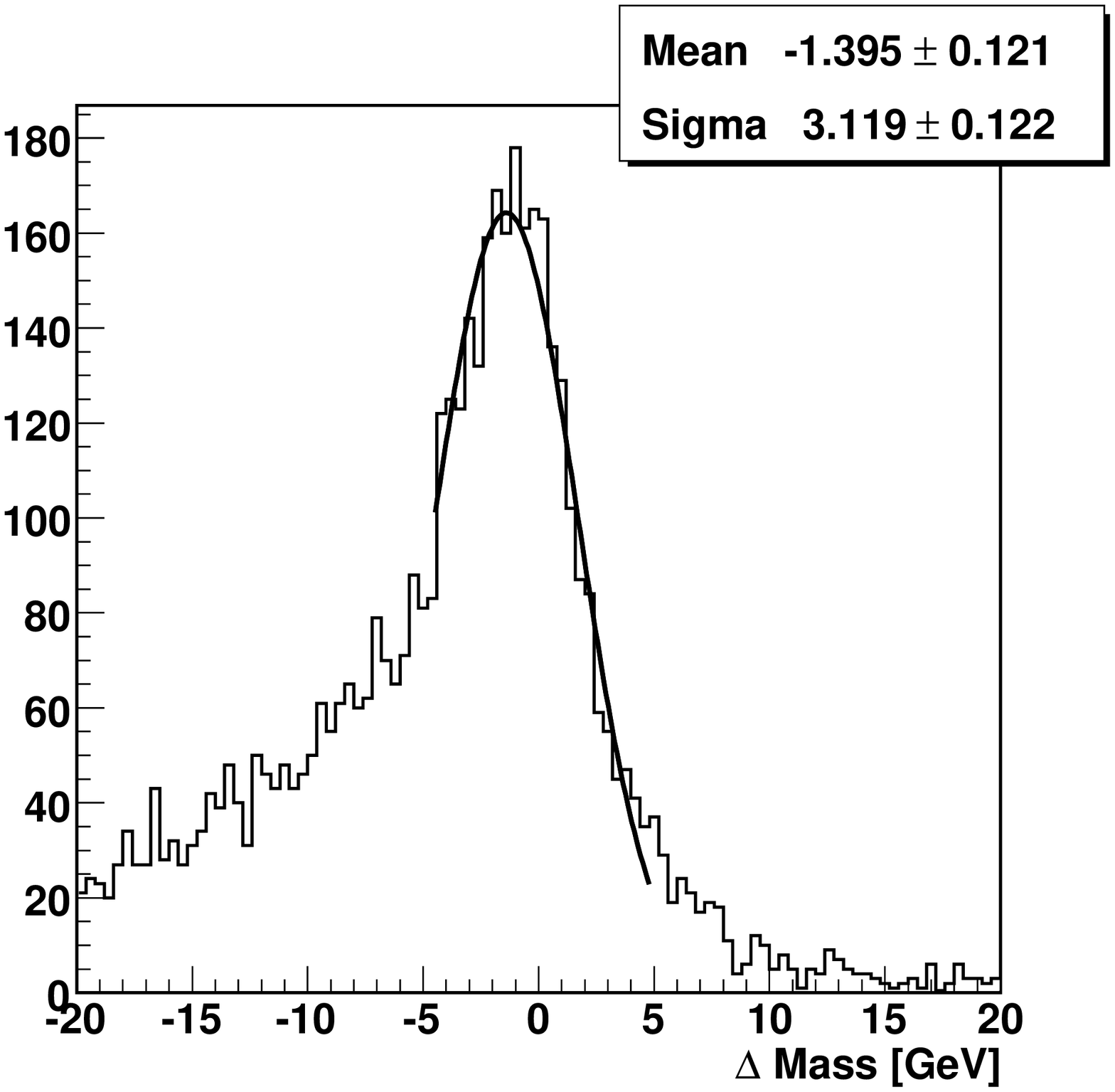,height=5cm}}
% <dmh> = -1.395 +/- 0.121
% sigma = 3.119 +/- 0.122
\end{minipage}
\hfill
\begin{minipage}[t]{5.2cm}
\centerline{
\epsfig{file=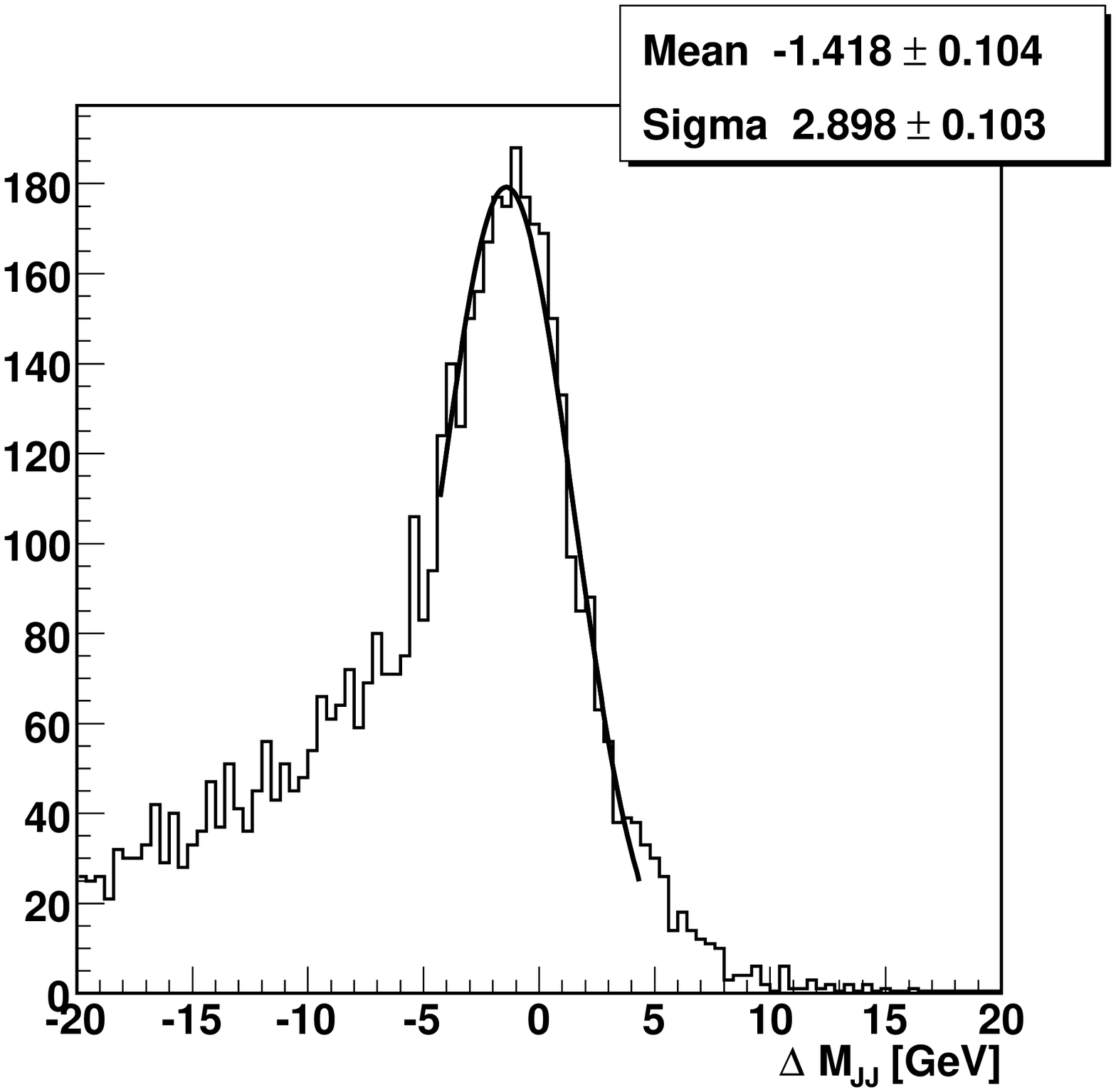,height=5cm}}
% <dmh> = -1.418 +/- 0.104
% sigma = 2.898 +/- 0.103
\end{minipage}
\hfill
}
\caption[Fig:dmh]{\label{Fig:dmh} \small \it
Jet invariant mass distributions for the reconstructed final-state jet systems for
$e^+e^- \to Z^0 H^0$ followed by $Z^0 \to \nu \bar{\nu}$
(a) that with the standard treatment, 
(b) that with double counted daughter PFOs eliminated by using the history keeper, and
(c) the same as (b) but with the photons from initial state radiations removed.
}
\end{minipage}\end{center}
\end{figure}

Fig.$\,$\ref{Fig:dmh} a) is the difference of the reconstructed final-sate
invariant mass from the nominal Higgs mass of $M_H = 120\,{\rm GeV}$.
%
% ------------------
%  Fig.16
% ------------------
%
One can see a huge lower tail there, which is coming
mostly from neutrinos from $b$-jet decays as we will see later.
To avoid the effect of this huge tail affecting the Gaussian fit,
we readjust the fit range here to $(-1, +2)$-$\sigma$s.
Nevertheless the resolution, $\sigma_{\Delta M} = 3.54\,{\rm GeV}$, 
is much worse than naive expectaion from the $Z^0$ boson cases studied above.
Though less dramatic, the higher tail seems also longer than the $Z^0$ boson cases.
Fig.$\,$\ref{Fig:dmh} b) is after eliminating double counted daughter PFOs.
The apparent width is reduced significantly to $\sigma_{\Delta M} = 3.12\,{\rm GeV}$
($\sqrt{3.54^2 - 3.12^2}=1.67\,$GeV: $(1.67/3.54)^2=22\%$ reduction)
but both the higher and the lower tails persist.
Some significant part of the higher tail turns out to be coming from
initial state radiations, which have been switched off artificially
for the $Z^0$ boson studies presented above.
Elimination of photons from the initial state radiations indeed 
suppresses the higher tail as shown in
Fig.$\,$\ref{Fig:dmh} c), reducing the sigma value
to $\sigma_{\Delta M} = 2.90\,$GeV ($\sqrt{3.12^2-2.90^2}=1.15\,$GeV: 11\% reduction).
\\

\begin{figure}[ht]
\begin{center}\begin{minipage}{\figurewidth}
\centerline{
\hfill
\begin{minipage}[t]{5.2cm}
\centerline{
\epsfig{file=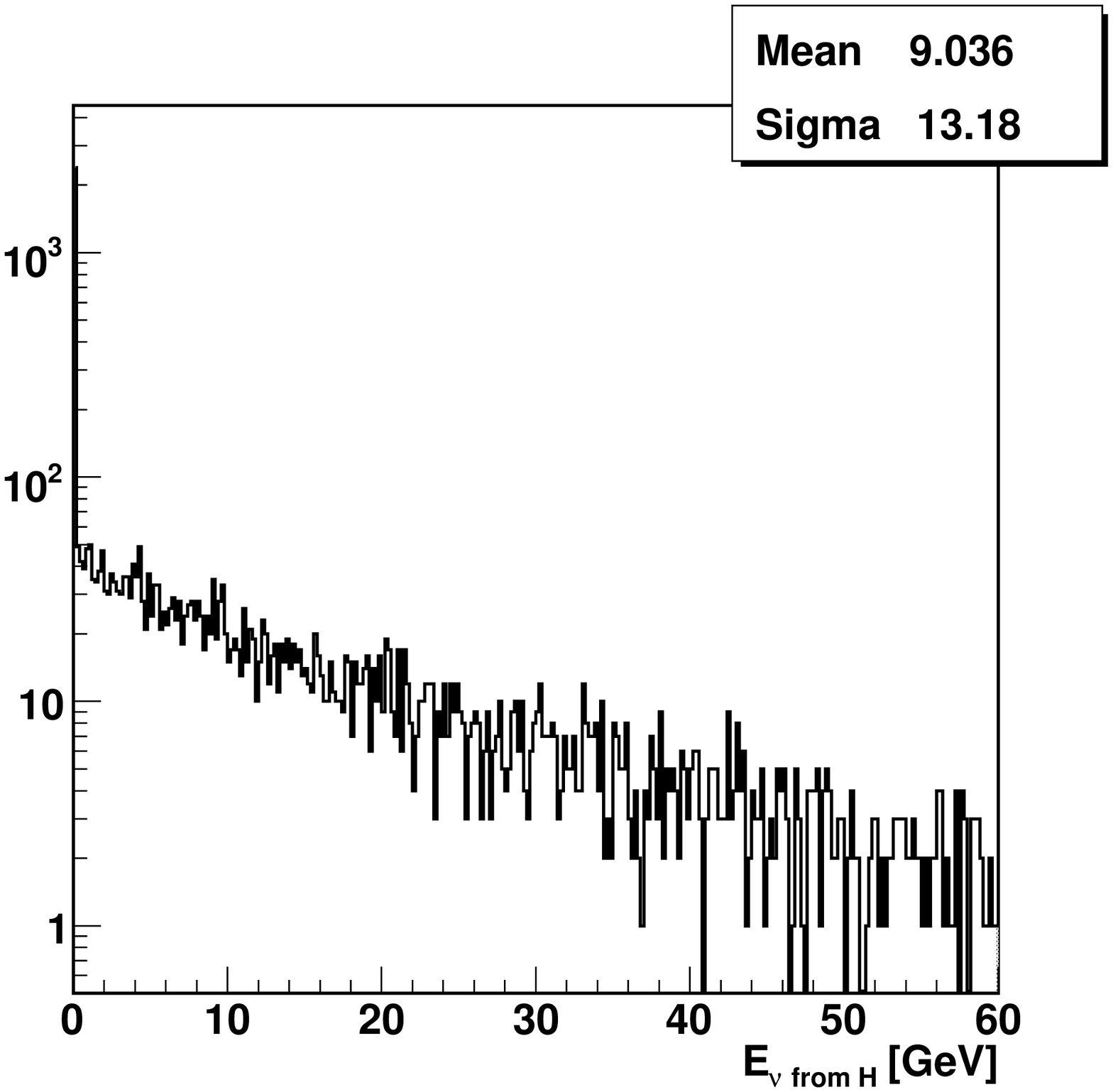,height=5cm}}
\end{minipage}
\hfill
\begin{minipage}[t]{5.2cm}
\centerline{
\epsfig{file=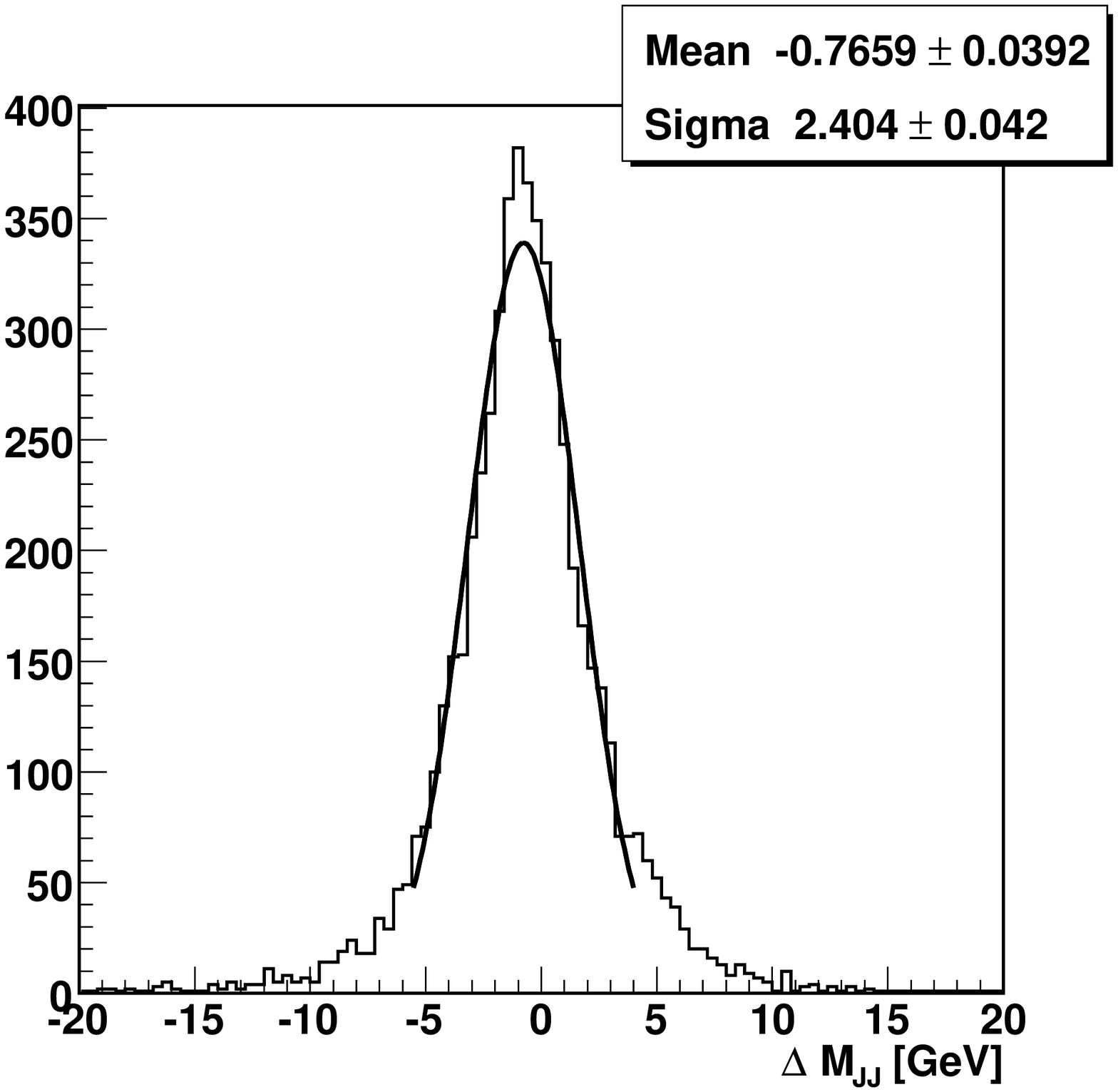,height=5cm}}
% <dmh> = -0.766 +/- 0.04
% sigma = 2.40 +/- 0.04
\end{minipage}
\hfill
\begin{minipage}[t]{5.2cm}
\centerline{
\epsfig{file=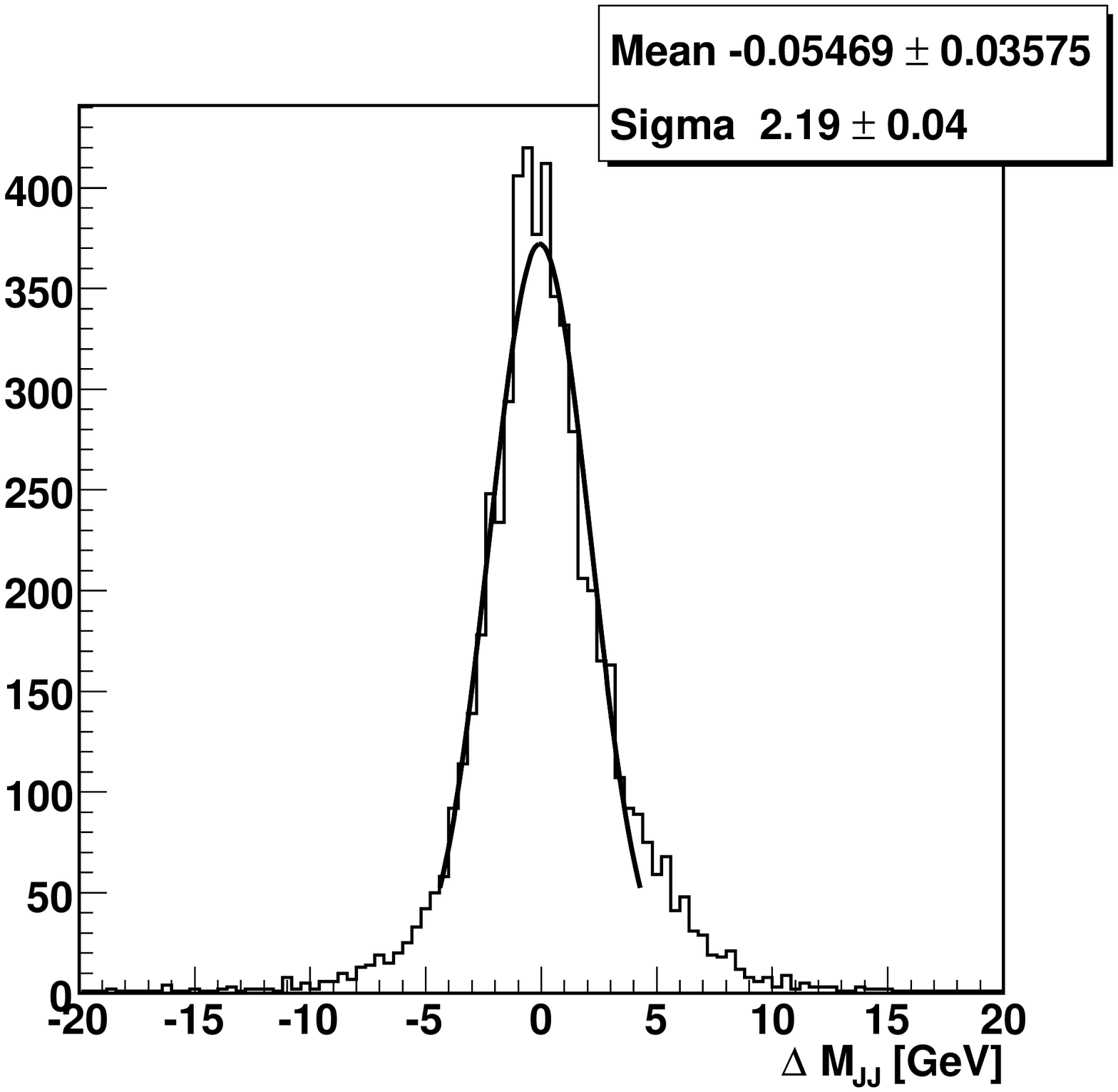,height=5cm}}
% <dmh> = -0.055 +/- 0.036
% sigma = 2.19 +/- 0.04
\end{minipage}
\hfill
}
\caption[Fig:dmnuh]{\label{Fig:dmnuh} \small \it
Distributions of
(a) energies of neutrinos from $H^0$ decays,
(b) the same as Fig.$\,$\ref{Fig:dmh} c) but 4-momenta of neutrinos from the
$H^0$ decays artificially added in.
(c) the same as (b) but with the other undetected particles also added in.
}
\end{minipage}\end{center}
\end{figure}

Let us now examine the neutrino effect.
Fig.$\,$\ref{Fig:dmnuh} a) shows the energy distribution
of neutrinos from $H^0$ decays.
%
% ------------------
%  Fig.17
% ------------------
%
Using the MC truths from the history keeper,
we can add the 4-momenta of these neutrinos 
and obtain Fig.$\,$\ref{Fig:dmnuh} b).
The huge lower tail disappeared as expected and the apparent width
now reduces to 
$\sigma_{\Delta M} = 2.40\,$GeV ($\sqrt{2.90^2-2.40^2}=1.63\,$GeV: 21\% reduction).
If we further add in the remaining undetected particles escaping into the acceptance holes,
we have Fig.$\,$\ref{Fig:dmnuh} c), yielding 
$\sigma_{\Delta M} = 2.19\,$GeV ($\sqrt{2.40^2-2.19^2}=0.98\,$GeV: 7.7\% reduction).
This remnant apparent width corresponding to $(2.19/3.54)^2=38\%$ 
of the initial apparent width of Fig.$\,$\ref{Fig:dmnuh} a) and
is consistent with that naively expected from the $Z^0$ boson result
in Fig.$\,$\ref{Fig:edepzz}.
Notice also that the mean value of this distribution is consistent with zero.

\subsection{Fundamental Limitations on the PFA Performance}

% On pole study
%   Kink
%   V0
%   Resolution (TK, EM, HD)
%   double counting (mostly 2ndary particles from materials)
%   acceptance hole
% 	----------------------------------------
% 	Table of contribution breakup
% 	----------------------------------------
%
%						* = tk from indirect method
%
%			Indirect		Direct(PBP)	Direct(BP)		Expected		DC indirect
%	--------------------------------------------------------------------------------------------------------------------
%	TK		0.39 (3%)		0.190		0.150		
%	EM		1.02 (21%)	0.846		0.703		0.71			0.570 (6.7%)
%	HD		1.73 (62%)	1.520		1.410		1.46			0.826 (14%)
%	--------------------------------------------------------------------------------------------------------------------
%	Subtotal	2.05 (86%)	1.75	(1.78*)	1.58 (1.62)
%	--------------------------------------------------------------------------------------------------------------------
%	DC		0.45 (4%)
%	AH		0.70 (10%)	0.81 (RMS)	
%	--------------------------------------------------------------------------------------------------------------------
%	Total		2.20 (100%)	2.00
%
%	acceptance hole from Pt cut of 0.23 GeV for TPC 
% 
% Energy Dependence
%   need for sophisticated calorimeter calibration 
%	sigma_E ~ 0.23 x sqrt(E)
%	slight increase in sigma_{HD} --> calibration problem
%
% Boost Dependence
%   mass energy of 2ndary protons and PID
%
% ZH Events
%   ISR photons
%   neutrinos
%   Real PFA
%     hit merging
%     mis-clustering
%     mis-linking 
% Problem with jet clustering

We have been examining the contributors to the jet energy resolution with the CPFA,
taking $e^+e^ \to q \bar{q}$, $Z^0 Z^0$, and $Z^0 H^0$ processes as benchmarks.
Let us now summarize the results and discuss their implications from the view point of
fundamental limits on the PFA performance.
In our on-pole $Z^0$ study we have emphasized the importance of appropriate treatment of kinks
and demonstrated that the kink mother scheme gives the best resolution in spite of possible
energy double counting.
If the neutral decay daughters can be identified by testing if their shower axes successfully
extrapolate to their corresponding kink positions,
the resolution could in principle be further improved
by removing such neutral decay daughter PFOs.
As compared with the dependence on the kink treatment schemes, 
the mistreatment of V$^0$s turned out to be less harmful,
making only a small difference to the jet energy resolution.
\\

Table \ref{Tab:breakup} summaries the contributions from various factors
to the jet energy resolution on the $Z^0$ pole
as estimated with the indirect ($\sigma_{\rm indir}$), the direct ($\sigma_{\rm dir}$), and 
the detector resolutions for single particles ($\sigma_{\rm det}$). 
\begin{table}[htb]
\begin{center}\begin{minipage}{\figurewidth}
\caption{\label{Tab:breakup} \small \it
Breakup of the jet energy resolution at the $Z^0$ pole into various contributors with
the indirect method ($\sigma_{\rm indir}$), that with the direct method using primary
break point information ($\sigma_{\rm dir}$),
and that expected from single particle detector resolution ($\sigma_{\rm det}$): 
TK, EM, HD, DC, and AH stand for the charged PFO contribution,
that from neutral electromagnetic PFOs, that from neutral hadronic PFOs,
that from double counted energies, and that from undetected primary particles due to
the acceptance holes of the detector.
Notice that the TK contribution obtained with the indirect method (marked by * in the table)
is the width of the sharp peak and does not reflect the width of the underlying broad component.
For the AH contribution with the direct method the r.m.s. value is shown, since the
distribution is far from being Gaussian.
}
\begin{center}
\begin{tabular}{| l | c r | c | c |} \hline
%~~~ & \multicolumn{2}{c |}{Indirect [GeV]} & Direct [GeV] & Expected [GeV]  \cr
~~~ & \multicolumn{2}{c |}{$\sigma_{\rm indir}$ [GeV]} & $\sigma_{\rm dir} [GeV]$ & $\sigma_{\rm det}$ [GeV]  \cr
\hline \hline
%$\sigma_{\rm TK}$ & 0.39 & 3\%    & 0.19 &  $< 0.1$ \cr
%$\sigma_{\rm EM}$ & 1.02 & 21\% & 0.85 & 0.70 \cr
%$\sigma_{\rm HD}$ & 1.73 & 62\% & 1.52   &  1.46 \cr
TK & 0.39 & 3\%    & 0.19$^*$ &  $-$ \cr
EM & 1.02 & 21\% & 0.85 & 0.70 \cr
HD & 1.73 & 62\% & 1.52   &  1.46 \cr
\hline \hline
subtotal & 2.05 & 86\%  & 1.83 & $-$ \cr
\hline \hline
%$\sigma_{\rm DC}$ & 0.45 & 4\% & - & \cr
%$\sigma_{\rm AH}$ & 0.70 & 10\% & 0.81 (rms) & \cr
DC & 0.45 & 4\% & $-$ & $-$ \cr
AH & 0.70 & 10\% & 0.81 (rms) & $-$ \cr
\hline \hline
total & 2.20 & 100\% & 2.00 & $-$ \cr
\hline
\end{tabular}
\end{center}
\end{minipage}\end{center}
\end{table}
The detector resolution contributions estimated with the indirect method include those from
the fluctuations of the double counted energies.
Accordingly, the contribution from the double counted energies of $0.45\,$GeV
in the indirect method does not include their fluctuations due to finite
detector resolutions.
This is the main reason why the indirect method gives larger values than the direct
method where the double counting is avoided by locking daughter PFOs.
On the other hand the difference between the values obtained with the direct method
and the naive estimates from the single particle resolution can be attributed to
the fact that the direct method compares the measured PFO energies with 
the energies of the corresponding primary particles which might decay before detection.
Some part of the daughter particles might escape into acceptance holes
or some part of the secondary particles might be protons kicked out from the detector
materials whose mass energies should have been subtracted.
\\

As we discussed in our $e^+e^- \to Z^0 Z^0$ study, the effect of the mass energy correction
is quite dramatic in particular when the parent $Z^0$ is highly boosted.
Without the mass energy correction, the reconstructed $Z^0$ invariant mass distribution has
a significant tail on the higher mass side.
Since the production points of the secondary protons from the detector materials show
a clear image of the material distribution in the detector, 
there is a possibility to do this mass energy correction, provided that we can
identify protons using, for instance, the energy loss information from the TPC.
\\

With these corrections made, the jet energy resolution can be 
given by the sum of the contribution from the detector resolutions and
the contribution from the undetected particles.
Among the detector resolutions the calorimeter resolutions, the HD contribution in particular, 
dominate obviously.
As mentioned earlier we have used a rather unsophisticated calibration method for
the calorimeters.
It is therefore of prime importance to optimize the calorimeter configuration
and to improve calibration methods.
Optimization of the calorimeter configuration and development of calibration methods
are, however, beyond the scope of this paper.
\\

Notice that the observations made above are from the Monte Carlo simulations with
complications like initial state radiations and neutrinos from heavy quark flavors switched off.
When these effects are turned on the situation becomes significantly more difficult
as demonstrated for the $e^+e^- \to Z^0 H^0$ process.
The missing neutrinos induce a huge tail on the lower side of the Higgs mass peak
and dominate the other contributors.
We also pointed out the existence of a higher tail caused by photons from initial state radiations.
Separating high energy isolated photons from initial state radiations is hence important.
\\

In any real PFA, we will have effects of cluster overlapping and subsequent confusion 
which will make significant additional contributors and might dominate the rest.
Actual development of a real PFA is beyond the scope of this paper but the results shown
here should set a clear goal for any real PFA.

\section{Summary and Conclusions}
\label{Sec:conclusions}

We have developed a set of C++ classes that work with Geant4
and facilitate history keeping of particle tracks thereby linking
calorimeter hits to their ancestor particles.
Using this software tool we have studied the fundamental limits
on the PFA performance that remain even with
a perfect particle flow algorithm.

We have shown that the jet energy resolution with a perfect PFA
can be estimated once we know the detector resolutions obtained from single particle performance studies
and the effect of undetected energies due to acceptance holes and neutrinos.
It should be emphasized, however, that this is true only after eliminating double counted energies
due to secondary PFOs and subtracting mass energies from the proton PFOs kicked out from
the detector materials.
The former requires the precise pointing of shower axes of neutral PFOs to  their production points
and the latter necessitates particle identifications of secondary protons kicked out from the detector materials,
both of which are challenging.

\section*{Acknowledgments}

The authors would like to thank T.~Yoshioka, H.~Ono, T.~Takeshita, 
and other members of the JLC-Software group
for useful discussions and helps.
%they are very grateful to
%Y.Nakamura for his great contribution in gas gain measurements,
%N.Ishihara and T.Ohama for useful discussions,
%and S.Iwata for continuous encouragements and supports.
This work was supported in part by the Creative Scientific Reserach 
Grant No. 18GS0202 of the Japan Society for Promotion of Science (JSPS)
and the JSPS Core University Program.


\begin{thebibliography}{99}

\bibitem{Ref:ILC}  http://www.linearcollider.org/ and references therein.

\bibitem{Ref:PFA} 
%Pandora PFA, 
M.A.Thomson, arXiv:physics/060726;
%WolfPFA, 
V.Morgunov and A.Raspereza, arXiv:physics/0412108;
%GLD-PFA, 
T.Yoshioka, ECONF C0508141:ALCPG1711,2005;
%Argonne-PFA, 
S.Magill and S.Kuhlmann, SLAC-PUB-12203,
in the Proceedings of 2005 International Linear Collider Workshop (LCWS 2005),
Stanford, California, 18-22 Mar 2005, pp 1015.

\bibitem{Ref:JLC-I} JLC group, KEK Report 92-16, December, 1992.

\bibitem{Ref:acfa}
ACFA Linear Collider Working Group, KEK Report 2001-11, August (2001),\\ 
http://www-jlc.kek.jp/subg/offl/jim/index-e.html.

\bibitem{Ref:Jupiter}
Proceedings of the APPI Winter Institute, KEK Proceedings 2002-08, July (2002).

\bibitem{Ref:geant4toolkit}
 http://wwwinfo.cern.ch/asd/geant4/G4UsersDocuments/\\
 UsersGuides/ForToolkitDeveloper/html/index.html
 
 \bibitem{Ref:JSF}
 http://acfahep.kek.jp/subg/sim/simtools/index.html.
 
 \bibitem{Ref:root}
 http://root.cern.ch/.
 
\bibitem{Ref:gld}
GLD Detector Outline Document, July (2006), arXiv:physics/0607154v1

\bibitem{Ref:pythia}
 http://www.thep.lu.se/\~{}torbjorn/Pythia.html

\end{thebibliography}
\end{document}